\documentclass[10pt,final,journal,twocolumn,letterpaper]{IEEEtran}

\usepackage{cite}
\ifCLASSINFOpdf
   \usepackage[pdftex]{graphicx}
\else
   \usepackage[dvips]{graphicx}
\fi

\usepackage{amsmath}
\allowdisplaybreaks[4]

\usepackage{array}

\ifCLASSOPTIONcompsoc
  \usepackage[caption=false,font=normalsize,labelfont=sf,textfont=sf]{subfig}
\else
  \usepackage[caption=false,font=footnotesize]{subfig}
\fi

\usepackage{url}

\usepackage[algoruled,linesnumbered]{algorithm2e}
\usepackage{url}
\usepackage{amsthm}
\usepackage{amssymb}
\usepackage{relsize}
\usepackage{mathtools}
\usepackage{cases}
\usepackage{booktabs}
\usepackage{multirow}
\usepackage{color}
\usepackage{graphics} 
\usepackage{epsfig} 
\usepackage{adjustbox}
\usepackage{blindtext}
\usepackage{graphicx}
\usepackage[justification=centering]{caption}
\DeclareMathOperator*{\argmax}{arg\,max}
\DeclareMathOperator*{\argmin}{arg\,min}

\begin{document}
%
\title{Thirty Years of Machine Learning:\\ The Road to Pareto-Optimal Wireless Networks}

\author{Jingjing~Wang,~\IEEEmembership{Member,~IEEE}, Chunxiao~Jiang,~\IEEEmembership{Senior Member,~IEEE},\\ Haijun~Zhang,~\IEEEmembership{Senior Member,~IEEE}, Yong~Ren,~\IEEEmembership{Senior Member,~IEEE},\\ Kwang-Cheng~Chen,~\IEEEmembership{Fellow,~IEEE,} and Lajos~Hanzo,~\IEEEmembership{Fellow,~IEEE}
\thanks{This work is partly supported by the National Natural Science Foundation of China (61922050), the Pre-research Fund of Equipments of Ministry of Education of China (6141A02022615), the Research Fund of China Academy of Space Technology (Co/Co-20180605-47), and also partly supported by the National Natural Science Foundation of China (61822104, 61771044), the Fundamental Research Funds for the Central Universities (RC1631, FRF-TP-19-002C1). Dr.~Wang would like to acknowledge the financial support of the Shuimu Tsinghua Scholar Program. (Corresponding author: Chunxiao Jiang)}
\thanks{J. Wang and Y. Ren are with the Department
of Electronic Engineering, Tsinghua University, Beijing, 100084, China. E-mail: chinaeephd@gmail.com, reny@tsinghua.edu.cn.}
\thanks{C. Jiang is with Tsinghua Space Center, Tsinghua University, Beijing, 100084, China. E-mail: jchx@tsinghua.edu.cn.}
\thanks{H. Zhang is with Institute of Artificial Intelligence, Beijing Advanced Innovation Center for Materials Genome Engineering, Beijing Engineering and Technology Research Center for Convergence Networks and Ubiquitous Services, University of Science and Technology Beijing, Beijing 100083, China. E-mail: haijunzhang@ieee.org.}
\thanks{K.-C. Chen is with the Electrical Engineering Department, University of South Florida, Tampa, FL 33620, USA. Email: kwangcheng@usf.edu.}
\thanks{L. Hanzo is with the School of Electronics and Computer Science, University of Southampton, Southampton, SO17 1BJ, UK. Email: lh@ecs.soton.ac.uk.}

}
\maketitle
\vspace{-1cm}
\begin{abstract}
Future wireless networks have a substantial potential in terms of supporting a broad range of complex compelling applications both in military and civilian fields, where the users are able to enjoy high-rate, low-latency, low-cost and reliable information services. Achieving this ambitious goal requires new radio techniques for adaptive learning and intelligent decision making because of the complex heterogeneous nature of the network structures and wireless services. Machine learning (ML) algorithms have great success in supporting big data analytics, efficient parameter estimation and interactive decision making. Hence, in this article, we review the thirty-year history of ML by elaborating on supervised learning, unsupervised learning, reinforcement learning and deep learning. Furthermore, we investigate their employment in the compelling applications of wireless networks, including heterogeneous networks (HetNets), cognitive radios (CR), Internet of things (IoT), machine to machine networks (M2M), and so on. This article aims for assisting the readers in clarifying the motivation and methodology of the various ML algorithms, so as to invoke them for hitherto unexplored services as well as scenarios of future wireless networks.
\end{abstract}

\section*{Nomenclature}
\noindent
\begin{tabular}{>{\raggedright}p{1.5cm}>{\raggedright}p{7cm}}
5G  & The 5th Generation Mobile Network \tabularnewline
\end{tabular}
\begin{tabular}{>{\raggedright}p{1.5cm}>{\raggedright}p{7cm}}
AI  & Artificial Intelligent     \tabularnewline
\end{tabular}
\begin{tabular}{>{\raggedright}p{1.5cm}>{\raggedright}p{7cm}}
AMC  & Automatic Modulation Classification \tabularnewline
\end{tabular}
\begin{tabular}{>{\raggedright}p{1.5cm}>{\raggedright}p{7cm}}
ANN  &  Artificial Neural Network  \tabularnewline
\end{tabular}
\begin{tabular}{>{\raggedright}p{1.5cm}>{\raggedright}p{7cm}}
AP  & Access Point     \tabularnewline
\end{tabular}
\begin{tabular}{>{\raggedright}p{1.5cm}>{\raggedright}p{7cm}}
AWGN  & Additive White Gaussian Noise     \tabularnewline
\end{tabular}
\begin{tabular}{>{\raggedright}p{1.5cm}>{\raggedright}p{7cm}}
BBU  & BaseBand processing Unit      \tabularnewline
\end{tabular}
\begin{tabular}{>{\raggedright}p{1.5cm}>{\raggedright}p{7cm}}
BS  &  Base Station    \tabularnewline
\end{tabular}
\begin{tabular}{>{\raggedright}p{1.5cm}>{\raggedright}p{7cm}}
CDF  & Cumulative Distribution Function     \tabularnewline
\end{tabular}
\begin{tabular}{>{\raggedright}p{1.5cm}>{\raggedright}p{7cm}}
CNN  & Convolutional Neural Network     \tabularnewline
\end{tabular}
\begin{tabular}{>{\raggedright}p{1.5cm}>{\raggedright}p{7cm}}
CogNet & Cognitive Network \tabularnewline
\end{tabular}
\begin{tabular}{>{\raggedright}p{1.5cm}>{\raggedright}p{7cm}}
CoMP  & Coordinated Multiple Points     \tabularnewline
\end{tabular}
\begin{tabular}{>{\raggedright}p{1.5cm}>{\raggedright}p{7cm}}
CR & Cognitive Radio \tabularnewline
\end{tabular}
\begin{tabular}{>{\raggedright}p{1.5cm}>{\raggedright}p{7cm}}
C-RAN & Cloud Radio Access Network \tabularnewline
\end{tabular}
\begin{tabular}{>{\raggedright}p{1.5cm}>{\raggedright}p{7cm}}
CRN & Cognitive Radio Network \tabularnewline
\end{tabular}
\begin{tabular}{>{\raggedright}p{1.5cm}>{\raggedright}p{7cm}}
CSI & Channel State Information     \tabularnewline
\end{tabular}
\begin{tabular}{>{\raggedright}p{1.5cm}>{\raggedright}p{7cm}}
CSMA/CA  & Carrier-Sense Multiple Access with Collision Avoidance     \tabularnewline
\end{tabular}
\begin{tabular}{>{\raggedright}p{1.5cm}>{\raggedright}p{7cm}}
CSMA/CD  & Carrier-Sense Multiple Access with Collision Detection     \tabularnewline
\end{tabular}
\begin{tabular}{>{\raggedright}p{1.5cm}>{\raggedright}p{7cm}}
C-S Mode & Client-Server Mode     \tabularnewline
\end{tabular}
\begin{tabular}{>{\raggedright}p{1.5cm}>{\raggedright}p{7cm}}
D2D & Device to Device  \tabularnewline
\end{tabular}
\begin{tabular}{>{\raggedright}p{1.5cm}>{\raggedright}p{7cm}}
DBN  &Deep Belief Network      \tabularnewline
\end{tabular}
\begin{tabular}{>{\raggedright}p{1.5cm}>{\raggedright}p{7cm}}
DNN  &Deep Neural Network      \tabularnewline
\end{tabular}
\begin{tabular}{>{\raggedright}p{1.5cm}>{\raggedright}p{7cm}}
DQN  &Deep Q-Network     \tabularnewline
\end{tabular}
\begin{tabular}{>{\raggedright}p{1.5cm}>{\raggedright}p{7cm}}
EA  & Energy Awareness      \tabularnewline
\end{tabular}
\begin{tabular}{>{\raggedright}p{1.5cm}>{\raggedright}p{7cm}}
EE  & Energy Efficiency      \tabularnewline
\end{tabular}
\begin{tabular}{>{\raggedright}p{1.5cm}>{\raggedright}p{7cm}}
EH & Energy Harvesting     \tabularnewline
\end{tabular}
\begin{tabular}{>{\raggedright}p{1.5cm}>{\raggedright}p{7cm}}
ELP  & Exponentially-weighted algorithm with Linear Programming      \tabularnewline
\end{tabular}
\begin{tabular}{>{\raggedright}p{1.5cm}>{\raggedright}p{7cm}}
EM  & Expectation Maximization      \tabularnewline
\end{tabular}
\begin{tabular}{>{\raggedright}p{1.5cm}>{\raggedright}p{7cm}}
eMBB  & enhanced Mobile Broad Band     \tabularnewline
\end{tabular}
\begin{tabular}{>{\raggedright}p{1.5cm}>{\raggedright}p{7cm}}
ERM & Empirical Risk Minimization     \tabularnewline
\end{tabular}
\begin{tabular}{>{\raggedright}p{1.5cm}>{\raggedright}p{7cm}}
EXP3  & EXPonential weights for EXPloration and EXPloitation      \tabularnewline
\end{tabular}
\begin{tabular}{>{\raggedright}p{1.5cm}>{\raggedright}p{7cm}}
FANET & Flying Ad Hoc Network    \tabularnewline
\end{tabular}
\begin{tabular}{>{\raggedright}p{1.5cm}>{\raggedright}p{7cm}}
FDA & Fisher Discriminant Analysis    \tabularnewline
\end{tabular}
\begin{tabular}{>{\raggedright}p{1.5cm}>{\raggedright}p{7cm}}
FDI & False Data Injection    \tabularnewline
\end{tabular}
\begin{tabular}{>{\raggedright}p{1.5cm}>{\raggedright}p{7cm}}
FSMC & Finite State Markov Channel    \tabularnewline
\end{tabular}
\begin{tabular}{>{\raggedright}p{1.5cm}>{\raggedright}p{7cm}}
GMM & Gaussian Mixture Model    \tabularnewline
\end{tabular}
\begin{tabular}{>{\raggedright}p{1.5cm}>{\raggedright}p{7cm}}
HetNet & Heterogeneous Network \tabularnewline
\end{tabular}
\begin{tabular}{>{\raggedright}p{1.5cm}>{\raggedright}p{7cm}}
HMM & Hidden Markov Model    \tabularnewline
\end{tabular}
\begin{tabular}{>{\raggedright}p{1.5cm}>{\raggedright}p{7cm}}
ICA & Independent Component Analysis     \tabularnewline
\end{tabular}
\begin{tabular}{>{\raggedright}p{1.5cm}>{\raggedright}p{7cm}}
IEEE  & Institute of Electrical and Electronics Engineers     \tabularnewline
\end{tabular}
\begin{tabular}{>{\raggedright}p{1.5cm}>{\raggedright}p{7cm}}
IoT & Internet of Things \tabularnewline
\end{tabular}
\begin{tabular}{>{\raggedright}p{1.5cm}>{\raggedright}p{7cm}}
ITS & Intelligent Transportation System     \tabularnewline
\end{tabular}
\begin{tabular}{>{\raggedright}p{1.5cm}>{\raggedright}p{7cm}}
KNN  & K-Nearest Neighbors     \tabularnewline
\end{tabular}
\begin{tabular}{>{\raggedright}p{1.5cm}>{\raggedright}p{7cm}}
LED  & Light Emitting Diode     \tabularnewline
\end{tabular}
\begin{tabular}{>{\raggedright}p{1.5cm}>{\raggedright}p{7cm}}
LOS  & Line of Sight     \tabularnewline
\end{tabular}
\begin{tabular}{>{\raggedright}p{1.5cm}>{\raggedright}p{7cm}}
LS  & Least Square     \tabularnewline
\end{tabular}
\begin{tabular}{>{\raggedright}p{1.5cm}>{\raggedright}p{7cm}}
LSTM  & Long Short Term Memory \tabularnewline
\end{tabular}
\begin{tabular}{>{\raggedright}p{1.5cm}>{\raggedright}p{7cm}}
LTE  & Long Term Evolution     \tabularnewline
\end{tabular}
\begin{tabular}{>{\raggedright}p{1.5cm}>{\raggedright}p{7cm}}
M2M &  Machine to Machine    \tabularnewline
\end{tabular}
\begin{tabular}{>{\raggedright}p{1.5cm}>{\raggedright}p{7cm}}
MANET  & Mobile Ad Hoc Network     \tabularnewline
\end{tabular}
\begin{tabular}{>{\raggedright}p{1.5cm}>{\raggedright}p{7cm}}
MAP & Maximum a Posteriori      \tabularnewline
\end{tabular}
\begin{tabular}{>{\raggedright}p{1.5cm}>{\raggedright}p{7cm}}
MDP  & Markov Decision Process     \tabularnewline
\end{tabular}
\begin{tabular}{>{\raggedright}p{1.5cm}>{\raggedright}p{7cm}}
MIMO  & Multiple-Input and Multiple-Output      \tabularnewline
\end{tabular}
\begin{tabular}{>{\raggedright}p{1.5cm}>{\raggedright}p{7cm}}
ML & Machine Learning      \tabularnewline
\end{tabular}
\begin{tabular}{>{\raggedright}p{1.5cm}>{\raggedright}p{7cm}}
MLE  & Maximum Likelihood Estimation     \tabularnewline
\end{tabular}
\begin{tabular}{>{\raggedright}p{1.5cm}>{\raggedright}p{7cm}}
mMTC  & massive Machine Type of Communication     \tabularnewline
\end{tabular}
\begin{tabular}{>{\raggedright}p{1.5cm}>{\raggedright}p{7cm}}
NB-IoT  & NarrowBand Internet of Things     \tabularnewline
\end{tabular}
\begin{tabular}{>{\raggedright}p{1.5cm}>{\raggedright}p{7cm}}
NB-M2M  & NarrowBand Machine to Machine     \tabularnewline
\end{tabular}
\begin{tabular}{>{\raggedright}p{1.5cm}>{\raggedright}p{7cm}}
NFV  & Network Function Virtualization     \tabularnewline
\end{tabular}
\begin{tabular}{>{\raggedright}p{1.5cm}>{\raggedright}p{7cm}}
NLOS  & Non-Line of Sight     \tabularnewline
\end{tabular}
\begin{tabular}{>{\raggedright}p{1.5cm}>{\raggedright}p{7cm}}
NOMA & Non-Orthogonal Multiple Access \tabularnewline
\end{tabular}
\begin{tabular}{>{\raggedright}p{1.5cm}>{\raggedright}p{7cm}}
OFDM & Orthogonal Frequency Division Multiplexing \tabularnewline
\end{tabular}
\begin{tabular}{>{\raggedright}p{1.5cm}>{\raggedright}p{7cm}}
OSPF & Open Shortest Path First  \tabularnewline
\end{tabular}
\begin{tabular}{>{\raggedright}p{1.5cm}>{\raggedright}p{7cm}}
P2P & Peer to Peer     \tabularnewline
\end{tabular}
\begin{tabular}{>{\raggedright}p{1.5cm}>{\raggedright}p{7cm}}
PCA & Principal Component Analysis     \tabularnewline
\end{tabular}
\begin{tabular}{>{\raggedright}p{1.5cm}>{\raggedright}p{7cm}}
POMDP  & \small{Partially Observable Markov Decision Process}     \tabularnewline
\end{tabular}
\begin{tabular}{>{\raggedright}p{1.5cm}>{\raggedright}p{7cm}}
PU  & Primary User      \tabularnewline
\end{tabular}
\begin{tabular}{>{\raggedright}p{1.5cm}>{\raggedright}p{7cm}}
QoE & Quality of Experience     \tabularnewline
\end{tabular}
\begin{tabular}{>{\raggedright}p{1.5cm}>{\raggedright}p{7cm}}
QoS &  Quality of Service    \tabularnewline
\end{tabular}
\begin{tabular}{>{\raggedright}p{1.5cm}>{\raggedright}p{7cm}}
RAT  & Radio Access Technology     \tabularnewline
\end{tabular}
\begin{tabular}{>{\raggedright}p{1.5cm}>{\raggedright}p{7cm}}
RBM & Restricted Boltzmann Machine    \tabularnewline
\end{tabular}
\begin{tabular}{>{\raggedright}p{1.5cm}>{\raggedright}p{7cm}}
RBF & Radial Basis Function    \tabularnewline
\end{tabular}
\begin{tabular}{>{\raggedright}p{1.5cm}>{\raggedright}p{7cm}}
RFID  & Radio Frequency IDentification     \tabularnewline
\end{tabular}
\begin{tabular}{>{\raggedright}p{1.5cm}>{\raggedright}p{7cm}}
RNN  & Recurrent Neural Network     \tabularnewline
\end{tabular}
\begin{tabular}{>{\raggedright}p{1.5cm}>{\raggedright}p{7cm}}
RRU  & Remote Radio Unit     \tabularnewline
\end{tabular}
\begin{tabular}{>{\raggedright}p{1.5cm}>{\raggedright}p{7cm}}
SDA  & Stacked Denoising Auto-encoder     \tabularnewline
\end{tabular}
\begin{tabular}{>{\raggedright}p{1.5cm}>{\raggedright}p{7cm}}
SDN  &  Software Defined Network    \tabularnewline
\end{tabular}
\begin{tabular}{>{\raggedright}p{1.5cm}>{\raggedright}p{7cm}}
SDR  & Software Defined Radio    \tabularnewline
\end{tabular}
\begin{tabular}{>{\raggedright}p{1.5cm}>{\raggedright}p{7cm}}
SE  & Spectrum Efficiency     \tabularnewline
\end{tabular}
\begin{tabular}{>{\raggedright}p{1.5cm}>{\raggedright}p{7cm}}
SG & Stochastic Geometry \tabularnewline
\end{tabular}
\begin{tabular}{>{\raggedright}p{1.5cm}>{\raggedright}p{7cm}}
SRM & Structural Risk Minimization \tabularnewline
\end{tabular}
\begin{tabular}{>{\raggedright}p{1.5cm}>{\raggedright}p{7cm}}
STBC  & Space Time Block Code     \tabularnewline
\end{tabular}
\begin{tabular}{>{\raggedright}p{1.5cm}>{\raggedright}p{7cm}}
SU  & Secondary User     \tabularnewline
\end{tabular}
\begin{tabular}{>{\raggedright}p{1.5cm}>{\raggedright}p{7cm}}
SVM & Support Vector Machine \tabularnewline
\end{tabular}
\begin{tabular}{>{\raggedright}p{1.5cm}>{\raggedright}p{7cm}}
TAS & Transmit Antenna Selection      \tabularnewline
\end{tabular}
\begin{tabular}{>{\raggedright}p{1.5cm}>{\raggedright}p{7cm}}
TCP  & Transmission Control Protocol      \tabularnewline
\end{tabular}
\begin{tabular}{>{\raggedright}p{1.5cm}>{\raggedright}p{7cm}}
TD & Temporal Difference     \tabularnewline
\end{tabular}
\begin{tabular}{>{\raggedright}p{1.5cm}>{\raggedright}p{7cm}}
TOA  & Time of Arrival     \tabularnewline
\end{tabular}
\begin{tabular}{>{\raggedright}p{1.5cm}>{\raggedright}p{7cm}}
UAV  & Unmanned Aerial Vehicle \tabularnewline
\end{tabular}
\begin{tabular}{>{\raggedright}p{1.5cm}>{\raggedright}p{7cm}}
UDN & Ultra Dense Network   \tabularnewline
\end{tabular}
\begin{tabular}{>{\raggedright}p{1.5cm}>{\raggedright}p{7cm}}
uRLLC  & ultra-Reliable Low-Latency Communication     \tabularnewline
\end{tabular}
\begin{tabular}{>{\raggedright}p{1.5cm}>{\raggedright}p{7cm}}
V2I  & Vehicle to Infrastructure    \tabularnewline
\end{tabular}
\begin{tabular}{>{\raggedright}p{1.5cm}>{\raggedright}p{7cm}}
V2V  & Vehicle to Vehicle     \tabularnewline
\end{tabular}
\begin{tabular}{>{\raggedright}p{1.5cm}>{\raggedright}p{7cm}}
V2X  & Vehicle to Everything     \tabularnewline
\end{tabular}
\begin{tabular}{>{\raggedright}p{1.5cm}>{\raggedright}p{7cm}}
VANET  & Vehicular Ad Hoc Network     \tabularnewline
\end{tabular}
\begin{tabular}{>{\raggedright}p{1.5cm}>{\raggedright}p{7cm}}
VLC  & Visible Light Communication   \tabularnewline
\end{tabular}
\begin{tabular}{>{\raggedright}p{1.5cm}>{\raggedright}p{7cm}}
VR  & Virtual Reality   \tabularnewline
\end{tabular}
\begin{tabular}{>{\raggedright}p{1.5cm}>{\raggedright}p{7cm}}
WANET  & Wireless Ad Hoc Network \tabularnewline
\end{tabular}
\begin{tabular}{>{\raggedright}p{1.5cm}>{\raggedright}p{7cm}}
WBAN  & Wireless Body Area Network \tabularnewline
\end{tabular}
\begin{tabular}{>{\raggedright}p{1.5cm}>{\raggedright}p{7cm}}
WLAN  & Wireless Local Area Network \tabularnewline
\end{tabular}
\begin{tabular}{>{\raggedright}p{1.5cm}>{\raggedright}p{7cm}}
WiMAX  & Worldwide Interoperability for Microwave Access     \tabularnewline
\end{tabular}
\begin{tabular}{>{\raggedright}p{1.5cm}>{\raggedright}p{7cm}}
Wi-Fi  & Wireless Fidelity     \tabularnewline
\end{tabular}
\begin{tabular}{>{\raggedright}p{1.5cm}>{\raggedright}p{7cm}}
WMAN  & Wireless Metropolitan Area Network \tabularnewline
\end{tabular}
\begin{tabular}{>{\raggedright}p{1.5cm}>{\raggedright}p{7cm}}
WPAN  & Wireless Personal Area Network \tabularnewline
\end{tabular}
\begin{tabular}{>{\raggedright}p{1.5cm}>{\raggedright}p{7cm}}
WSN & Wireless Sensor Network      \tabularnewline
\end{tabular}
\begin{tabular}{>{\raggedright}p{1.5cm}>{\raggedright}p{7cm}}
WWAN  & Wireless Wide Area Network     \tabularnewline
\end{tabular}

\section{Introduction}
\label{Introduction}
\IEEEPARstart{W}{ireless} networks have supported a variety of military services, intelligent transportation, healthcare, etc. To elaborate briefly, next-generation mobile networks are expected to support high date rate communication~\cite{agiwal2016next}. As a complement, wireless sensor networks (WSN) support sustained monitoring in unmanned or hostile environments relying on widely dispersed operating sensors~\cite{rawat2014wireless}. Furthermore, the popular Wi-Fi network provides convenient Internet access for various devices in indoor scenarios~\cite{deng2017ieee}. With the rapid proliferation of portable mobile devices and the demand for a high quality of service (QoS) and quality of experience (QoE), future wireless networks will continue to support a broad range of compelling applications, where the users benefit from high-rate, low-latency, low-cost and reliable information services.

\subsection{Motivation}

In contrast to the conventional wireless networks, future wireless networks have the following evolutionary tendency~\cite{larsson2014massive,kalil2017framework}:
\begin{itemize}
  \item \emph{Network Scale}: The network is associated with a tremendous network size including all kinds of entities, each of which has different service capabilities as well as requirements. Furthermore, interactions among these entities result in a diverse variety of traffic, such as text, voice, audio, images, video, etc.
  \item \emph{Network Structure}: On one hand, the future wireless network tends to have a self-configuring element, where each entity cooperatively completes tasks. This characteristic is termed as ``being ad hoc''. On the other hand, the future wireless network is heterogeneous and hierarchical, having different network slices\footnote{In our paper, network slices are multiple logical networks running on the top of a shared physical network infrastructure and operated by a control center.}. Furthermore, the mobility of entities results in a complex time-variant network structure, which requires dynamic time-space association.
  \item \emph{Network Control}: Future wireless networks facilitate convenient reconfiguration by software-based network management, hence improving network flexibility and efficiency.
\end{itemize}

Machine learning (ML) was first introduced as a popular technique of realizing artificial intelligence in the late 1950's~\cite{samuel1959some}. ML algorithms can learn from training data without being explicitly programmed. It is beneficial for classification/regression, prediction, clustering and decision making~\cite{alpaydin2014introduction,michalski2013machine,jordan2015machine}, whilst relying on the following three basic elements~\cite{vapnik1999overview}:
\begin{itemize}
  \item \emph{Model}: Mathematical or signal models are constructed from training data and expert knowledge, in order to statistically describe the characteristics of the given data set. Then again, relying on these trained models, ML can be used for classification, prediction and decision making. In case the appropriate models are not available, techniques on the feature extraction or knowledge discovery can be developed to achieve the same goal.
  \item \emph{Strategy}: The criteria used for training mathematical models are called strategies. How to select an appropriate strategy is closely associated with training data. Empirical risk minimization~\cite{vapnik1992principles} and structural risk minimization~\cite{guyon1992structural} constitute a pair of fundamental strategies, where the latter can beneficially avoid the notorious ``over-fitting'' phenomenon.
  \item \emph{Algorithm}: Algorithms are constructed to find solutions based on predetermined model and strategy selected, which can be viewed as an optimization process. A powerful algorithm can find a globally optimal solution with high probability at a low computational complexity and storage.
\end{itemize}

In the last thirty years, ML has been successfully applied to the field of computer vision~\cite{sebe2005machine}, automatic control~\cite{fu1971learning}, bioinformatics~\cite{baldi2001bioinformatics}, etc.
Considering the aforementioned characteristics of future wireless networks, data-driven ML is also likely to become a powerful technique of network association for substantially improving the network performance. This is achieved by accurately learning the near-real-time physical operating scenario, which allows them to outperform the traditional model-driven optimization algorithms based on more-restrictive assumptions detailed in~\cite{jiang2017machine}. More specifically,
\begin{itemize}
  \item The wireless data torrance may be conveniently managed by the \emph{big data processing capability} of ML~\cite{qiu2016survey}. For example,  the tele-traffic volume generated by on-demand information and entertainment is predicted to substantially increase over the next decade, and an average smart phone may generate as much as 4.4 GB data per month by the year 2020~\cite{osseiran2014scenarios,index2015cisco,cao2017towards}. The massive amount of data constitutes a large training set, which can be statistically exploited for data-mining as well as for classification and for prediction with the aid of ML algorithms.
  \item Future wireless networks require both individual node intelligence and swarm intelligence~\cite{kennedy2011particle}. Moreover, as for resource allocation and management, we tend to strike a trade-off among numerous factors, such as the capacity, power consumption, latency, complexity, etc. rather than only considering a single-component objective function. Thanks to learning from trial and error experiments, ML is conducive to supporting \emph{intelligent multi-objective decision making} in the context of multi-agent collaborative network management. Future wireless networks may hence be expected to benefit from intelligent multi-agent systems.
  \item \emph{Modeling and parameter estimation} play an important role in wireless networks. For instance, in massive multiple-input and multiple-output (MIMO) systems, an accurate estimate of the channel state information (CSI) potentially allows us to approach the system's capacity. Given that traditional mathematical models often fail to accurately describe the system in typical time-varying scenarios, ML provides an alternative technique of adaptive modeling and parameter estimation relying on learning from the recorded history.
  \item Future wireless networks are also expected to take into account the \emph{human behavior}, for example by taking into account the geographic position of access points (AP) in an ultra dense network (UDN), where user-centric designs have been conceived for reducing the cluster-edge effects. By mimicking human intelligence, ML may be deemed to be the most appropriate tool for adapting the network's structure to the human behavior observed~\cite{pantic2007human,pentland1999modeling}.
\end{itemize}
Next-generation wireless network optimization relying on ML has emerged as an important research topic, so much so that the standard body ITU-T has formed a dedicated focus group to study this subject from 2018 to 2020. When incorporating ML functionalities into the network architecture, there are two fundamental mechanisms of exploiting ML algorithms, namely online and offline ML. The online ML family represents ML functionalities embedded into networking algorithms or protocols. By contrast, offline ML may be executed by a co-located computing facility connected to the corresponding network entities. However, offline ML can also be supported by remote computing facilities.

In recent years, a range of surveys have been conceived on ML paradigms. Some of them focused their scope on a specific wireless scenario, such as WSNs~\cite{alsheikh2014machine,kulkarni2011computational}, cognitive radio networks (CRN)~\cite{bkassiny2013survey,gavrilovska2013learning,he2010survey}, Internet of Things (IoT)~\cite{park2016learning}, wireless ad hoc networks (WANET)~\cite{forster2007machine}, self-organizing cellular networks~\cite{klaine2017survey}, etc. Specifically, Alsheikh \textit{et al.}~\cite{alsheikh2014machine} provided an extensive overview of ML methods applied to WSNs which improved the resource exploitation and prolonged the lifespan of the network.
Kulkarni \textit{et al.}~\cite{kulkarni2011computational} surveyed some common issues of WSNs solved by computational intelligence algorithms, such as data fusion, routing, task scheduling, localization, etc. Moreover, Bkassiny \textit{et al.}~\cite{bkassiny2013survey} investigated decision-making and feature classification problems solved by both centralized
and decentralized learning algorithms in CRN in a non-Markovian environment. Gavrilovska \textit{et al.}~\cite{gavrilovska2013learning} studied the nature of the CRN's capability of reasoning and learning.
Park \textit{et al.}~\cite{park2016learning} reviewed a range of learning aided frameworks designed for adapting to the heterogeneous resource-constrained IoT environment. Forster~\cite{forster2007machine} portrayed the advantages of using ML for the data routing problem of WANETs. Furthermore, a detailed literature review of the past fifteen years of ML techniques applied
to self-configuration, self-optimization and self-healing, was provided by Klaine \textit{et al.}~\cite{klaine2017survey}.

Some of the literature were restricted to a specific application~\cite{al2015application,fadlullah2017state,nguyen2008survey,buczak2016survey,Xie2019A,Pacheco2019Towards,Sun2019Application}, whilst others considered a single learning technique~\cite{usama2017unsupervised,yau2012reinforcement,alsheikh2016mobile,ota2017deep,Mao2018Deep,Chen2019Artificial}. To elaborate, Al-Rawi \textit{et al.}~\cite{al2015application} presented an overview of the features, methods and performance enhancement of learning-assisted routing schemes in the context of distributed wireless networks.
Additionally, Fadlullah \textit{et al.}~\cite{fadlullah2017state} provided an overview of the state-of-the-art in learning aided network traffic control schemes as well as in deep learning aided intelligent routing strategies, while Nguyen \textit{et al.}~\cite{nguyen2008survey} focused their attention on the ML techniques conceived for Internet traffic classification. ML and data mining assisted cyber intrusion detection were surveyed in~\cite{buczak2016survey}, including the complexity comparison of each algorithm and a set of recommendations concerning the best methods applied to different cyber intrusion detection problems. Moreover, ML techniques applied to software defined networking (SDN) were investigated in~\cite{Xie2019A}, from the perspective of traffic classification, routing optimization, resource management, etc.
Pacheco \textit{et al.}~\cite{Pacheco2019Towards} surveyed the ML techniques based on several steps to achieve traffic classification.
Sun \textit{et al.}~\cite{Sun2019Application} focused on the the recent advances of ML techniques in the MAC layer, network layer, and application layer.

\begin{figure*}
  \centering
 \includegraphics[width=0.9\textwidth]{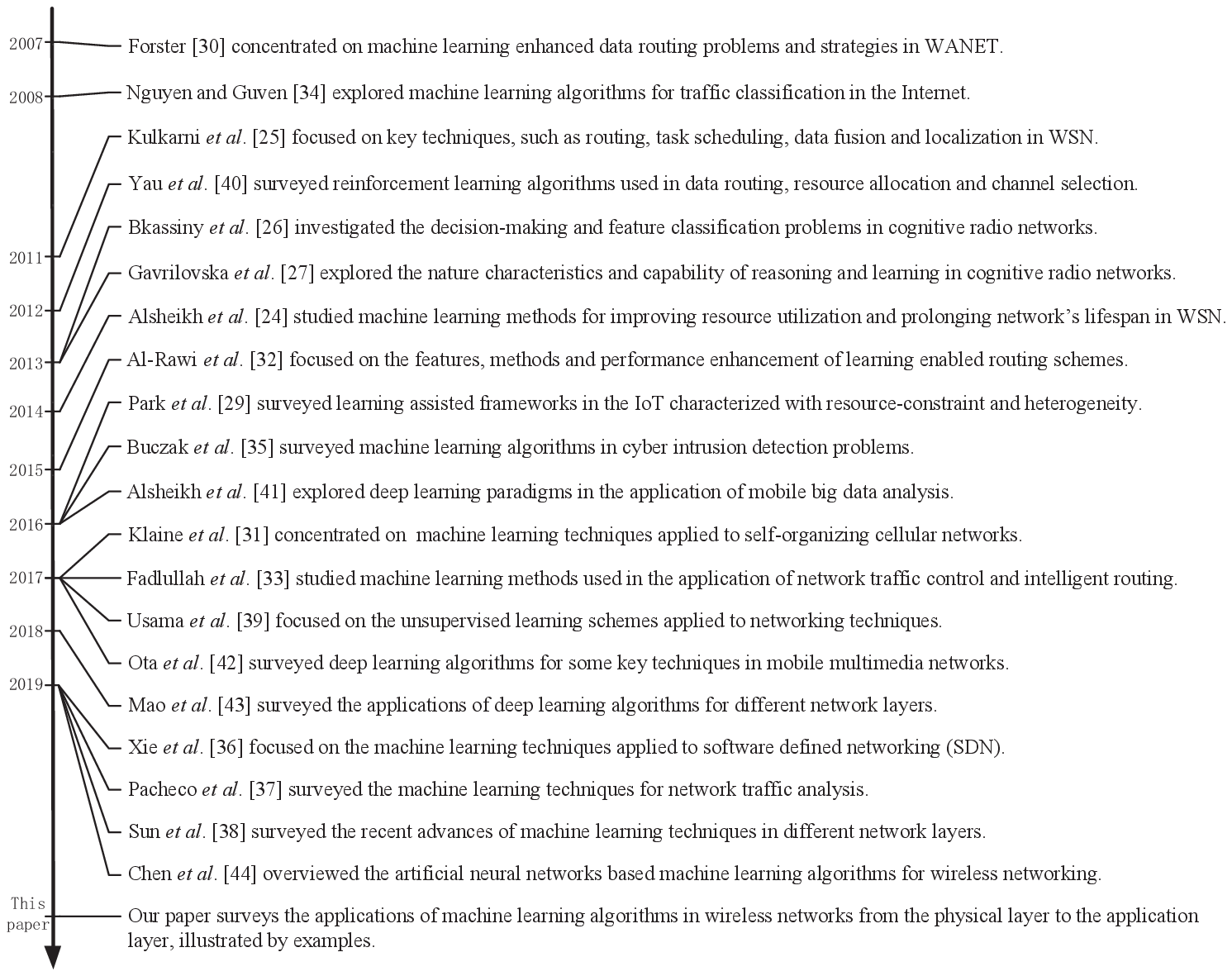}\\
  \caption{The timeline of survey papers on the application of different ML paradigms in wireless networks.}\label{timeline}
\end{figure*}

As for exploring learning techniques, Usama \textit{et al.}~\cite{usama2017unsupervised} provided an overview
of the recent advances of unsupervised learning in the context of networking, such as traffic classification, anomaly detection, network optimization, etc. Yau \textit{et al.}~\cite{yau2012reinforcement} investigated the employment of reinforcement learning invoked for achieving context awareness and intelligence in a variety of wireless network applications such as data routing, resource allocation and dynamic channel selection.
The authors of~\cite{alsheikh2016mobile} and~\cite{ota2017deep} focused their attention on the benefit of deep learning in wireless multimedia network applications, including ambient sensing, cyber-security, resource optimization, etc.
Mao~\textit{et al.}~\cite{Mao2018Deep} provided a comprehensive survey of the applications of deep learning algorithms in terms of different network layers, including physical layer, data link layer and routing layer.
Additionally, Chen~\textit{et al.}~\cite{Chen2019Artificial} overviewed the artificial neural networks based ML algorithms conceived for various wireless networking problems.
The main contributions of the existing ML aided wireless networks survey and tutorial papers are contrasted in Fig.~\ref{timeline} and Table~\ref{topic} to this survey.

\begin{table}[t!]
  \centering
  \caption{The Topics of Survey Papers on Different ML Paradigms in Wireless Networks}
    \begin{tabular}{|l|l|}
    \hline
    \multicolumn{1}{|c|}{Application-oriented} & \multicolumn{1}{c|}{ML-oriented} \\
    \hline
    [24],[29]~resource allocation &     [27]~capability of learning                \\
    \hline
    [25],[30],[32]~routing schemes & [45]~supervised learning \\
    \hline
    [33]~traffic control & [39]~unsupervised learning \\
    \hline
    [26],[34],[37]~traffic classification & [44]~artificial neural networks \\
    \hline
    [35]~intrusion detection &  [40]~reinforcement learning\\
    \hline
    [36]~software defined networking & [41],[42],[43]~deep learning\\
    \hline
    [31],[38]~networking techniques&  \\
    \hline
    \end{tabular}%
  \label{topic}%
\end{table}%

\subsection{Contributions}

Hence, our focus is on the comprehensive survey of ML aided wireless networks. Inspired by above-mentioned challenges, in this article we review the development of ML aided wireless networks. We commence by investigating a series of popular learning algorithms and their compelling applications in wireless networks and then provide some specific examples based on some recent research results, followed by a range of promising open issues in the design of future networks.
Our original contributions are summarized as follows:
\begin{itemize}
  \item We critically review the thirty-year history of ML. Depending on how we use training data, we classify ML algorithms into three categories, i.e. supervised learning~\cite{suthaharan2016supervised}, unsupervised learning~\cite{barlow1989unsupervised} and reinforcement learning~\cite{sutton1998reinforcement}. In addition, we highlight the family of deep learning algorithms, given their success in the field of signal processing.
  \item The development of wireless networks is reviewed from their birth to the future wireless networks. Moreover, we summarize the evolution of wireless networking techniques, and characterize a variety of representative scenarios for future wireless networks.
  \item We appraise a range of typical supervised, unsupervised, reinforcement learning as well as deep learning algorithms. Moreover, their compelling applications in wireless networks are surveyed for assisting the readers in refining the motivation of ML in  wireless networks, all the way from the physical layer to the application layer.
  \item Relying on recent research results, we highlight a pair of examples conceived for wireless networks, which can help the readers to gain the insight into hitherto unexplored scenarios and into their applications in wireless networks.
\end{itemize}

\subsection{Organization}

\begin{figure}
  \centering
 \includegraphics[width=0.47\textwidth]{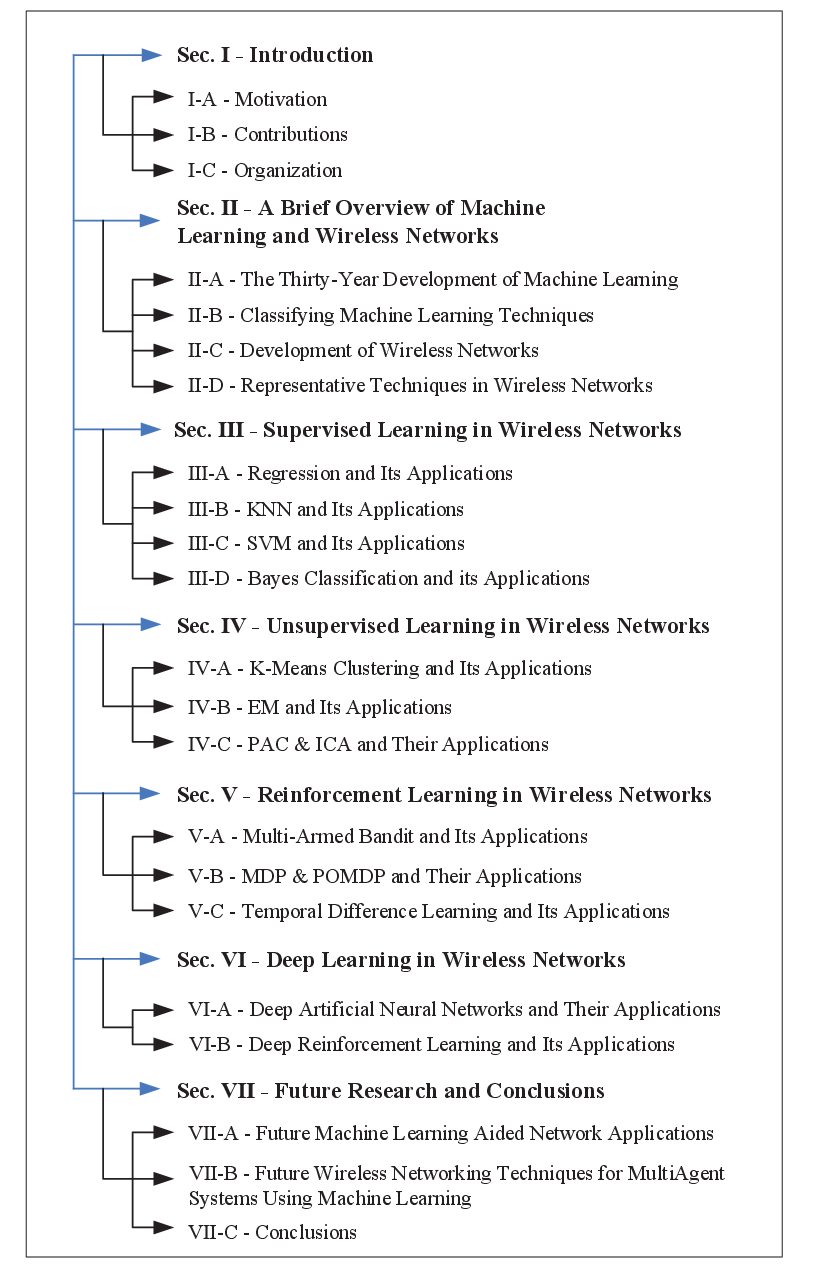}\\
  \caption{The structure of this treatise.}\label{skeleton}
\end{figure}

The remainder of this article is outlined as follows. In Section~\ref{A Brief Overview of Machine Learning and Wireless Networks}, we provide a brief overview of the history of ML and of the development of wireless networks. In Section~\ref{Supervised Learning in NGWN}, we introduce a range of typical supervised learning algorithms and highlight their compelling applications in wireless networks. In Section~\ref{Unsupervised Learning in NGWN}, we investigate the family of unsupervised learning algorithms and their related applications. Some popular reinforcement learning algorithms are elaborated on in Section~\ref{Reinforcement Learning in NGWN}. Moreover, we present two examples of how these reinforcement learning algorithms can improve the performance of wireless networks. In Section~\ref{Deep Learning in NGWN}, we introduce some typical deep learning algorithms and their applications in wireless networks.
Some future research ideas and our conclusions are provided in Section~\ref{Future Research and Conclusions}. The structure of this treatise is summarized at a glance in Fig.~\ref{skeleton}.

\section{A Brief Overview of Machine Learning and Wireless Networks}
\label{A Brief Overview of Machine Learning and Wireless Networks}
\subsection{The Thirty-Year Development of Machine Learning}

The term ``machine learning'' was first proposed by Arthur Samuel in 1959~\cite{samuel1959some}, which referred to computer systems having the capability of learning from their large amounts of previous tasks and data, as well as of self-optimizing computer algorithms. Hard-programmed algorithms are difficult to adapt to dynamically fluctuating demands and constantly renewed system states. By contrast, relying on learning from previous experiences, ML aided algorithms are beneficial for scientific decision making and task prediction, which is achieved by constructing a self-adaption model from sample inputs. To elaborate a little further, as for the concept of ``learning'', Tom M. Mitchell~\cite{mitchell1997does} provided the widely quoted description: ``\emph{A computer program is said to learn from experience $E$ with respect to some class of tasks $T$ and performance measure $P$, if its performance at tasks in $T$, as measured by $P$, improves with experience $E$}.''

ML began to flourish in the 1990s~\cite{michalski2013machine}. Before this era, logic- and knowledge-based schemes, such as inductive logic programming, expert systems, etc. dominated the artificial intelligence scene relying on high-level human-readable symbolic representations of tasks and logic. Thanks to the development of statistics theory and stochastic approximation, ML schemes regained researchers' attention leading to a range of beneficial probabilistic models. Researchers embarked on creating date-driven programs for analyzing a large amount of data and tried to draw conclusions or to learn from the data. During this era, ML algorithms such as neural networks as well as kernel methods became mature. During the 2000s, researchers gradually renewed their interest in deep learning with the aid of the advances in hardware-based computational capability, which made ML indispensable for supporting a wide range of services and applications.

Given the development of progressive learning techniques~\cite{su2003relevance}, at present, the research focus of ML has shifted from ``learning being the purpose'' to ``learning being the method''. Specifically, ML algorithms no longer blindly pursue to imitate the learning capability of human beings, instead they focus more on the task-oriented intelligent data-driven analysis. Nowadays, thanks to the abundance of raw data and to the frequent interaction between exploration and exploitation, ML algorithms have prospered in the fields of computer vision, data mining, intelligent control, etc. Future wireless networks aim for providing ubiquitous information services for users in a variety of scenarios. However, the rapid growth in the number of users and the resulted explosive growth of tele-traffic data pushes the limits of network-capacity. As a remedy, ML aided network management and control can be viewed as a corner stone of future wireless networks in view of their limited power, spectrum and cost.

\subsection{Classifying Machine Learning Techniques}

\begin{figure*}[htbp]
  \begin{adjustbox}{addcode={\begin{minipage}{\width}}
  {\caption{A comprehensive survey of ML algorithms.}\label{learningalgorithm}\end{minipage}},rotate=90,center}
  \includegraphics[scale=.9]{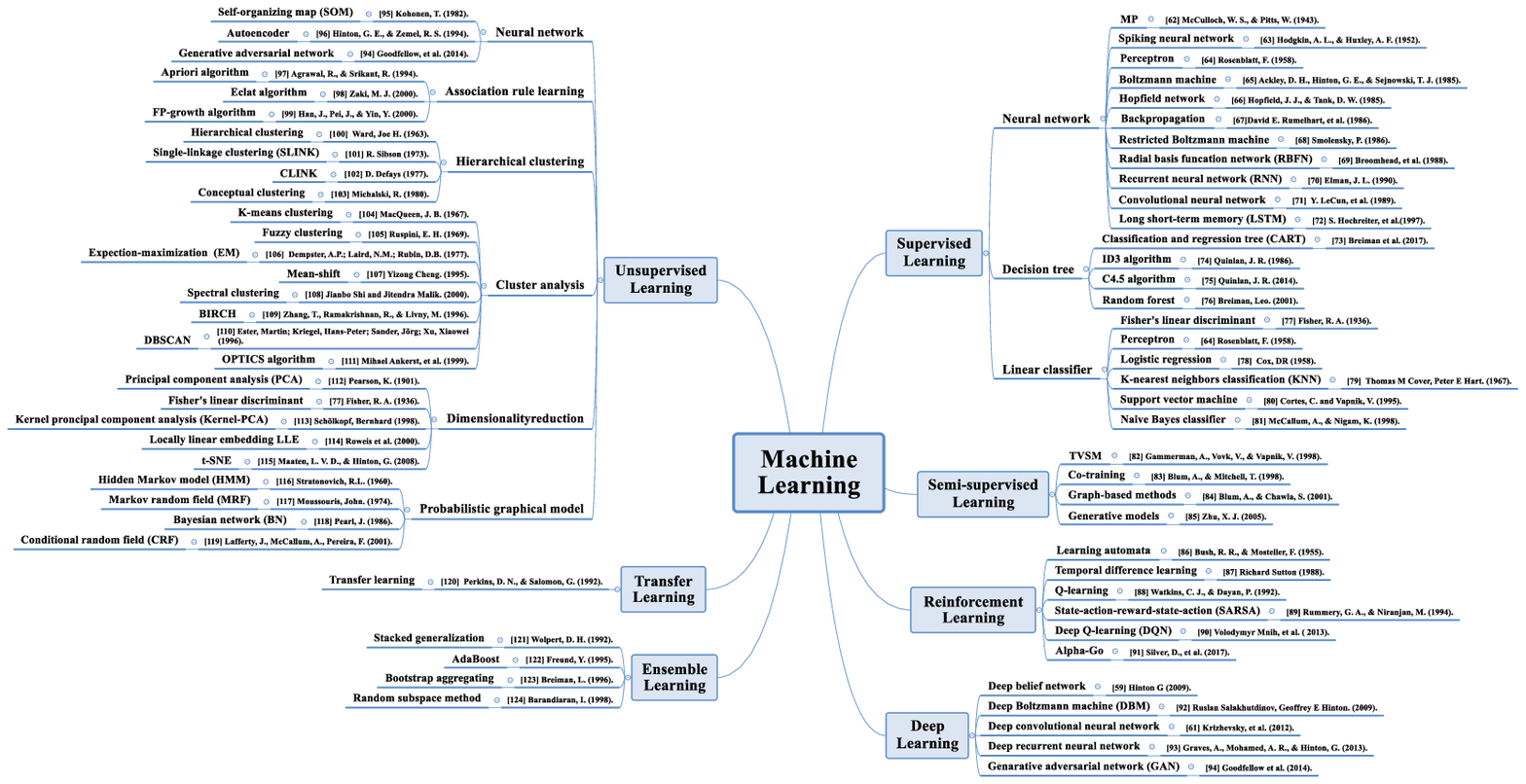}%
  \end{adjustbox}
\end{figure*}

\subsubsection{A General Taxonomy}
Again, depending on how training data is used, ML algorithms can be grouped into three categories, i.e. supervised learning, unsupervised learning and reinforcement learning~\cite{mohri2012foundations,christopher2016pattern}.
In the following, we will provide a brief description of the three types of algorithms.
\begin{itemize}
  \item \emph{Supervised Learning}: The algorithms are trained on a certain amount of labeled data~\cite{suthaharan2016supervised}. Both the input data and its desired label are known to the computer, resulting in a data-label pair. Their goal is to infer a function that maps the input data to the output label relying on the training of sample data-label pairs. Specifically, considering a set of $N$ sample data-label pairs in the form of $\{(x_1, y_1), (x_2, y_2),\dots (x_N, y_N)\}$, where $x_n$ is the $n$-th sample input data and $y_n$ represents its label. Let $\mathbb{X}=\{x_1, x_2, \dots, x_N\}$ denote the input data set and $\mathbb{Y}=\{y_1, y_2, \dots, y_N\}$ represent the output label set. Usually, these sample pairs are independent and identically distributed (i.i.d.). The learning algorithms aim for seeking a function $g(x)$ that yields the highest value of the score function $f(x,y)$, hence we have $g(x)=\argmax\limits_{y} f(x,y)$. As a special case, if only part of the sample data-label pairs are known to the computer and some of the desired output labels of input data are missing, the corresponding learning algorithms are termed as semi-supervised learning\footnote{In this paper, semi-supervised learning algorithms are viewed as a specific category of supervised learning algorithms. However, in some of the literature, semi-supervised learning is listed as a separate member of the ML family}. These supervised learning algorithms can be widely used in the context of classification, regression and prediction.
  \item \emph{Unsupervised Learning}: Relying on unlabeled input data, unsupervised learning algorithms try to explore the hidden features or structure of the data~\cite{barlow1989unsupervised,hastie2009unsupervised}. Given the lack of sample data-label pairs, there is no standard accuracy evaluation for the output of unsupervised learning algorithms, which is the main difference compared to its supervised learning aided counterpart. By analyzing $N$ input data $\mathbb{X}=\{x_1, x_2, \dots, x_N\}$, a pair of popular methods has been conceived for revealing the underlying unknown features of $N$ input data, namely density estimation~\cite{silverman2018density} as well as feature extraction~\cite{guyon2006introduction}. To elaborate, density estimation aided methods are characterized by explicitly building statistical models of how the underlying features might create the input. By contrast, feature extraction based techniques aim for directly extracting statistical regularities or even sometimes irregularities from the input data set.
  \item \emph{Reinforcement Learning}: In contrast to the aforementioned two learning techniques, reinforcement learning algorithms are conceived for decision making by learning from interaction with the environment, which are trained by the data on the basis of trial and error~\cite{sutton1998reinforcement,kaelbling1996reinforcement}. They neither try to identify a category as supervised learning algorithms do, nor do they aim for finding hidden structures as unsupervised learning algorithms do. Specifically, at each time step, the system or environment is in some state $S$, and the agent selects a legitimate action $A$. The system responds at the next time step by moving into a new state $S'$ with a certain probability influenced both by the specific action chosen as well as by the system's inherent transitions. Meanwhile, the agent receives a corresponding reward $r(S,A)$ from the system, as time evolves. Reinforcement learning algorithms aim for learning how to map situations $S$ into actions $A$ in order to attain the maximal cumulative weighted reward within the horizon in such a closed-loop fashion.
\end{itemize}

\subsubsection{Deep Learning}
As an important member of the ML family, \emph{deep learning} has been booming since 2010, because it was found to be capable of handling the soaring growth of training data volume facilitated by the rapid development of computing hardware~\cite{lecun2015deep,schmidhuber2015deep}. Deep learning algorithms rely on
a multiple-layer ``network'' consisting of inter-connected nodes for feature extraction and transformation, which is inspired by the biological nervous system, namely the neural network. Each layer utilizes the output of the previous layer as its input. The term ``deep'' refers to having multiple layers in the network.
Generally, relying on the way the training data is exploited, deep learning algorithms can also be classified into deep supervised learning, deep unsupervised learning as well as deep reinforcement learning~\cite{lecun2015deep}. Moreover, some deep learning architectures, such as deep neural networks (DNN)~\cite{mnih2015human}, deep belief networks (DBN)~\cite{hinton2009deep}, recurrent neural networks (RNN) with LSTM units~\cite{pascanu2013construct} and convolutional neural networks (CNN)~\cite{krizhevsky2012imagenet}, have had success in a range of fields including computer vision, natural language processing (NLP), speech recognition, etc. They have also been invoked in compelling applications of wireless networks.

\begin{figure*}
  \centering
 \includegraphics[width=0.9\textwidth]{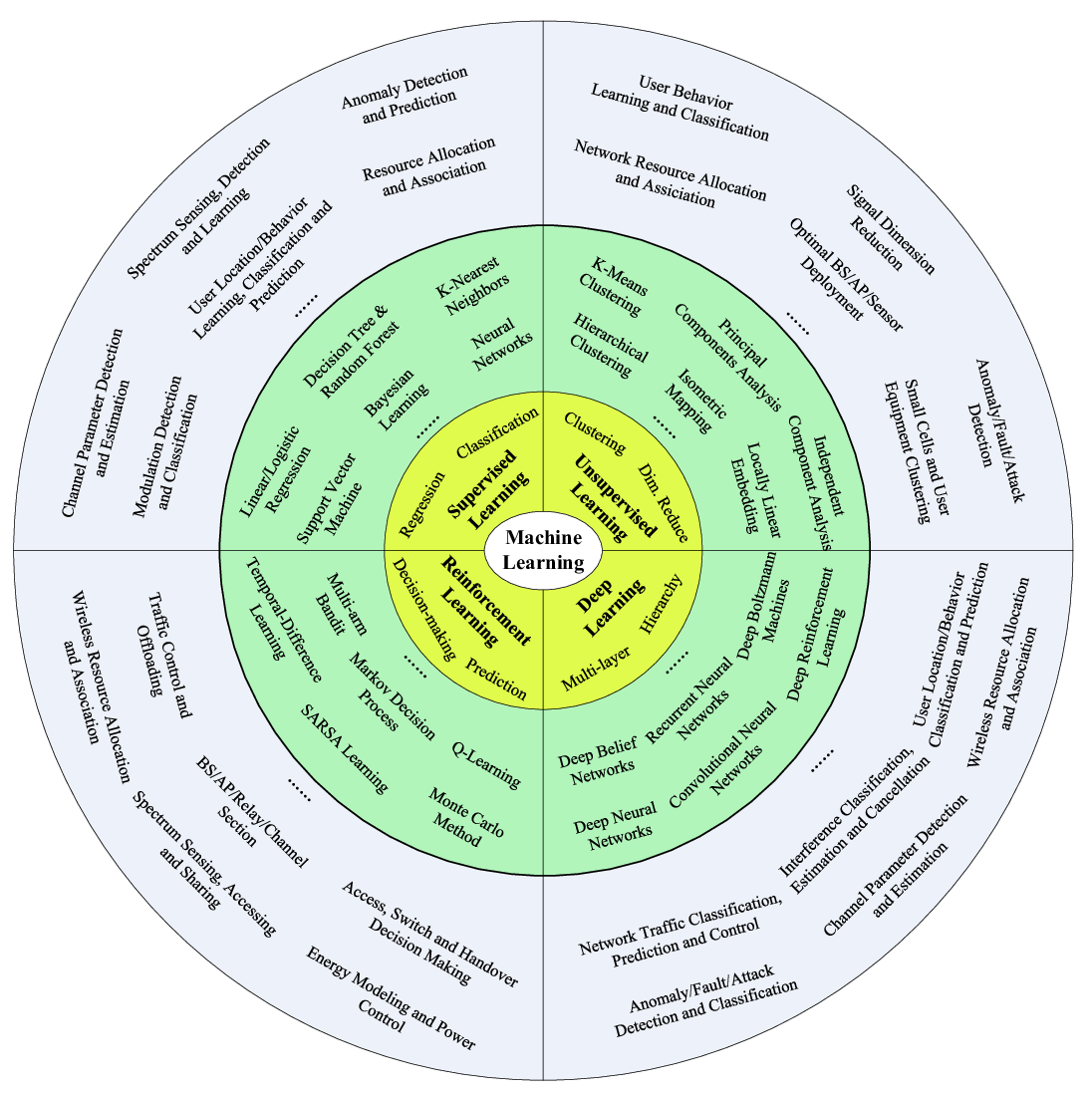}\\
  \caption{The categories, characteristics and applications of ML algorithms.}\label{machine learning}
\end{figure*}

Fig.~\ref{learningalgorithm} comprehensively summarizes ML algorithms even including above-mentioned supervised learning, unsupervised learning, reinforcement learning and deep learning algorithms as well as some non-mainstream branches~\cite{mcculloch1943logical,hodgkin1952quantitative,rosenblatt1958perceptron,ackley1985learning,hopfield1985neural,rumelhart1988learning,smolensky1986information,broomhead1988radial,elman1990finding,lecun1989backpropagation,hochreiter1997long,breiman2017classification,quinlan1986induction,quinlan2014c4,breiman2001random,fisher1936use,cox1958regression,cover1967nearest,cortes1995support,mccallum1998comparison,gammerman1998learning,blum1998combining,blum2001learning,zhu2005semi,bush1955stochastic,sutton1988learning,watkins1992q,rummery1994line,mnih2013playing,silver2017mastering,salakhutdinov2009deep,graves2013speech,goodfellow2014generative,kohonen1982self,hinton1994autoencoders,agrawal1994fast,zaki2000scalable,han2000mining,ward1963hierarchical,sibson1973slink,defays1977efficient,michalski1980knowledge,macqueen1967some,ruspini1969new,dempster1977maximum,cheng1995mean,shi2000normalized,zhang1996birch,ester1996density,ankerst1999optics,pearson1901liii,scholkopf1998nonlinear,roweis2000nonlinear,maaten2008visualizing,stratonovich1960conditional,moussouris1974gibbs,pearl1986fusion,lafferty2001conditional,perkins1992transfer,wolpert1992stacked,freund1995boosting,breiman1996bagging,barandiaran1998random}. Furthermore, Fig.~\ref{machine learning} shows the involvement of ML in wireless networks based on the aforementioned four categories. Below we list a variety of popular learning algorithms and highlight their applications in future wireless networks.

\subsection{Development of Wireless Networks}

\subsubsection{The Birth and Development of Wireless Networks}

\begin{figure*}
  \centering
 \includegraphics[width=0.90\textwidth]{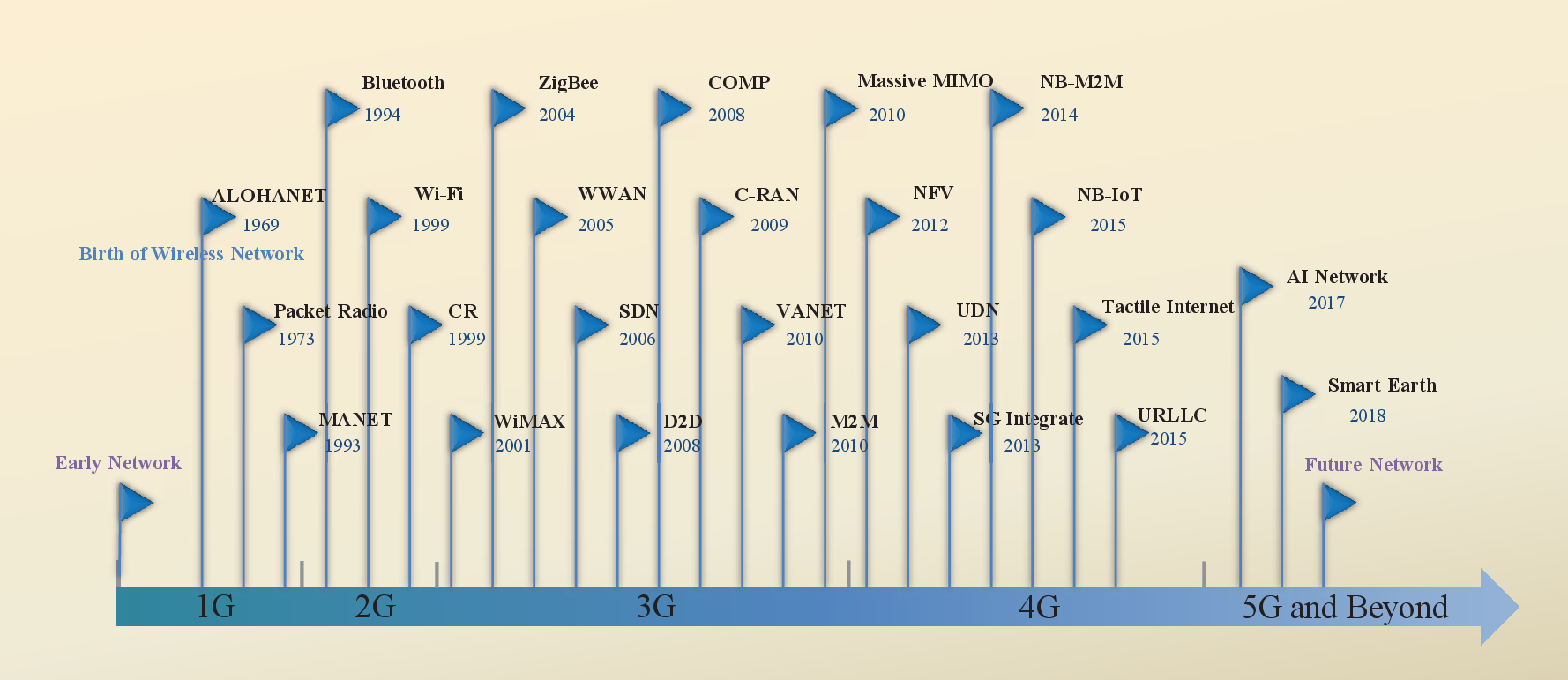}\\
  \caption{The development of wireless networks interspersed with ever-accelerated milestone techniques.}\label{wireless develop}
\end{figure*}

Just as the terminology implies, wireless networks connect various network nodes via electromagnetic waves.
Relying on their coverage, wireless networks can be roughly classified into four categories, such as wireless personal area networks (WPAN)~\cite{gutierrez2004low}, wireless local area networks (WLAN)~\cite{crow1997ieee}, wireless metropolitan area networks (WMAN)~\cite{eklund2006wirelessman} and wireless wide area networks (WWAN)~\cite{hansen2008wide}. Correspondingly, a family of networking standards and their variants that cover most of the physical layer specifications have been established by the IEEE 802 Working groups, including the IEEE 802.15 for WPAN, IEEE 802.11 for WLAN, IEEE 802.16 for WMAN and IEEE 802.20 for WWAN standards. Furthermore, when considering the network's functions, some popular representatives of wireless networks include cellular networks~\cite{gupta2015survey}, WSNs~\cite{baronti2007wireless}, WANETs~\cite{ilyas2017handbook}, wireless body area networks (WBAN)~\cite{movassaghi2014wireless}, etc.

The first wireless network, namely ALOHANET, was developed at the University of Hawaii in 1969 and came into operation in 1971, which for the first time transmitted wireless data packets over a network~\cite{abramson1985development}. The first commercial wireless network was the WaveLAN product family designed by the NCR Corporation in 1986. In 1997, the first IEEE 802.11 protocol was released for WLAN~\cite{crow1997ieee}. Afterwards, the emergence and progress of reliable and low-cost Wi-Fi marked the maturity of wireless networking technologies at the end of the 20th century, which facilitated Internet access for a range of Wi-Fi compatible devices including personal computers, smart phones, etc. Future wireless networks aim for providing high-rate, low-latency, full-coverage and low-cost yet reliable information services. Compared to traditional wireless networks connecting humans and their devices, future wireless networks are expected to interconnect everything under the umbrella of the `Internet of Everything'. Fig.~\ref{wireless develop} demonstrates the development of wireless networks in terms of their milestone techniques.

Wireless networks have evolved from the simple client-server (C-S) mode to the distributed dense multi-layer C-S mode, and finally to the ad hoc peer-to-peer (P2P) mode.
The decentralization of network architectures grant more freedom both for the network nodes and their protocols, which requires more sophisticated techniques for supporting efficient and reliable implementations. Furthermore, the soaring growth of both the type and the amount of data provides a promising field of applications for ML algorithms, which are beneficial for self-organized and self-adaptive network architectures.

\subsubsection{Pareto-Optimal Future Wireless Networks}
\begin{figure*}
  \centering
 \includegraphics[width=0.45\textwidth]{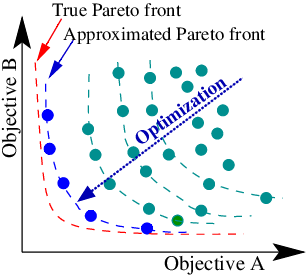}\\
  \caption{A simple example of a two-parameter Pareto-optimization problem.}\label{pareto_front}
\end{figure*}

The challenging real-world optimization problems encountered in future wireless networks have to meet multiple objectives in order to arrive at an attractive solution~\cite{fei2017survey}. In contrast to conventional single-objective optimization where we find the global optimum relying on a single metric, multi-objective optimization aims for finding the globally optimal solution relying on the notion of Pareto optimality~\cite{censor1977pareto}. The aim of multi-objective optimization in wireless networks is that of generating a diverse set of Pareto-optimal solutions, where by definition it is only possible to improve any of the metrics considered at the cost of degrading at least one of the others. The collection of Pareto-optimal points is referred to as the Pareto front. Fig.~\ref{pareto_front} shows a simple Pareto optimization problem having a pair of objective parameters, where the light blue circles represent legitimate operating
points, while the dark blue circles denote the approximated Pareto optimal points, which are not dominated by any other solutions. The arrow intimates that the complexity of optimization is gradually increased as we gradually approach the full-search complexity, which defines the `true Pareto front' of Fig.~\ref{pareto_front}.

More specifically, let us briefly reminisce by recalling the past few decades of wireless history. Explicitly, the wireless community has invested decades of research efforts into making near-capacity single-user operation a reality~\cite{gupta2000capacity}, which is however only possible at the cost of an ever-increasing delay, complexity and power consumption. When designing a powerful wireless network, which includes the physical layer, MAC layer, network layer, transmission layer as well as application layer, we face substantial challenges, since we often have to meet conflicting design objectives. Moreover, these conflicting optimization metrics are often coupled with each other. It may also be a substantial challenge
to carry out a fair comparison among locally
optimal solutions, hence
we often strike a tradeoff~\cite{zhang2019aeronautical}. For example, in future wireless networks we would like to be more ambitious than 'only' optimize the network's capacity - for delay-sensitive services we would like to reduce the latency and/or reduce the total energy consumption, as well as to improve the system's reliability and the user's QoS. By contrast, in wireless sensor networks we may concentrate on optimizing both the connectivity and the network's life time, just to name a few of these conflicting practical design objectives.

\begin{figure*}
  \centering
 \includegraphics[width=0.45\textwidth]{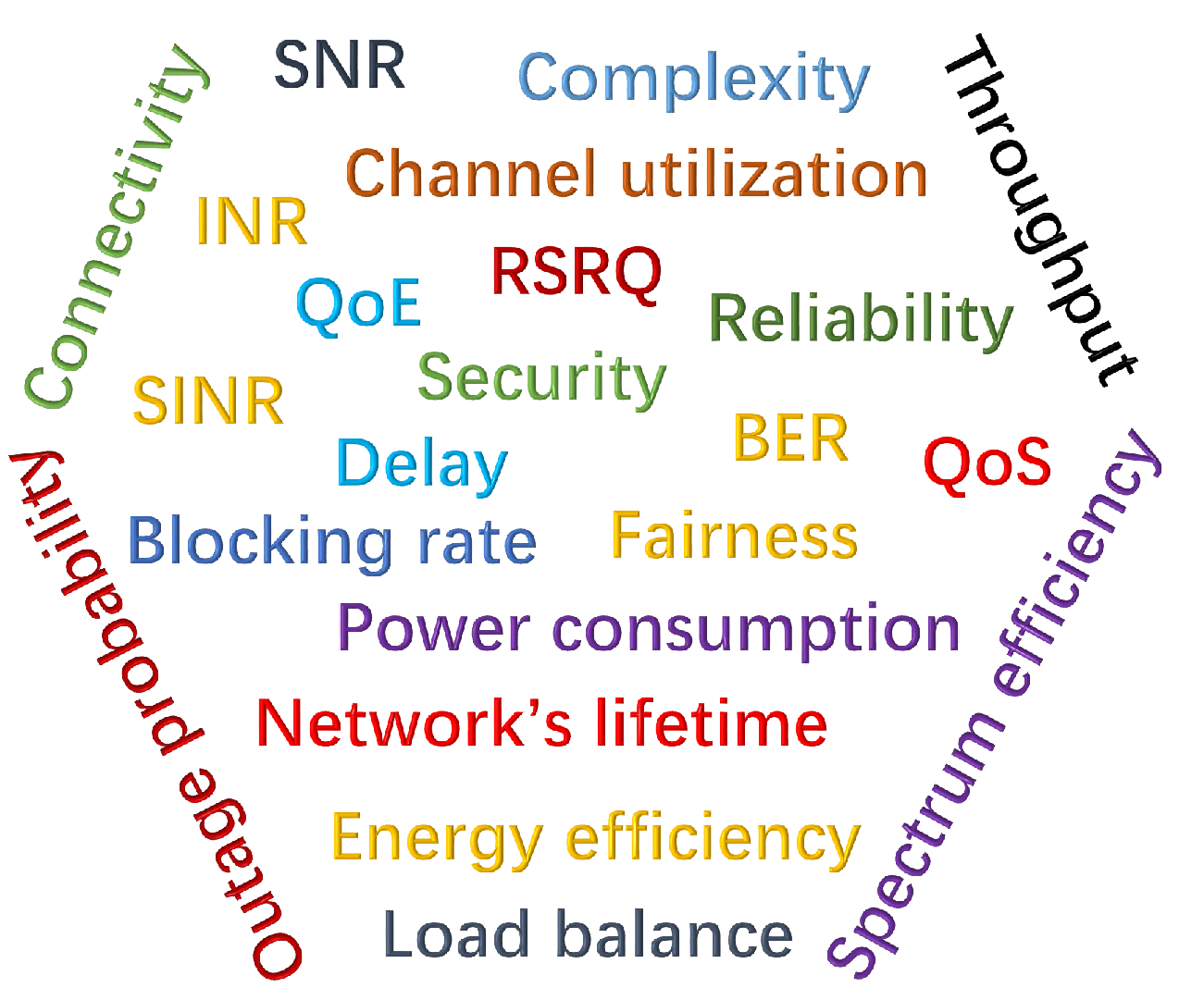}\\
  \caption{Useful metrics commonly used in constructing multi-objective optimization problems.}\label{metrics}
\end{figure*}

In this context, the family of ML techniques may be viewed as an attractive set of optimization tools for finding Pareto-optimal solutions of multi-objective optimization problems in future wireless networks, which tend to have a large search-space. To expound a little further, it is plausible that every time we incorporate an additional parameter into the objective function, the search-space is expanded and the surface of optimal solutions may exhibit numerous locally optimal solutions. Hence traditional gradient-based techniques routinely fail to find the global optimum. In this context Fig.~\ref{metrics} portrays some popular metrics commonly used in constructing multi-objective optimization problems in future wireless networks.

\subsection{Representative Techniques in Wireless Networks}
\begin{figure*}
  \centering
 \includegraphics[width=0.90\textwidth]{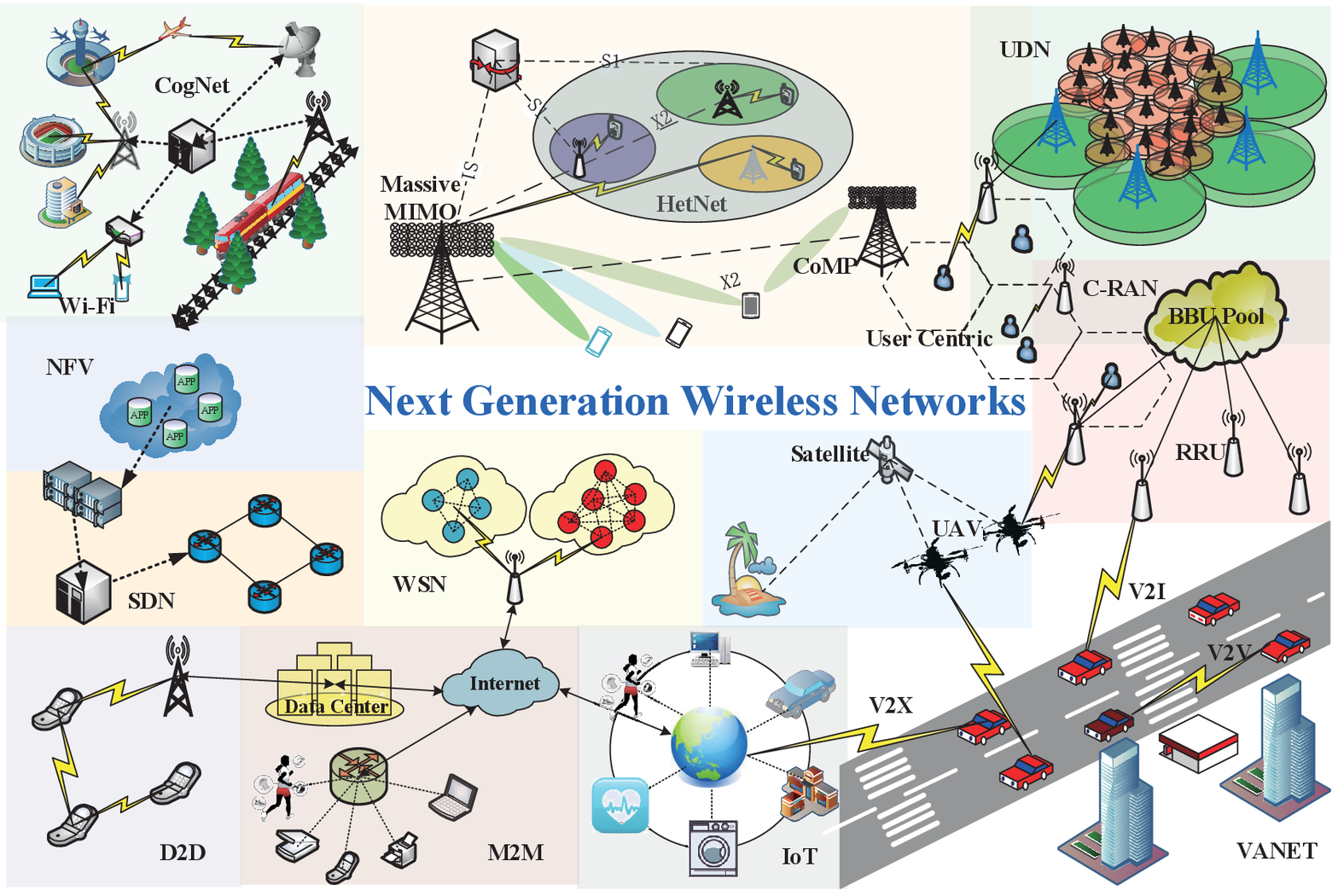}\\
  \caption{The representative application scenarios and related techniques of future wireless networks.}\label{wireless network}
\end{figure*}

\begin{figure*}
  \centering
 \includegraphics[width=0.90\textwidth]{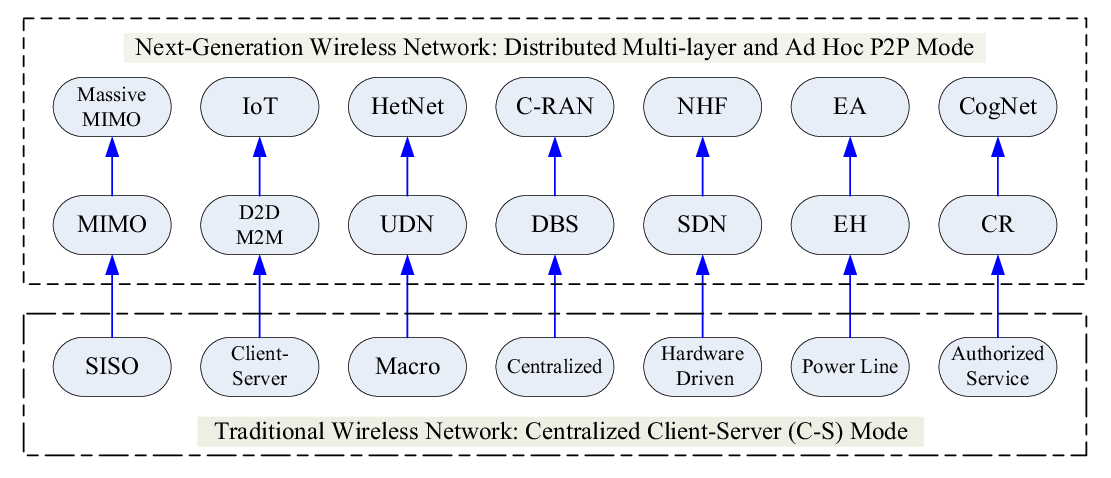}\\
  \caption{The development trends of wireless networks.}\label{development tendency}
\end{figure*}

As shown in Fig.~\ref{wireless network}, we first of all portray the representative application scenarios and techniques of future wireless networks.
In the following, we will briefly introduce a range of compelling techniques and their development trends in future wireless networks, which is summarized in Fig.~\ref{development tendency}.
\subsubsection{From MIMO to Massive MIMO}
The MIMO technology relying on multiple antennas in both the transmitter and receiver can be viewed as a breakthrough in terms of multiplying the capacity of a radio link compared to the single-transmit single-receive antenna aided wireless system having a variety of cost, technology and regulatory constraints~\cite{pan2013capacity}. Both single-user MIMO (SU-MIMO) and Multi-user MIMO (MU-MIMO) schemes have been proposed. To elaborate, multiple data streams of the same source are sent to a single user in SU-MIMO, while a transmitter simultaneously serves multiple users on the same channel resource in MU-MIMO~\cite{liu2012downlink,wang2014pair}.

Generally, invoking more antennas is beneficial in terms of improving the data rate and/or link reliability~\cite{larsson2014massive,rusek2013scaling,pan2018many,wang2015design}.
However, the performance degradation caused by inaccurate CSI and the high computational complexity of channel estimation constitute substantial challenges~\cite{goldsmith2003capacity}.

\subsubsection{From D2D, M2M to IoT}
In the spirit of direct communication between nearby mobile devices without traversing base stations (BS) or core networks, device-to-device (D2D) communication networks have been widely investigated in recent years, which can be deemed to be important milestones on the road towards self-organization and P2P collaboration. In D2D networks, the same resource slots can be reused both by the D2D links as well as the cellular links, which is capable of substantially improving the network capacity. Moreover, it is potentially beneficial in terms of enhancing the energy efficiency (EE), also reducing the transmission delay and improving the network's fairness to users~\cite{wang2016mobile,feng2013device}, which is also closely related to
machine-to-machine (M2M) communications. The corresponding massive machine type of communication (mMTC)~\cite{lien2011toward} mode of the 5G network\footnote{In 2015, International Telecommunication Union (ITU) officially defined three application scenarios of 5G network, i.e. enhanced mobile broad band (eMBB), massive machine type communication (mMTC) and ultra reliable low latency communication (uRLLC).} is capable of supporting sensing, transmitting, fusion and processing sensory data. Furthermore, M2M is also capable of supporting the smart home~\cite{zhang2011home}, smart grid~\cite{niyato2011machine}, etc.

Aiming for ``connecting everything'', IoT was first defined for enabling objects to connect and exchange data in 1999~\cite{al2015internet}. Furthermore, the IoT allows objects to be sensed and controlled remotely, creating opportunities for direct interaction between the physical world and computer-based virtual systems, which is beneficial in terms of improving operational efficiency and of reducing human intervention.
Both WSNs and M2M communications can be viewed as a part of the IoT. Although the IoT faces a range of reliability, robustness and security challenges, there is no doubt that it will make our world ever smarter~\cite{chiang2016fog,kaur2018edge}.

\subsubsection{From UDN to HetNet}
In order to meet the demand of supporting massive data traffic, the so-called UDN architecture has been defined where the density of BSs or APs potentially reaches or even exceeds the density of users~\cite{zhang2017energy,zhang2015resource}. The UDN architecture is conducive to increasing the network capacity as well as simultaneously improving the user experience. However, the interference encountered in UDNs tends to be more severe and of higher volatility than that in traditional cellular networks because of the dense deployment of BSs and APs. Hence, the joint consideration of resource allocation, interference management and traffic routing are essential for UDNs~\cite{gupta2015survey,andrews2016we}.

Considering a wide area network scenario, heterogeneous networks (HetNet) are characterized by the employment of multiple types of radio access technologies (RAT)~\cite{hoadley2012enabling}. Upon combining macrocells, microcells, picocells~\cite{zhang2015practical} and femtocells~\cite{zhang2014resource,jiang2014optimal}, HetNets are capable of providing a seamless wireless coverage ranging from outdoor environments to office buildings and even to underground areas by selecting another RAT when a RAT fails, and HetNets can also provide load-balancing in the face of non-uniform spatial distribution of users~\cite{zhang2016self}.

\subsubsection{From DBS to C-RAN}
Compared to the traditional BS, which integrates baseband processing units (BBU) and remote radio units (RRU)\footnote{In some works, RRU is also called remote radio head (RRH)} in a single cabinet, distributed base station (DBS) aided systems separate the BBU as well as the RRU and connects them with optical fiber.
The DBS system allows more flexibility in network planning and deployment, where  RRUs can be placed a few hundred meters or a few kilometres away for enhancing network's edge-coverage.

Cloud-radio access networks (C-RAN) can be viewed as an evolution of the aforementioned DBS system, which is a centralized processing and cloud computing aided radio access network architecture~\cite{wu2015cloud}. The principle of C-RAN relies on gathering the BBUs from several BSs into a centralized BBU pool, whilst allowing hundreds of RRUs to connect to the centralized BBU pool~\cite{checko2015cloud}. Hence, resources can be allocated to each user based on joint dynamic scheduling. By exploiting coordination and virtualization, the spectral efficiency (SE), the system's flexibility and the load balancing capability are substantially improved. Moreover, the centralized management of resources reduces the cost of the system's operation and maintenance.

\subsubsection{From SDN to NFV}
Software-defined networking (SDN) is employed as a programmable network architecture in order to achieve cost-effective dynamic network configuration and monitoring~\cite{xia2015survey,liang2015wireless}.
The SDN philosophy suggests to centralize network intelligence in a single network component by
decoupling the control plane and the data plane, which disassociates network control and its forwarding functions. The two planes can communicate with the aid of the OpenFlow protocol\footnote{The OpenFlow protocol is a communication protocol that gives access to the forwarding plane of a switcher or router over the network.}, and the network resources can be managed logically and efficiently. A SDN connects decentralized users to cloud computing through a ``network pipeline''~\cite{kim2013improving}~\cite{jain2013network}.

Relying on IT virtualization techniques, network function virtualization (NFV) transforms the entire set of network node functions into different building blocks, which separates the networking functions from specific hardware blocks~\cite{han2015network}. Hence, NFV is eminently suitable for service diversification and promotes the standardization of networking equipment~\cite{li2015software}. Explicitly, NFV can be viewed as a beneficial hardware-agnostic design in the application layer of SDN architectures.

\subsubsection{From EH to EA}
Energy harvesting (EH) is an environmentally friendly process, which captures and stores ambient energy, such as solar power, thermal energy, wind energy, etc. for low-power wireless devices~\cite{sudevalayam2011energy}, especially in WSNs and WBANs, for example.

In future wireless networks, energy optimization is a significant concern motivated by mitigating climate change. However, energy consumption is related to both the network's throughput and to its entire lifetime with a trade-off between them.
As a remedy, instead of only focusing on EH, energy awareness (EA) at every stage of the network's design and management is the most promising approach to striking a trade-off amongst the conflicting objectives of reducing energy consumption, improving the system's throughput as well as prolonging its lifetime, especially in energy-constrained networks~\cite{raghunathan2002energy,saleh2014survey}.

\subsubsection{From CR to CogNet}
Cognitive radio (CR) constitutes a technique that allows us to dynamically and efficiently exploit the wireless spectral resources~\cite{haykin2005cognitive,jiang2014dynamic,yucek2009survey,jiang2013joint}. By relying on spectrum sensing, CR is capable of achieving dynamic spectrum access and spectrum sharing. Specifically, in the process of spectrum sensing, the secondary user (SU) detects an empty slicer of spectrum, for example, based on energy detection schemes. Then, in the process of spectrum access, power control is invoked by the SU for maximizing its capacity, whilst observing the interference power constraint in order to protect the primary user (PU). As a benefit, CR dynamically and flexibly exploits the scarce wireless spectral resources, hence substantially improving the spectrum efficiency~\cite{jiang2014jsac}.

In contrast to CR techniques, which only deal with the issues of physical-layer spectrum sensing and data link-layer access, cognitive networks (CogNet) are characterized by a cognitive cross-layer process according to their end-to-end goals, where the overall network conditions are monitored, and then decisions are made based on the perceived conditions as well as on the feedback and experience gleaned from previous actions~\cite{thomas2006cognitive}. The network's cognitive capability relies on a range of advanced techniques, such as knowledge representation and ML, which exploit a wealth of information generated within the network improving both the network management, the resource efficiency~\cite{manoj2008cognet} and the energy efficiency~\cite{jiang2014energy}.

\subsubsection{Interference Management}
{\color{black}
Interference constitutes the fundamental limiting factor of the overall wireless system performance, hence it is a key challenge faced by designers. Therefore substantial efforts have been dedicated to exploiting the communication channel's state information (CSI) either at the transmitter (CSIT) or at the receiver (CSIR) for mitigating the effects of interference. Hence diverse time/frequency/space division multiple access based resource allocation schemes have been conceived for avoiding interference by creating orthogonal resource units~\cite{turkboylari1998efficient,ma2010interference,kountouris2012downlink}. Creative efforts have also been dedicated to the conception of non-orthogonal access systems, as exemplified by a large variety of cognitive radio~\cite{zhang2016interference} and non-orthogonal multiple access (NOMA) schemes~\cite{liu2017nonorthogonal} relying on sophisticated transceiver designs. Additionally, multi-antenna based techniques, such as joint/partial pre/postcoding and antenna selection, have also been proposed for ameliorating the effects of interference by exploiting the benefits of spatial diversity~\cite{paulraj2004overview}.

A closely related issue in future wireless networks is interference management, which is a particularly critical task in ultra-dense networks in the face of their stringent throughput, delay and reliability specifications. Hence sophisticated resource allocation and interference management schemes are required. Therefore a range of ML algorithms have also been invoked for interference management relying on their environmental awareness and learning capability~\cite{attar2011interference,deb2015learning,bernardo2011intercell}.
}

\section{Supervised Learning in Wireless Networks}
\label{Supervised Learning in NGWN}
Having covered the networking basics, in this section, we will introduce some rudimentary supervised learning algorithms, such as regression, K-nearest neighbors (KNN), support vector machines (SVM) and Bayes classification including their applications in wireless networks.
Table~\ref{tbsup} summarizes some of the typical applications of the above-mentioned four supervised learning algorithms in wireless networks.
\subsection{Regression and Its Applications}
\subsubsection{Methods}
Regression analysis is capable of estimating the relationships among variables. Relying on modeling the functional relationship between a dependent variable (objective) and one or more independent variables (predictors), regression constitutes a powerful statistical tool of predicting and forecasting a continuous-valued objective given a set of predictors.

In regression analysis, there are three variables, namely the
\begin{itemize}
  \item \emph{Independent variables} (predictors): $X$
  \item \emph{Dependent variable} (objective): $Y$
  \item \emph{Other unknown parameters} that affect the estimated value of the dependent variable: $\varepsilon$
\end{itemize}
The regression function $f$ models the functional $Y$ vs $X$ relationship perturbed by $\varepsilon$, which can be formulated as: $Y=f(X, \varepsilon)$. Usually, we characterize this relationship in terms of a specific regression function with the aid of its probability distribution. Moreover, the approximation is often modeled as $E=[Y\mid X]=f(X, \varepsilon)$.
When conducting regression analysis, first of all we have to determine the specific form of the regression function $f$, which relies on both the common knowledge about the dependent vs independent variables as well as on its convenient evaluation.
Based on the specific form of regression function, regression analysis methods can be classified as ordinary linear regression~\cite{seber2012linear}, logistic regression~\cite{hosmer2013applied}, polynomial regression~\cite{max1976segmented}, etc.

In linear regression, the dependent variable is a linear combination of the independent variables or unknown parameters. Let us assume having $N$ random training samples and $M$ independent variables, formulated as $\{y_n, x_{n1}, x_{n2},\dots, x_{nM}\}, n=1,2,\dots,N$. Then the linear regression function can be formulated as:
\begin{equation}\label{1inear}
y_n=\varepsilon_0+\varepsilon_1x_{n1}+\varepsilon_2x_{n2}+\dots+\varepsilon_Mx_{nM}+e_n,
\end{equation}
where $\varepsilon_0$ is termed as the regression intercept, while $e_{n}$ is the error term and $n=1,2,\dots,N$. Hence, Eq.~(\ref{1inear}) can be rewritten in the form of a matrix as $\pmb y=\pmb X\pmb\varepsilon+\pmb e$, where $\pmb y=[y_1,y_2,\dots,y_N]^{T}$ is an observation vector of the dependent variable and $\pmb e=[e_1,e_2,\dots,e_N]^{T}$, while $\pmb \varepsilon=[\varepsilon_0,\varepsilon_2,\dots,\varepsilon_M]^{T}$ and $\pmb X$ represents the observation matrix of independent variables, given by:
\begin{equation*}
  \pmb X =\left[ \begin{matrix}
   1 & {{x}_{11}} & \cdots  & {{x}_{1M}}  \\
   1 & {{x}_{21}} & \cdots  & {{x}_{2M}}  \\
   \vdots  & \vdots  & \ddots  & \vdots   \\
   1 & {{x}_{N1}} & \cdots  & {{x}_{NM}}  \\
\end{matrix} \right].
\end{equation*}

Linear regression analysis~\cite{seber2012linear} aims for estimating the unknown parameter ${\widehat{\pmb \varepsilon}}$ relying on the least squares (LS) criterion. The corresponding solution can be expressed as:
\begin{equation}\label{linear solution}
{\widehat{\pmb \varepsilon}}=({\pmb X}^{T}\pmb X)^{-1}{\pmb X}^{T}\pmb y.
\end{equation}

By contrast, in logistic regression~\cite{hosmer2013applied}, the dependent variable is binary. In order to facilitate our analysis, in the following we consider the case of a binary dependent variable, for example. The goal of the binary logistic regression is to model the probability of the dependent variable having the value of $0$ or $1$, given the training samples. To elaborate a little further, let the binary dependent variable $y$ depend on $M$ independent variables $\pmb x=[x_{1}, x_{2},\dots, x_{M}]$. The conditional distribution of $y$ under the condition of $\pmb x$ obeys a Bernoulli distribution. Hence, the probability of $\Pr(y=1 \mid \pmb x)$ can be expressed in the form of a standard logistic function\footnote{The logistic function is a common ``S'' shape function, which is the cumulative distribution function (CDF) of the logistic distribution.}, also termed as a sigmoid function:
\begin{equation}\label{logistic}
P\triangleq \Pr(y=1 \mid \pmb x)=\frac{1}{1+e^{-g({\pmb x})}},
\end{equation}
where $g({\pmb x})=w_0+w_1x_{1}+w_2x_{2}+\dots+w_Mx_{M}$ and $\pmb w=[w_{0}, w_{1},\dots, w_{M}]$ represents the regression coefficient vector. Similarly, we have:
\begin{equation}\label{logistic2}
\Pr(y=0 \mid \pmb x)=1-P=\frac{1}{1+e^{g({\pmb x})}}.
\end{equation}
Relying on the aforementioned definitions, we have $g({\pmb x})=\ln (\frac{P}{1-P})$.
Hence, for a given dependent variable, the probability of its value being $y_{n}$ can be expressed  by $P(y_{n})=P^{y_n}(1-P)^{1-y_n}$.
Given a set of training samples $\{y_n, x_{n1}, x_{n2},\dots, x_{nM}\}, n=1,2,\dots,N$, we are capable of estimating the regression coefficient vector $\pmb w=[w_{0}, w_{1},\dots, w_{M}]$ with the aid of the maximum likelihood estimation (MLE) method. Explicitly, logistic regression can be deemed to form a special case of the generalized linear regression family using kernel model.

\begin{figure*}
  \centering
 \includegraphics[width=0.90\textwidth]{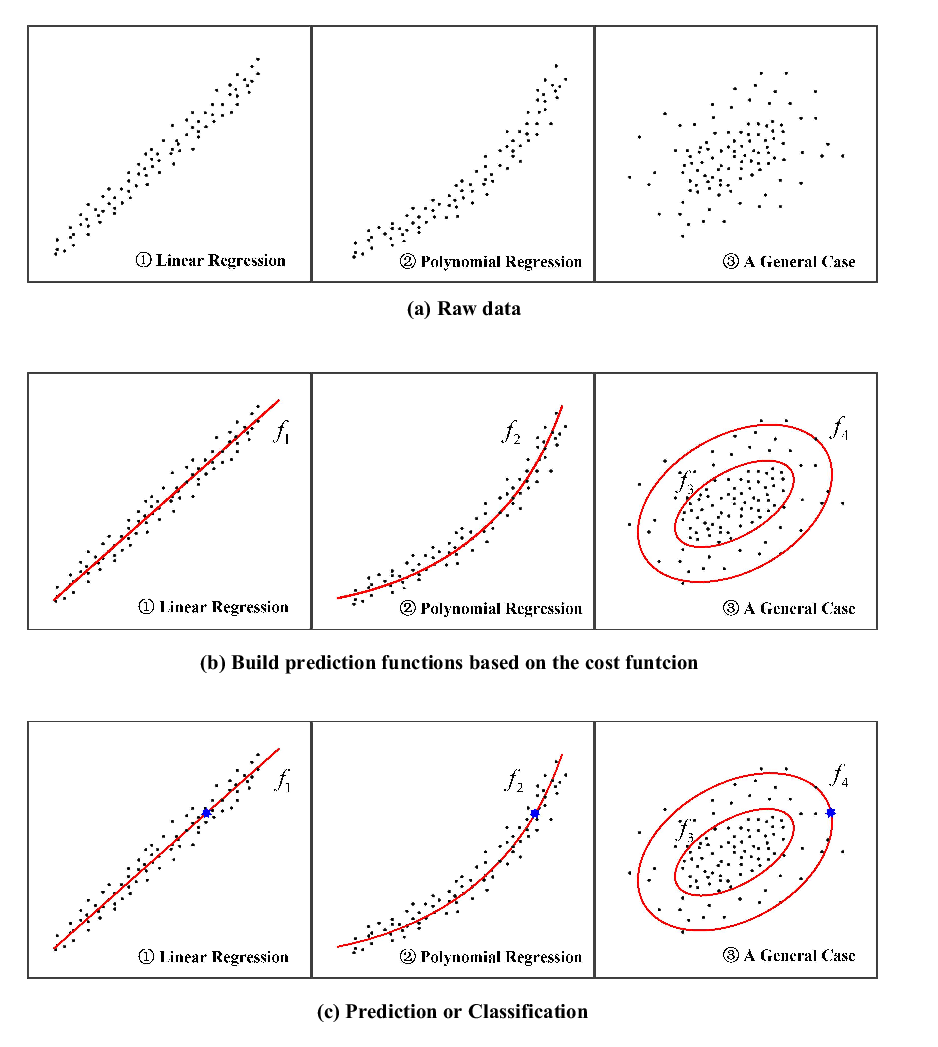}\\
  \caption{The basic flow of a regression model.}\label{regression fig}
\end{figure*}

Furthermore, there exist numerous other useful regression models~\cite{max1976segmented,hoerl1970ridge,tibshirani1996regression,zou2005regularization}. When the dependent variable is a polynomial function of the independent variables, we refer to it as polynomial regression~\cite{max1976segmented}, where the best-fit line is a curve. Moreover, ridge regression~\cite{hoerl1970ridge}, least absolute shrinkage and selection operator (LASSO) regression~\cite{tibshirani1996regression} and ElasticNet regression~\cite{zou2005regularization} are widely applied, when independent variables are of multi-collinear nature and highly correlated.
Fig.~\ref{regression fig} demonstrates the basic flow of a regression model.

\subsubsection{Applications}

The regression models can be used for estimating, detecting and predicting physical layer radio parameters related to wireless network scenarios. Specifically, Chang \textit{et al}.~\cite{chang2016accuracy} proposed a novel regression-aided interference model, which characterized the relationship between the SINR and the packet reception ratio, and evaluated its accuracy relying on the statistics.
Based on this model, they constructed an analytic framework for striking a trade-off between the overhead imposed and the accuracy of interference measurement attained. In~\cite{umebayashi2018efficient}, Umebayashi \textit{et al}. used regression analysis for formulating a deterministic-stochastic hybrid model for detecting the spectrum usage by PUs, which had a reduced number of parameters and yet maintained a high detection accuracy. In~\cite{al2017estimating}, Al Kalaa \textit{et al}. used logistic regression for estimating the likelihood of Wi-Fi and ZigBee wireless coexistence in the context of medical devices. Furthermore, Xiao \textit{et al}.~\cite{xiao2018phy} constructed a logistic regression-aided physical layer authentication model for detecting spoofing attacks in wireless networks without relying on a known channel model, which exhibited a high detection accuracy, despite its low computational complexity.

The regression models can also be employed for solving both estimation and detection problems in the upper layers of the seven-layer OSI model. For example, Chang \textit{et al}. derived a regression-based analytical model for the sake of estimating the contention success probability considering heterogeneous sensor-traffic demands, which beneficially improved the channel's exploitation in IoT~\cite{chang2018traffic}.
Moreover, in~\cite{chen2017learning}, Chen \textit{et al}. employed a regression model for reconstructing the radio map with the aid of signal strength models for the path planning and UAV-location design in UAV-assisted wireless networks.
As a further advance, Lei \textit{et al}.~\cite{lei2018fingerprint} employed a logistic regression classifier for device-free localization relying on fingerprint signals, which yielded a low localization error.

\subsection{KNN and Its Applications}
\subsubsection{Methods}
\begin{figure*}
  \centering
 \includegraphics[width=0.90\textwidth]{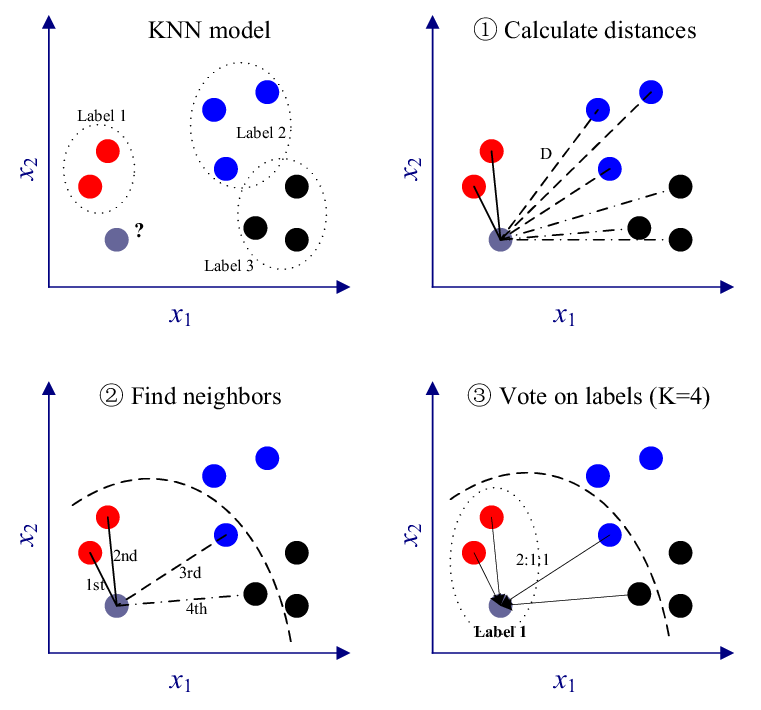}\\
  \caption{The illustration of the unweighted KNN mechanism with $K=4$, for example.}\label{knnmodel}
\end{figure*}

KNN constitutes a non-parametric instance-based learning method, which can be used both for classification and regression. Proposed by Cover and Hart in 1968, the KNN algorithm is one of the simplest of all ML algorithms. By relying on the distance between the object and training samples in a feature space, the KNN algorithm determines which class of the object belongs to. Specifically, in a classification scenario,
an object is categorized into a specific class by a majority vote of its $K$ nearest neighbors. If $K=1$, the category of the object is the same as that of its nearest neighbor. In this case, it is termed as the one nearest neighbour classifier. By contrast, in a regression scenario, the output value of the object is calculated by the average of the value of its $K$ nearest neighbors. Fig.~\ref{knnmodel} shows the illustration of the unweighted KNN mechanism associated with $K=4$.

Let us assume that there are $N$ training sample pairs of $\{(\pmb x_1,y_1),(\pmb x_2,y_2),\dots,(\pmb x_N,y_N)\}$, where $y_n$ is the property value or class label of the sample $\pmb x_n$, $n=1,2,\dots,N$. Typically, we use the Euclidean distance or the Manhattan distance~\cite{sherwood2002automatically} for calculating the similarity between the object $\overline{\pmb x}$ and the training samples. Let ${\pmb x_n}=[x_{n1},x_{n2},\dots,x_{nM}]$ contain $M$ different features. Hence, the Euclidean distance between $\overline{\pmb x}$ and $\pmb x_n$ can be expressed by:
\begin{equation}\label{knn1}
d_e=\sqrt {\sum \limits_{m=1}^{M}{(\overline{x}_m-x_{nm})^2}},
\end{equation}
while their Manhattan distance is calculated as~\cite{sherwood2002automatically}:
\begin{equation}\label{knn2}
d_m=\sum\limits_{m=1}^{M}{|\overline{x}_m-x_{nm}|}.
\end{equation}
Relying on the associated similarity, the class label or property value of $\overline{\pmb x}$ can be voted on or first weighted and then voted on by its $K$ nearest neighbors, which is formulated:
\begin{equation}\label{knn3}
\overline{y}\leftarrow\textrm{VOTE}\ \{K\ \textrm{nearest}\ (\pmb x_k,y_k)\}.
\end{equation}

The performance of the KNN algorithm critically depends on the value of $K$, whilst
the best choice of $K$ hinges upon the training samples. In general, a large $K$ is conducive to resisting the harmful influence of noise, but it fuzzifies the class boundary between different categories. Fortunately, an appropriate value of $K$ can be determined by a variety of heuristic techniques based on the true characteristics of the training data set.

\subsubsection{Applications}
In KNN, an object can be classified into a specific category by a majority vote of the object's neighbours, with the object being assigned to the class that is the most common one among its $K$ nearest neighbors. Hence, as a kind of simple and efficient classification algorithms, KNN is beneficial in terms of, for example, traffic prediction~\cite{feng2017proactive}, anomaly detection~\cite{xie2013scalable,onireti2016cell}, missing data estimation~\cite{pan2010k}, modulation classification~\cite{aslam2012automatic}, interference elimination~\cite{yu20145}, etc.

To elaborate, for the sake of capturing the dynamic characteristics of wireless resource demands, Feng \textit{et al}. constructed a weighted KNN model by learning from a large-scale historical data set generated by cellular operators' networks, which was used for exploring both the temporal and spatial characteristics of radio resources~\cite{feng2017proactive}. In~\cite{xie2013scalable}, Xie \textit{et al}. proposed a novel KNN aided online anomaly detection scheme based on hypergrid intuition in the context of WSN applications for overcoming the `lazy-learning' problem~\cite{bontempi1999lazy} especially when the computational resource and the communication cost quantified in terms of bandwidth and energy were constrained.
Moreover, in~\cite{onireti2016cell}, Onireti \textit{et al}. proposed a KNN based anomaly detection algorithms for improving the outage detection accuracy in dense heterogeneous networks.
As for missing data estimation, a KNN assisted missing data estimation algorithm was conceived on the basis of the temporal and spatial correlation feature of sensor data, which jointly utilized the sensor data from multiple neighbor nodes~\cite{pan2010k}. Furthermore, Aslam \textit{et al}.~\cite{aslam2012automatic} combined genetic programming and the KNN in order to improve the modulation classification accuracy, which can be viewed as a reliable modulation classification scheme for the SU in cognitive radio networks. In~\cite{yu20145}, the KNN algorithm was used both for extracting the environmental interference imposed by 5G Wi-Fi signals and for reducing the computational complexity and yet improving the performance of indoor localization.

\subsection{SVM and Its Applications}
\subsubsection{Methods}
Being constructed purely by mathematical theory, SVM is another supervised learning model conceived for classification and regression relying on constructing a hyperplane or a set of hyperplanes in a high-dimensional space.
The best hyperplane is the one that results in the largest margin amongst the classes. However, the training data set may often be linearly non-separable in a finite dimensional space. To address this issue, SVM is capable of mapping the original space into a higher dimensional space, where the training data set can be more easily discriminated.

Considering a linear binary SVM, for example, there are $N$ training samples in the form of $\{(\pmb x_1,y_1),(\pmb x_2,y_2),\dots,(\pmb x_N,y_N)\}$, where $y_n=\pm 1$ indicates the class label of the point $\pmb x_n$. SVM aims for searching for a hyperplane having the maximum possible separation from the training samples, which best discriminates the two classes of $\pmb x_n$ associated with $y_n=1$ and $y_n=-1$. Here, the maximum separation implies having the maximum possible distance between the nearest point and the hyperplane. The hyperplane is represented by:
\begin{equation}\label{svm1}
\pmb \omega^{T} \pmb x + b=0.
\end{equation}
Hence, we can quantify the separation of the training sample $(\pmb x_n,y_n)$ as:
\begin{equation}\label{svm2}
\gamma_{n}=y_n(\pmb \omega^{T} \pmb x_n + b).
\end{equation}
Moreover, we assume having the correct classification if $\pmb \omega^{T} \pmb x_n + b\geq 0$ when $y_n=1$, while $\pmb \omega^{T} \pmb x_n + b\leq 0$ when $y_n=-1$.
Because we have $y_n(\pmb \omega^{T} \pmb x_n + b)\geq 0$, a higher separation implies a more reliable classification. Again, the SVM tries to find the optimal hyperplane that maximizes the minimum separation between the training samples and the hyperplane considered. Given a set of linearly separable training samples, after the operation of normalization, the SVM based classification can be formulated as the following optimization problem:
\begin{equation}\label{svm3}
\begin{aligned}
\max_{\pmb \omega,b} \ &\min_{n=1,\dots,N} y_n\left(\left(\frac{\pmb \omega}{\|\pmb \omega \|}\right)^{T} \pmb x_n + \frac{b}{\| \pmb \omega \|}\right) \\
\mathrm{s.t.} \ &y_n(\pmb \omega^{T} \pmb x_n + b)\geq \gamma, n=1,2,\dots,N, \\
      &\| \pmb \omega \|=1,
\end{aligned}
\end{equation}
where we have $\gamma=\min\limits_{n=1,\dots,N} y_n\left(\left(\frac{\pmb \omega}{\|\pmb \omega \|}\right)^{T} \pmb x_n + \frac{b}{\| \pmb \omega \|}\right)$. After some further mathematical manipulations, the problem in (\ref{svm3}) can be reduced to an optimization problem having a convex quadratic objective function and linear constraints, which can be expressed by:
\begin{equation}\label{svm4}
\begin{aligned}
\min_{\pmb \omega,b} \ &\frac{1}{2} (\| \pmb \omega \|)^2 \\
\mathrm{s.t.} \ &y_n(\pmb \omega^{T} \pmb x_n + b)\geq 1, n=1,2,\dots,N.\\
\end{aligned}
\end{equation}
Problem~(\ref{svm4}) is a typical convex optimization problem. Taking advantage of Lagrange duality~\cite{boyd2004convex}, we can obtain the optimal $\pmb \omega$ and $b$.

Again, if the training samples are linearly non-separable, SVM is capable of mapping data to a high dimensional feature space with a high probability of being linearly separable. This may result in a non-linear classification or regression in the original space. Fortunately, kernel functions play a critical role in avoiding the ``curse of dimensionality'' in the above-mentioned dimensionality ascending procedure~\cite{bergman1950kernel,scholkopf2001learning}. To elaborate a little further, given the original input samples $\pmb x$, we may be interested in learning some features $\phi(\pmb x)$. Let us assume $\pmb x_m, \pmb x_n \in \mathbb{R}^n$, hence the corresponding kernel function $K(\pmb x_m, \pmb x_n)$ is defined as:
\begin{equation}\label{kernel}
K(\pmb x_m, \pmb x_n)=\phi(\pmb x_m)^{T} \phi(\pmb x_n).
\end{equation}
Fortunately, even though the high dimensional feature mapping $\phi(\pmb x_n)$ may be expensive to calculate, the kernel function calculated relying on their inner product can be easy obtained after some further mathematics manipulations.

There are a variety of alternative kernel functions, such as linear kernel function, polynomial kernel function, radial basis kernel function, neural network kernel function, etc. Furthermore, some regularization methods haven been conceived in order to make SVM be less sensitive to outlier points.

The specific choice of the kernel function plays a key role in ML~\cite{hofmann2008kernel}, hence we have to beneficially design the kernel function. The construction of kernels can be generally developed by the inner product operations of feature mappings $\phi(\pmb x_n)$ between the input samples over the Hilbert space, whose infinite number of dimensions allow the appropriate representation of big data to exploit their geometric properties. Such a Hilbert space associated with a kernel invoked for producing functions by calculating the inner product of the feature mappings is known as the reproducing kernel Hilbert space (RKHS)~\cite{parzen1962extraction}, and has been applied in diverse learning contents~\cite{fukumizu2004dimensionality,kivinen2004online}. The RKHS therefore serves a critical foundation in statistical learning theory. Fig.~\ref{kernel} provides a graphical illustration of the kernel-based method.

On the other hand, we may rely on statistical learning theory for appropriately constructing the signal space in order to identify sufficient statistics for reliable signal detection and estimation in statistical communication theory~\cite{chhabra2006principles}. Inspired by Parzen~\cite{parzen1962extraction}, Kailath observed that RKHS may also be beneficially invoked both for detection and estimation~\cite{kailath1971rkhs} by exploiting the one-to-one relationship between RKHS and finite-variance linear functionals of a random process. Corresponding to the simplest setup of signal detection in additive white Gaussian noise (AWGN) using the Karhunen-Loeve expansion~\cite{fukunaga1970application}, the RKHS representation associated with the noise covariance function is capable of providing an equivalent theoretical framework of statistical communication theory. After a series of efforts inverted into different areas of signal detection and estimation, Kailath and Poor~\cite{kailath1998detection} conceived the RKHS approach for the detection of stochastic signals.

\begin{figure*}[!t]
  \centering
  \includegraphics[width=0.6\textwidth]{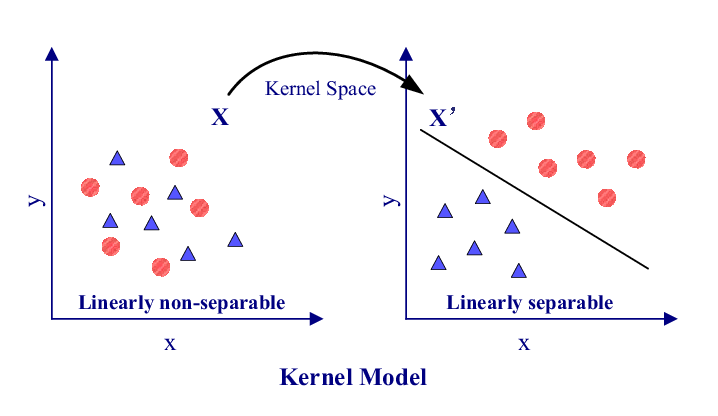}\\
  \caption{An illustration of the kernel-based methods.}
  \label{kernel}
\end{figure*}

\subsubsection{Applications}
\begin{figure*}
  \centering
 \includegraphics[width=0.90\textwidth]{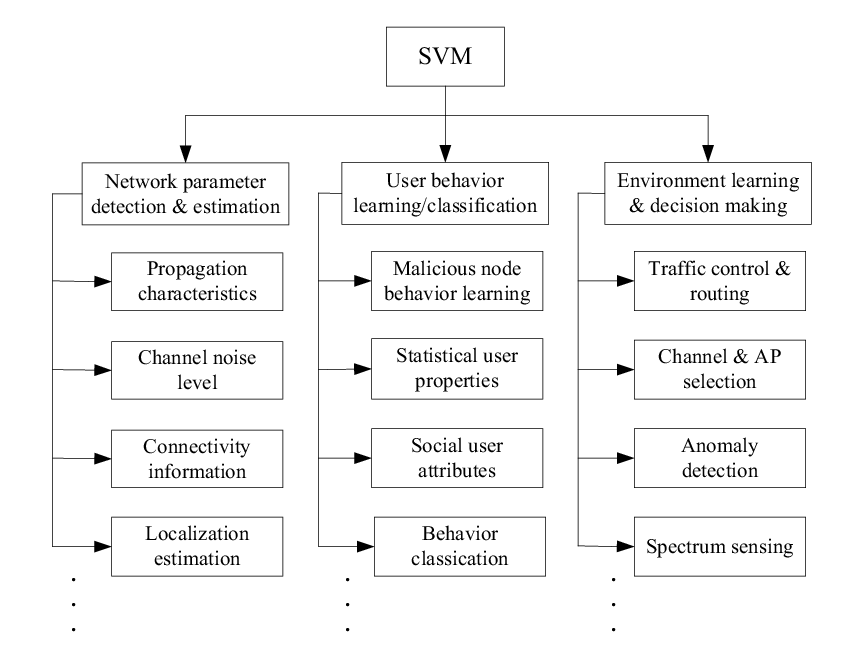}\\
  \caption{The applications of SVM aided learning models for wireless networks.}\label{svmmodel}
\end{figure*}

As mentioned before, SVM hinges on a mapping that can transform the original training data into a higher dimension, where the events to be classified do become linearly separable. Then it searches for the optimal separating hyperplane
for delineating one class from another in this higher dimension considered.
As highlighted in Fig.~\ref{svmmodel}, in the spirit of this, SVM aided learning models can be used for detecting and estimating network parameters, for learning and classifying environmental signals and the user's behavior, as well as for guiding decision making concerning channel selection and anomaly detection, for example~\cite{feng2012determination,tran2008localization,sun2005robust,donohoo2014context,joseph2011cross,pianegiani2008energy,thilina2016dccc,yang2013detection,rajasegarar2008anomaly}.

As for detecting and estimating the network parameters, Feng and Chang~\cite{feng2012determination} constructed a hierarchical SVM (H-SVM) structure for multi-class data estimation. The H-SVM was constructed by a number of levels and each level was composed by a finite number of SVM classifiers. Feng and Chang used their H-SVM model both for estimating the physical locations of nodes in an indoor wireless network and the Gaussian channel's noise level in a MIMO-aided wireless network. Thanks to its hierarchical structure, the H-SVM was capable of providing an efficient distributed estimation procedure. Furthermore, Tran \textit{et al}. proposed an SVM model for estimating the geographic location of sensor nodes in WSNs whilst only relying on their connectivity information, more precisely the hop counts~\cite{tran2008localization}. It yielded fast convergence in a distributed manner. The final estimation error can be upper bounded by any small threshold upon relying on a sufficiently large training dataset. Moreover, Sun and Guo~\cite{sun2005robust} conceived a least square-SVM (LS-SVM) algorithm for estimating the user's position by correlating the time-of-arrival (TOA) of radio frequency signals at the BSs without any detailed knowledge about the base station's location as well as about the propagation characteristics.

SVM can also be used for learning a user's behavior and for classifying environmental signals considering the complex spatio-temporal context and the diverse selection of devices.
In~\cite{donohoo2014context}, Donohoo \textit{et al.} studied the context-aware energy-efficiency improvement options for smart devices.
These solutions may become beneficial in terms of configuring their location-specific interface for heterogeneous networks (HetNets) constituted by diverse cells.
In~\cite{joseph2011cross}, by combining the SVM and Fisher discriminant analysis (FDA) Joseph \textit{et al.} learned the malicious sinking behavior in wireless ad hoc networks for finding the security vulnerabilities and for designing novel intrusion detection scheme. Moreover, features such as delay between data and acknowledgement, number of re-transmits, etc. gleaned from the MAC layer were jointly considered with those from other layers, which constituted a correlated feature set.
Furthermore, Pianegiani \textit{et al.}~\cite{pianegiani2008energy} proposed an SVM-based binary classification solution for classifying acoustic signals emitted by vehicles relying on spectral analysis aided feature extraction, which was beneficial in terms of improving the classification accuracy, despite reducing the implementation complexity.

As for the SVM's benefit in assisting decision making, in~\cite{thilina2016dccc}, a common control channel selection mechanism was conceived for SUs during a given frame relying on an SVM-based learning technique proposed for a cognitive radio network, which was capable of implicitly and cooperatively learning the surrounding environment cooperatively in an online way.
Moreover, Yang \textit{et al.}~\cite{yang2013detection} investigated the spoofing attack detection problem based on the spatial correlation of received signal strength gleaned from network nodes, where a cluster-based SVM mechanism was
developed for determining the number of attackers. Relying on carefully designed certain training data, the SVM algorithm employed further improved the accuracy of determining the number of attackers.
Rajasegarar \textit{et al.}~\cite{rajasegarar2008anomaly} also investigated the malicious
activity detection issues of WSNs invoking a variety of SVM based algorithms.

\subsection{Bayes Classification and Its Applications}
\subsubsection{Methods}
The Bayes classifier, a popular member of the probabilistic classifier family relying on Bayes' theorem, operates by computing the \textit{posteriori} probability distribution of the objective function values given a set of training samples. As a widely-used classification method, the naive Bayes classifier can be trained for example conditioned on a simple but strong independence assumption in features. Furthermore, the complexity of training a naive Bayes model is linearly proportional to the training set size.

To elaborate a little further, let the vector $\pmb x=[x_1, x_2,\dots, x_M]$ represent $M$ independent features for a total of $K$ classes $\{y_1,y_2,\dots,y_K\}$. For each of the $K$ possible class labels $y_k$, we have the conditional probability of $p=(y_k | x_1,\dots, x_M)$.
Relying on Bayes' theorem, we decompose the conditional probability to yield the form of:
\begin{equation}\label{nb1}
p=(y_k | x_1,\dots, x_M)=\frac{p(y_k)p(x_1,\dots, x_M | y_k )}{p(x_1,\dots, x_M)},
\end{equation}
where $p=(y_k | x_1,\dots, x_M)$ is the \textit{posteriori} probability, whilst $p(y_k)$ is the \textit{priori} probability of $y_k$.
Given that $x_i$ is conditionally independent of $x_j$ for $i\neq j$, we have:
\begin{equation}\label{nb2}
p=(y_k | x_1,\dots, x_M)=\frac{p(y_k)}{p(x_1,\dots, x_M)}\prod\limits_{m=1}^{M}{ p(x_m | y_k )},
\end{equation}
where $p(x_1,\dots, x_M)$ only depends on $M$ independent features, which can be viewed as a constant.

The maximum a \textit{posteriori} probability (MAP) is used as the decision making rule for the naive Bayes classifier. Given a feature vector $\overline{\pmb x}=(\overline{x}_1, \overline{x}_2,\dots, \overline{x}_M)$, its label $\overline{y}$ can be determined according to:
\begin{equation}\label{nb3}
\overline{y}= \argmax\limits_{y_k \in \{y_1,\dots,y_K\}} \ p(y_k)\prod\limits_{m=1}^{M}{ p(\overline{x}_m | y_k )}.
\end{equation}
Despite idealized simplifying assumptions, naive Bayes classifiers have enjoyed popularity in numerous complex real-world situations, such as outlier detection~\cite{agrawal2015survey}, spam filtering~\cite{feng2016support}, etc.

\subsubsection{Applications}
Based on the Bayes' theorem, Bayes classifier techniques are particularly applicable to the context where the dimensionality of the input is high. Despite their simplicity, they can often outperform other sophisticated classification methods. As for their applications in wireless networks, in the following, we will elaborate on some typical examples in different wireless scenarios, such as antenna selection, network association, anomaly detection, indoor location and QoE prediction.

Specifically, in~\cite{he2018transmit}, He \textit{et al.} modeled the transmit antenna selection (TAS) problem of MIMO wiretap channels as a multi-class classification problem. Then, they used the naive Bayes-based classification scheme to select the optimal antenna for enhancing the physical layer security of the system considered. In contrast to conventional TAS schemes, simulation results showed that the proposed scheme resulted in a reduced feedback overhead at a given secrecy performance.
In~\cite{abouzar2011action}, Abouzar \textit{et al.} proposed an action-based network association technique for wireless body area networks (WBANs).
Relying on the level of received signal strength indicator of the on-body link, the naive Bayes algorithm was employed to recognize the ongoing action, which was beneficial in terms of scheduling the time slot assignment in the context of fixed power allocation on various links by the sink node under a specific data rate constraint.
Moreover, Klassen \textit{et al.}~\cite{klassen2012anomaly} used the naive Bayes classifier for detecting
anomaly in \textit{ad hoc} wireless network involving the black hole attack, the denial of service (DoS) attack and the selective forwarding attack.

Bayes classifier can also be applied to the indoor location estimation. For example, in~\cite{ouyang2012indoor}, a probabilistic model was conceived for characterizing the relationship between the received signal strength and location with the aid of the naive Bayes generative learning method, which was used for learning the parameters of an initial probabilistic model, given a limited number of labeled samples. The proposed indoor location estimation method was capable of both reducing the off-line calibration efforts required, whilst maintaining a high location estimation accuracy.
Furthermore, as for QoE prediction, in order to evaluate the impact of different networking and channel conditions on the QoE attained in the context of different network services, Charonyktakis \textit{et al.}~\cite{charonyktakis2016user} proposed a modular algorithm for user-centric QoE prediction. They integrated multiple ML algorithms, including the Gaussian naive Bayes classifier and conceived a nested cross validation protocol for selecting the optimal classifier and its corresponding optimal hyper-parameter value for the sake of accurate QoE prediction.

\begin{table*}[t!]
\begin{center}
\scriptsize
\caption{Compelling Applications of Supervised Learning in Wireless Networks}
\label{tbsup}
\begin{tabular}{ | c | l | c | r |}
\hline
Paper &  Application &  Method & Description  \\ \hline
\cite{chang2016accuracy} & interference estimate &  regression  &  strike a trade-off between the overhead and accuracy of interference measurement \\ \hline
\cite{umebayashi2018efficient} & spectrum sensing & regression &  reduce the number of parameters and maintain a high detection accuracy\\ \hline
\cite{al2017estimating} &  wireless coexistence & regression & estimate the likelihood of the wireless coexistence of Wi-Fi and ZigBee\\ \hline
\cite{xiao2018phy} &  PHY authentication & regression & do not need the assumption on the accurate known channel model\\ \hline
\cite{chang2018traffic} &  traffic estimation & regression & estimate the contention success probability considering sensors' heterogeneous traffic demands\\ \hline
\cite{chen2017learning} & map reconstruction & regression & reconstruct the wireless radio map for UAV path planning and location design\\ \hline
\cite{lei2018fingerprint} &  wireless localization &  regression &  logistic regression classifier for counteracting the negative influence relying on fingerprint signals\\ \hline
\cite{feng2017proactive} & traffic prediction & KNN & explore both the temporal and spatial characteristics of radio resources\\ \hline
\cite{xie2013scalable} & anomaly detection & KNN &  rely on the hypergrid intuition in the context of WSN applications \\ \hline
\cite{pan2010k} &  missing data estimation & KNN & rely on the temporal and spatial correlation feature of sensor data \\ \hline
\cite{aslam2012automatic} & modulation classification & KNN & combine the genetic programming and KNN for improving the modulation classification accuracy \\ \hline
\cite{yu20145} &  interference elimination & KNN & extract environmental interference from Wi-Fi signal and reduce computational complexity \\ \hline
\cite{feng2012determination} &  data estimation & SVM &  provide an efficient estimation procedure in a distributed manner \\ \hline
\cite{tran2008localization} & localization estimation & SVM & yield fast convergence performance and efficiently use the communication resources\\ \hline
\cite{sun2005robust} &  user location & SVM &  without knowledge about base station location and environmental propagation characteristics \\ \hline
\cite{donohoo2014context} &  data prediction & SVM &  provide location-specific interface configuration for HetNets\\ \hline
\cite{joseph2011cross} &  behavior learning & SVM & combine both the superior accuracy of SVM and fast convergence speed of FDA \\ \hline
\cite{pianegiani2008energy} & signal classification & SVM & classify acoustic signals emitted by vehicles rely on feature extraction \\ \hline
\cite{thilina2016dccc} &  channel selection & SVM & propose a control channel selection mechanism for a cognitive radio network \\ \hline
\cite{yang2013detection} &  attacker counting & SVM & develop a cluster-based SVM mechanism for determining the number of attackers \\ \hline
\cite{he2018transmit} &  antenna selection & Bayes &  enhance the physical layer security relying on Bayes-based optimal antenna selection \\ \hline
\cite{abouzar2011action} &  network association & Bayes & schedule time slot assignment and fixed power allocation under data rate constraint \\ \hline
\cite{klassen2012anomaly} & anomaly detection & Bayes & detect
anomaly involving black hole attack, DoS attack and selective forwarding attack\\ \hline
\cite{ouyang2012indoor} &  indoor location & Bayes & characterize the relationship between the received signal strength and location\\ \hline
\cite{charonyktakis2016user} &  QoE prediction & Bayes & accurate QoE prediction by selecting optimal classifier and optimal hyper-parameter values\\ \hline
\end{tabular}
\end{center}
\end{table*}

\section{Unsupervised Learning in Wireless Networks}
\label{Unsupervised Learning in NGWN}
In this section, we will highlight some typical unsupervised learning algorithms, such as $K$-means clustering~\cite{hartigan1979algorithm}, expectation-maximization (EM)~\cite{moon1996expectation}, principal component analysis (PCA)~\cite{wold1987principal} and independent component analysis (ICA)~\cite{comon1994independent} in terms of their methodology and their applications in wireless networks.
Table~\ref{tbunsup} summarizes some typical applications of the above-mentioned unsupervised learning algorithms in wireless networks.

\subsection{$K$-Means Clustering and Its Applications}
\subsubsection{Methods}
$K$-means clustering is a distance based clustering method that aims for partitioning $N$ unlabeled training samples into $K$ different cohesive clusters, where each sample belongs to one cluster. To elaborate a little further, $K$-means clustering measures the similarity between two samples in terms of their distance and it has two main steps, namely assigning each training sample to one of $K$ clusters in terms of the closest distance between the sample and the cluster centroids, and then updating each cluster centroid according to the mean of the samples assigned to it. The whole algorithm is hence implemented by repeatedly carrying out the above-mentioned pair of steps until convergence is achieved.

To elaborate a little further, given a set of samples $\{\pmb x_1, \pmb x_2,\dots,\pmb x_N\}$, where $\pmb x_n=[x_{n1},x_{n2},\dots,x_{nM}]$ is a $M$-dimensional vector, let $\mathbb{S}=\{s_1, s_2,\dots,s_K\}$ represent the above-mentioned cluster set, and $\pmb \mu_k$ the mean of the samples in $s_k$.
$K$-means clustering intends to find an optimal cluster-based segmentation, which solves the following optimization problem:
\begin{equation}\label{kc1}
\mathbb{S}^{\ast}= \argmin\limits_{\{s_1, s_2,\dots,s_K\}}\ \sum_{k=1}^{K}{\sum_{\pmb x\in s_k} {{\|\pmb x -\pmb \mu_k\|}^2}}.
\end{equation}
However, problem (\ref{kc1}) is a non-deterministic polynomial-time hardness (NP-hard) problem~\cite{paz1981non}. Fortunately, there are a range of efficient heuristic algorithms, which converge quickly to a local optimum.

One of the popular low-complexity iterative refinement algorithms suitable for $K$-means clustering is Lloyd's algorithm~\cite{kanungo2002efficient}, which often yields satisfactory performance after a low number of iterations. Specifically, given $K$ initial cluster centroid $\pmb \mu_k, k=1,\dots,K$, Lloyd's algorithm arrives at the final cluster segmentation result by alternating between the following two steps,
\begin{itemize}
  \item Step~1: In the iterative round $r$, assign each sample to a cluster.
  For $n=1,2 \dots, N$ and $i,k=1,2 \dots, K$, if we have:
  \begin{equation}\label{kc2}
   s_i^{(r)}=\{\pmb x_n:\ {\|\pmb x_n- \pmb \mu_i^{(r)}\|}^2 \leq {\|\pmb x_n- \pmb \mu_k^{(r)}\|}^2, \forall k\},
  \end{equation}
  then we assign the sample $\pmb x_n$ to the cluster $s_i$, even if it could potentially be assigned to more than one cluster.
  \item Step~2: Update the new centroids of the new clusters formulated in the iterative round $r$ relying on:
  \begin{equation}\label{kc3}
   \pmb \mu_i^{(r+1)}=\frac{1}{|s_i^{(r)}|}\sum\limits_{\pmb x_j \in s_i^{(r)}}{\pmb x_j},
  \end{equation}
  where $|s_i^{(r)}|$ denotes the number of samples in cluster $s_i$ in iterative round $r$.
\end{itemize}
Convergence is deemed to be obtained when the assignment in Step~1 is stable. Explicitly, reaching convergence means that the clusters formulated in the current round are the same as those formed in the last round.
Since this is a heuristic algorithm, there is no guarantee that it can converge to the global optimum. Hence, the result of clustering largely relies on specific choice of the initial clusters and on their centroids.

\subsubsection{Applications}
$K$-means clustering aims for partitioning $N$ samples into $K$ clusters. Each sample belongs to the closest cluster. The clustering algorithm proceeds in an iterative manner, where the in-cluster differences are minimized by iteratively updating the cluster centroid, until convergence is achieved.

Clustering functioning under uncertainty or incomplete information is a common problem in wireless networks, especially in the scenarios associated with numerous small traffic cells, heterogeneous large and small cell structures relying on diverse carrier frequencies, diverse time-varying tele-traffic, etc.
First of all, the small cells have to be carefully clustered for avoiding excessive interference using coordinated multi-point transmission. Moreover, the devices and users should be beneficially clustered for the sake of achieving a high energy efficiency, maintaining an optimal access point association, obeying an efficient offloading policy, and of guaranteeing a high network security.
In~\cite{xia2012optical}, a mixed integer programming problem was formulated for jointly optimizing both the gateway deployment and the virtual-channel allocation for optical/wireless hybrid networks, where Xia \textit{et al.} designed an efficient $K$-means clustering based solution for iteratively solving this problem, which beneficially reduced the delay, as well as improved the network throughput. Moreover, in~\cite{hajjar2017hybrid}, Hajjar \textit{et al.} proposed a $K$-means based relay selection algorithm for creating small cells under the umbrella of an oversailing LTE macro cell within a multi-cell scenario under the constraint of low power clusters. Relying on the proposed relay selection algorithm, the total capacity was increased by reusing the frequency in each low power cluster, which had the benefit of supporting high data rate services.
Additionally, Cabria and Gondra~\cite{cabria2017potential} proposed a so-called potential-$K$-means scheme for partitioning data collection sensors into clusters and then for assigning each cluster to a storage center.
The proposed $K$-means solution had the advantage of both balancing the storage center loads and minimizing the total network cost (optimizing the total number of sensors).
Parwez \textit{et al.}~\cite{parwez2017big} invoked both $K$-means clustering and hierarchical clustering algorithms for their user-activity analysis and user-anomaly detection in a mobile wireless network, which verified genuine identity of users in the face of their dynamic spatio-temporal activities. Furthermore, El-Khatib~\cite{el2010impact} designed a $K$-means classifier for selecting the optimal set of features of the MAC layer bearing in mind the specific relevance of each feature, which beneficially improved the accuracy of intrusion detection, despite reducing the learning complexity.

Clustering can also be used in signal detection for the sake of both reducing the detection complexity and for improving the energy efficiency attained.
In~\cite{liang2016coding}, the $K$-means clustering algorithm was invoked in a blind transceiver, where the training process was completely dispensed within the transmitter for reducing its energy dissipation, since no pilot power was required. Furthermore, Zhao \textit{et al.}~\cite{zhao2006k} conceived an efficient $K$-means clustering algorithm for optical signal detection in the context of burst-mode data transmission.

\subsection{EM and Its Applications}
\subsubsection{Methods}
\begin{figure*}
  \centering
 \includegraphics[width=0.60\textwidth]{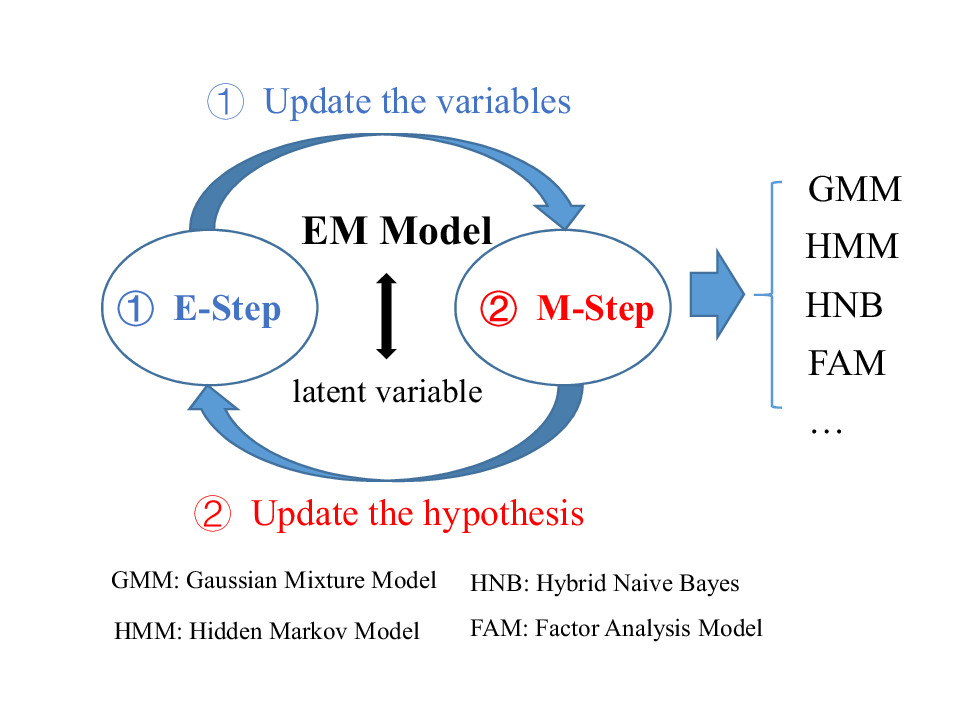}\\
  \caption{The operation process of the EM model.}\label{emmodel}
\end{figure*}
The EM algorithm is an iterative method conceived for searching for the maximum likelihood estimate of parameters in a statistical model. Typically, in addition to unknown parameters whose existence has been ascertained, the statistical model also has some latent variables. In this scenario it is an open challenge to derive a closed-form solution, because we are unable to find the derivatives of the likelihood function with respect to all the unknown parameters and latent variables.
The iterative EM algorithm consists of two steps, as shown in Fig.~\ref{emmodel}. During the expectation step (E-step), it calculates the expected value of the log likelihood function conditioned on the given parameters and latent variables, while in the maximization step (M-step), it updates the parameters by maximizing the specific log-likelihood expectation function considered.

More explicitly, upon considering a statistical model with observable variables $\pmb X$ and latent variables $\pmb Z$, the unknown parameters are represented by $\pmb \theta$. The log-likelihood function of the unknown parameters is given by:
\begin{equation}\label{em1}
l(\pmb \theta;\pmb X, \pmb Z)=\log p(\pmb X, \pmb Z; \pmb \theta).
\end{equation}
Hence, the EM algorithm can be described as follows~\cite{moon1996expectation}:
\begin{itemize}
  \item E-step: Calculate the expected value of the log likelihood function under the current estimate of $\overline{\pmb \theta}$, i.e.
      \begin{equation}\label{em2}
       Q(\pmb \theta | \overline{\pmb \theta})=E_{\pmb Z |\pmb X, \overline{\pmb \theta}}[\log p(\pmb X, \pmb Z; \pmb \theta)].
      \end{equation}
  \item M-step: Maximize Eq.~(\ref{em2}) with respect to $\pmb \theta$ for generating an updated estimate of $\overline{\pmb \theta}$, which can be formulated as:
       \begin{equation}\label{em3}
       \overline{\pmb \theta}'= \argmax_{\pmb \theta} Q(\pmb \theta | \overline{\pmb \theta}).
      \end{equation}
\end{itemize}

The EM algorithm plays a critical role in parameter estimation based on many of the popular statistical models, such as the Gaussian mixture model (GMM), hidden Markov model (HMM), etc. which are beneficial both for clustering and prediction.
\subsubsection{Applications}
The EM model can be readily invoked for a variety of parameter
learning and estimation problems routinely encountered in wireless networks. Specifically, Wen \textit{et al}.~\cite{wen2015channel} estimated both the channel parameters of the desired links in a target cell and those of the interfering links in the adjacent cells relying on constructing a GMM, which was estimated with the aid of
the EM algorithm. Choi \textit{et al}.~\cite{choi2013estimation} modeled the cognitive radio system as a HMM, where the secondary users (SUs) estimated the channel parameters such as the primary user's (PU) sojourn time, signal strength, etc. based on the standard EM algorithm.
Moreover, Assra \textit{et al}.~\cite{assra2016approach} also adopted the EM algorithm to jointly estimate the channel unknown frequency domain responses as well as the noise variance and detected the PU's signal in a cooperative wide-band cognitive system, which was shown to converge to the upper bound solution based on maximum likelihood estimation under the idealized assumption of having perfect channel parameter estimation. Additionally, Zhang \textit{et al}.~\cite{zhang2008joint} proposed an EM aided joint symbol detection and channel estimation algorithm for MIMO-OFDM systems in the presence of frequency selective fading, which provided a distribution-estimate for both the hidden symbol and unknown channel parameters in an iterative manner. Li and Nehorai~\cite{li2015joint} built an asynchronous state-space model for connecting asynchronous observations with the most likely target state transition in the context of multi-sensor WSNs. Then, they adopted the EM algorithm for jointly estimating the sequential target state as well as the network's synchronization state under the assumption of knowing the temporal order of sensor clocks.
Furthermore, Zhang \textit{et al}.~\cite{zhang2017novel} used a variational EM iterative algorithm to recover the transmitted signals and to identify the active users in a low-activity code division multiple access based M2M communications without the knowledge of the user activity factor. The EM algorithm can also be invoked for target or source localization, which can be viewed as a joint sparse signal recovery and parameter estimation problem~\cite{meng2011efficient}~\cite{sun2017multiple}.

\subsection{PCA \& ICA and Their Applications}
\subsubsection{Methods}
Feature reduction and extraction constitutes the preparatory phase of ML, because the initially acquired raw data may contain some irrelevant or redundant features. The feature reduction and extraction processing phase results in a reduced set of features with the aid of feature transforms, conceived for reducing the feature space dimension, by eliminating correlated features.
However, given the diverse definitions of ``features'', it is only realistic to discuss feature-extraction in specific application-oriented contexts.

PCA and ICA constitute sophisticated dimensionality reduction methods in ML, which are capable of reducing both the computational complexity and the storage requirements.

PCA utilizes an orthogonal transformation for converting a set of potentially correlated features of the training samples into a set of uncorrelated features, which are termed as the ``principal components''. The number of principal components is expected to be lower than the number of the original features of the training samples, which hence provide a more compact representation of the original samples. More explicitly, less principle components can be used for representing the original samples in the transformed domain. In PCA, the first principal component tends to have the largest variance, which indicates that it encapsulates the most information of the original features provided that these features were correlated. Similarly, each succeeding component tends to have the next highest variance. These principal components can be generated by invoking the eigenvectors of the normalized covariance matrix.

Specifically, let us consider $N$ training samples of $\{\pmb x_1, \pmb x_2,\dots,\pmb x_N\}$, where $\pmb x_n=[x_{n1},x_{n2},\dots,x_{nM}]^{T}$ is composed of $M$ different features. Let us first pre-process the samples by normalizing their mean and variance. Given a unit vector $\pmb u$, $\pmb x_n^{T} \pmb u$ can be interpreted as the length of the projection of $\pmb x_n$ onto the direction $\pmb u$. The PCA attempts to maximize the variance of the projections, which is formulated as:
\begin{equation}\label{pca1}
\max_{\pmb u} \frac{1}{N}\sum\limits_{n=1}^{N}{(\pmb x_n^{T} \pmb u)^2}=\max_{\pmb u} \pmb u^T \left(\frac{1}{N}\sum\limits_{n=1}^{N}{(\pmb x_n \pmb x_n^{T})} \right)\pmb u.
\end{equation}
Given the covariance matrix $\pmb \Lambda=\frac{1}{N}\sum\limits_{n=1}^{N}{(\pmb x_n \pmb x_n^{T})}$, the solution of problem (\ref{pca1}) is given by the eigenvector of the covariance matrix $\pmb \Lambda$. If we denote the top $K$ eigenvectors of $\pmb \Lambda$ by $\pmb u_1, \pmb u_2,\dots,\pmb u_K$ and $K<M$, a dimensionality reduction expression of $\pmb x_n$ can be formulated as:
\begin{equation}\label{pca2}
\pmb y_n=[\pmb u_1, \pmb u_2, \dots, \pmb u_K]^{T}  \pmb x_n,
\end{equation}
where $\pmb u_1, \pmb u_2,\dots,\pmb u_K$ are the first $K$ principle components of the training samples.

By contrast, the ICA attempts to find a new basis for representing original samples that are assumed to be a linear weighted superposition of some unknown latent variables. It aims for decomposing multivariate variables into a set of additive subcomponents, which are non-Gaussian variables and are statistically independent from each other. As for the independent components, also termed as the latent variables, they exhibit the maximum possible ``statistical independence'', which can be commonly characterized by either the minimization of their mutual information quantified in terms of the Kullback-Leibler divergence metric and the maximum entropy criterion, or by the maximization of what is termed in parlance as the non-Gaussianity relying on kurtosis and negentropy, for example.

Let us consider the linear noiseless ICA model in a simple example, where the multivariate training variables are denoted by $\pmb x=[x_1, x_2,\dots, x_N]^{T}$. Its latent independent component vector is represented by $\pmb s=[s_1, s_2,\dots, s_M]^{T}$. Each component of $\pmb x$ can be generated by a linearly weighted sum of independent components, i.e. we have $x_n=a_{n1}s_1+a_{n2}s_2+\dots+a_{nM}s_M$, where $a_{nm}$ is the weighting coefficient. The vectorial form of $\pmb x$ can be expressed as:
\begin{equation}\label{ica1}
\pmb x=\sum\limits_{m=1}^{M}s_m \pmb a_{m},
\end{equation}
where $\pmb a_{m}=[a_{1m}, a_{2m},\dots, a_{Nm}]^{T}$. Furthermore, let $\pmb A=(\pmb a_{1},\dots,\pmb a_{M})$. Then the original multivariate training variables can be rewritten as:
\begin{equation}\label{ica2}
\pmb x=\pmb A \pmb s,
\end{equation}
where the unknown matrix $\pmb A$ is referred to as the mixing matrix. ICA algorithms attempt to estimate both the mixing matrix $\pmb A$ and the independent component vector $\pmb s$ relying on setting up a cost function, which again, either maximizes the non-Gaussianity or minimizes the mutual information. Thus, we can recover the independent component vector by computing $\pmb s=\pmb A^{-1} \pmb x$, where $\pmb A^{-1}$ is termed as the `unmixing' matrix. Usually, we assume that $N=M$ and that the mixing matrix $\pmb A$ is a square-shaped matrix. Moreover, the \textit{apriori} knowledge of the probability distribution of $\pmb s$ is beneficial in terms of formulating the cost function.

\subsubsection{Applications}
As for the application of PCA and ICA in wireless networks, Shi \textit{et al.}~\cite{shi2018accurate} utilized PCA to extract the most relevant feature vectors from fine-grained subchannel measurements for improving the localization and tracking accuracy in an indoor location tracking system.
Moreover, Morell \textit{et al.}~\cite{morell2016data} designed an efficient data aggregation method for WSNs based on PCA amalgamated with a non-eigenvector projection basis, while keeping the reconstruction error below a pre-defined threshold. Quer \textit{et al.}~\cite{quer2012sensing}
exploited PCA for inferring the spatial and temporal features of a range of signals monitored by a WSN. Based on this they recovered the large original data set from a small observation set.

Additionally, Qiu~\cite{qiu2011cognitive} combined ICA with PCA in a smart grid scenario for recovering smart meter data, which were jointly capable of enhancing the transmission efficiency both by avoiding the channel estimation in each frame and by eliminating wide-band interference or jamming signals. A semi-blind received signal detection method based on ICA was proposed by Lei \textit{et al.}~\cite{shen2017ica}, which additionally estimated the channel information of a multicell multiuser massive MIMO system. Moreover, Sarperi \textit{et al.}~\cite{sarperi2007blind} proposed an ICA based blind receiver structure for MIMO OFDM systems, which approached the performance of its idealized counterpart relying on perfect CSI. ICA was also used for digital self-interference cancellation in a full duplex system~\cite{li2017digital}, which relied on a reference signal used for estimating the leakage into the receiver. More explicitly, in full duplex systems the high-power transmit signal leaks into the receiver through a nonlinear leakage path and drowns out the low-power received signal. Hence its cancellation requires at least $120$~dB interference rejection. Furthermore, in~\cite{nguyen2013binary},
the Boolean ICA concept was proposed based on the integration of Boolean functions of binary signals for inferring the activities of the underlying latent signal sources. Specifically, it was shown that given $m$ SUs, the activities of up to $(2m-1)$ PUs can be determined.

\begin{table*}[t!]
\begin{center}
\scriptsize
\caption{Compelling Applications of Unsupervised Learning in Wireless Networks}
\label{tbunsup}
\begin{tabular}{ | c | l | c | r |}
\hline
Paper &  Application &  Method & Description  \\ \hline
\cite{xia2012optical} & gateway deployment &  $K$-means  &  reduce delay and improve network throughput for optical/wireless hybrid networks \\ \hline
\cite{hajjar2017hybrid} & relay selection & $K$-means &  create small cells in an LTE macro cell with low power cluster constraint\\ \hline
\cite{cabria2017potential} &  sensor partitioning& $K$-means & balance the load of storage centers and minimize the total network cost\\ \hline
\cite{parwez2017big} &  anomaly detection & $K$-means & verify spatio-temporal varying users' genuineness relying on ground truth information\\ \hline
\cite{el2010impact} &  intrusion detection & $K$-means & improve intrusion detection accuracy and reduce the learning complexity\\ \hline
\cite{liang2016coding} &  blind transceiver & $K$-means & not require pilot duration and pilot power for saving energy consumption\\ \hline
\cite{zhao2006k} &  signal detection &  $K$-means &  burst-mode data transmission with an unbalanced ratio of bits zero and bits one\\ \hline
\cite{wen2015channel} & channel estimation & EM algorithm & construct a GMM to estimate channel parameters in both target cell and adjacent cells\\ \hline
\cite{choi2013estimation} & PU detection & EM algorithm &  SUs estimate PU's sojourn time and signal strength relying on a HMM model \\ \hline
\cite{assra2016approach} &  channel state detection & EM algorithm & jointly estimate channel frequency responses, noise variance and PU's signal \\ \hline
\cite{zhang2008joint} &  symbol detection & EM algorithm & joint symbol detection and channel estimation for MIMO-OFDM systems \\ \hline
\cite{li2015joint} &  network state detection & EM algorithm & joint estimate the sequential target state and network synchronization state \\ \hline
\cite{zhang2017novel} &  active user detection & EM algorithm &  detect active user for the low-activity CDMA based M2M communications \\ \hline
\cite{meng2011efficient} & source localization & EM algorithm & formulate localization as a joint sparse signal recovery and parameter estimation problem\\ \hline
\cite{shi2018accurate} &  indoor location & PCA &  extract relevant feature vectors from fine-grained subchannel measurements \\ \hline
\cite{morell2016data} &  data aggregation & PCA &  limit the reconstruction error based on a non-eigenvector projection basis\\ \hline
\cite{quer2012sensing} &  data recovery & PCA & exploit PCA to extract spatial and temporal features of real signals\\ \hline
\cite{qiu2011cognitive} &  data recovery & ICA \& PCA &  enhance transmission efficiency by avoiding channel estimation and eliminating jamming signals\\ \hline
\cite{shen2017ica} &  channel estimation & ICA &  differentiate and decode the received signal, and estimate the channel information \\ \hline
\cite{sarperi2007blind} &  blind receiver & ICA & yield an ideal performance close to that with perfect CSI \\ \hline
\cite{li2017digital} &  interference cancellation & ICA &  digital interference cancellation based on the reference signal from transmitter power amplifier \\ \hline
\cite{nguyen2013binary} &  signal detection & ICA & infer the activities of latent signal sources based on the Boolean functions\\ \hline
\end{tabular}
\end{center}
\end{table*}

\section{Reinforcement Learning in Wireless Networks}
\label{Reinforcement Learning in NGWN}
Reinforcement learning deals with an agent interacting with the environment. Three specific aspects of reinforcement learning,
multi-arm bandit problem, Markov decision process (MDP) and temporal-difference (TD) learning can be very useful for wireless networks. Then, we explore further on these algorithms of reinforcement learning to wireless networks.

\subsection{Multi-Armed Bandit and Its Applications}
\subsubsection{Methods}
The multi-armed bandit technique, also called $K$-armed bandit, models a decision making problem, where an agent is faced with a dilemma of $K$ different actions. After each choice, the agent receives a reward relying on a stationary probability distribution that is associated with its decision. The agent attempts to maximize its expected total reward over a series of decision making rounds relying on a balance striking a trade-off between consulting existing knowledge and acquiring new knowledge when optimizing its decisions. The action of referring to existing knowledge to make decisions is termed as ``exploitation'', while the trial of acquiring new knowledge is referred to as ``exploration''. Striking a trade-off between exploration and exploitation is also sought by other reinforcement learning algorithms, where exploitation is the plausible action for maximizing the expected reward within the current round, while exploration may produce a greater reward in the long run.

In a $K$-armed bandit model, $K$ possible actions, $a_1, a_2,\dots, a_K$, yield different rewards associated with the $K$ unknowns of the problem at hand, which may have different distributions with $K$ mean values of $\mu_1, \mu_2,\dots,\mu_K$, respectively. The agent iteratively chooses an action $A_i$ at the round $i$ and receives the corresponding reward of $R_i$. Up to the round $i$, the expected reward of an action $a$ can be expressed as $Q_i(a)=E[R_i | A_i=a ]$. Upon striking a balance between the exploration and the exploitation, we may arrive at a simple bandit algorithm as follows, for example. In each decision-making round, we greedily opt for the action $A= \argmax\limits_{a} Q(a)$ relying on the probability of $(1-\varepsilon)$, whilst riskily embarking on a random action selection based on the probability of $\varepsilon$, where $\varepsilon$ is the probability of a brave attempt for exploring new knowledge.

In contrast to the above-mentioned $\varepsilon$-greedy bandit algorithm, there are also more complex bandit algorithms, such as the gradient aided bandit algorithm, associative-search bandit, non-stationary bandit, etc~\cite{sutton1998reinforcement}. Moreover, the multi-armed bandit problem can be extended into a multi-play and multi-armed bandit problem~\cite{zhou2017budget}, where the reward of each agent depends on others' actions, and each agent tries to find its optimal decision by predicting the future actions of the other agents relying on previous decision making strategies.
\subsubsection{Applications}
As mentioned before, multi-armed bandit based techniques are capable of dealing with uncertainties in the context of future wireless networks because of limited prior knowledge and the associated resource-thirsty feedback. Moreover, it is beneficial to model the selfishness and the decision conflicts of/among the users during the decision making process. Hence, the multi-armed bandit based algorithms have  become powerful tools for rational decision making in wireless networks both for distributed users and APs as well as for the central control center.
Specifically, Maghsudi \textit{et al}.~\cite{maghsudi2016multi} proposed a small cell activation scheme relying on the multi-armed bandit philosophy
given only limited information about the available energy of the small cell BS as well as the number of users to be served. The overall heterogeneous network's throughput was improved with the aid of an energy-efficient small cell on-off switching regime controlled by the macro BS, while the inter-interference level was reduced.
Another compelling application of the multi-armed bandit regime in the heterogeneous network is constituted by the dynamic network selection in the context of uncertain heterogeneous network state information. Wu \textit{et al}.~\cite{wu2016traffic} formulated the optimal network selection problem as a continuous-time multi-armed bandit problem considering diverse traffic types. Moreover, the network access cost function and the QoE reward were defined as the metrics of evaluating the proposed network selection schemes.
In~\cite{si2008distributed}, given the time-varying and user-dependent fading channels of wireless peer-to-peer (P2P) networks, a multi-armed bandit aided optimal distributed transmitter scheduling policy was conceived for multi-source multimedia transmission, which was beneficial of maximizing the data transmission rate and reducing the related power consumption in the light in terms of the realistic energy constraints of wireless mobile devices.
In addition to transmitter scheduling, Maghsudi and Sta\'{n}czak applied the covariate multi-armed bandit regime~\cite{maghsudi2013dynamic} for solving the relay selection problem in the wireless network, where the
geographical location of relay nodes was assumed to be known by the source node, but no knowledge was assumed about the corresponding
fading gains. The proposed covariate multi-armed bandit model is capable of dealing with the exploitation-exploration dilemma of the relay selection process. Lee \textit{et al}.~\cite{lee2015frequency} proposed a $K\epsilon$-greedy multi-armed bandit based framework for exploiting the gains provided by
frequency diversity in Wi-Fi channels. They struck a trade-off between the achievable gain stemming from frequency diversity and the resource consumption imposed by channel estimation and coordination.

Given the open broadcast nature of the wireless channel environment and the access contention mechanism among multi-priority users, multi-armed bandit based techniques have played a special role in cognitive networks~\cite{zhao2008myopic,gwon2013optimizing,li2014almost,li2015adaptive,avner2016multi,zhou2016toward}.
For example, Zhao \textit{et al}.~\cite{zhao2008myopic} formulated a multi-armed restless bandit model for opportunistic multi-channel access, which approached the maximum attainable throughput by accurately predicting which is next idle channel likely to become. In~\cite{li2015adaptive}, a channel selection scheme was investigated which was capable of adapting to the link quality and hence finding the optimal channel for avoiding interferences and deep fading.
Moreover, Gwon \textit{et al}.~\cite{gwon2013optimizing} and Zhou \textit{et al}.~\cite{zhou2016toward} further considered the choice of access strategy in the presence of both legitimate desired users and jamming cognitive radio nodes, which was resilient to adaptive jamming attacks with different strengths spanning from near no-attack to the full-attack across the entire spectrum. In contrast to only sensing and accessing a single channel, considering the correlated rewards of different arms, a sequential multi-armed bandit regime was conceived by Li \textit{et al}.~\cite{li2014almost} for identifying multiple channels to be sensed in a carefully coordinated order. Furthermore, Avner and Mannor~\cite{avner2016multi} studied multi-user coordination in cognitive networks, where each user's successful channel selection relies on both the channel state as well as on the decisions of the other users.

\subsubsection{An Example}
Visible light communication (VLC) systems have the compelling benefit of a wide unlicensed communication bandwidth as well as innate security in downlink (DL) transmission scenarios, hence they may find their way into the construction of future wireless networks. However, considering the limited coverage and dense deployment of light-emitting diodes (LED), traditional network association strategies are not readily applicable to VLC networks. Hence by exploiting the power of online learning algorithms, in~\cite{wang2017learning}, the authors focused their attention on sophisticated multi-LED access point selection strategies conceived for hybrid indoor LiFi-WiFi communication systems with the aid of a multi-armed bandit model. Explicitly, since light-fidelity (LiFi) VLC transmissions are less suitable for uplink (UL) transmissions, a classic WiFi UL was used in this study.

\begin{figure}[!t]
  \begin{minipage}[t]{0.48\linewidth}
  \centering
  \includegraphics[width=0.9\textwidth]{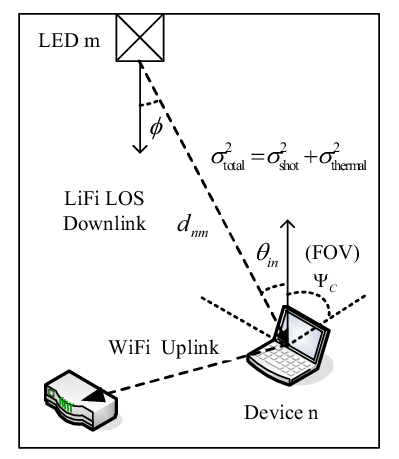}
  \centerline{\footnotesize{(a) LOS path}}
  \end{minipage}
  \begin{minipage}[t]{0.48\linewidth}
  \centering
  \includegraphics[width=0.9\textwidth]{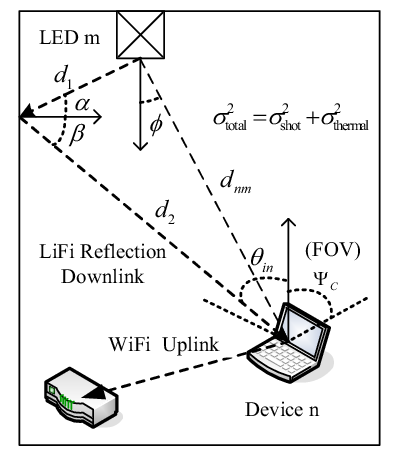}
  \centerline{\footnotesize{(b) Reflected and LOS path}}
  \end{minipage} \\
\caption{The model of the indoor VLC link including both the LOS path and the one-hop reflected path, where FOV represents the field of view, while $\sigma^2_{\mathrm{slot}}$ and $\sigma^2_{\mathrm{sthermal}}$ denote the  variance of the shot noise as well as the thermal noise, respectively~\cite{wang2017learning}~\copyright IEEE.}\label{vlcmodel}
\end{figure}

To elaborate, in the indoor VLC system, the communication between the devices and the backbone network relies on the VLC DL as well as on the RF WiFi UL, which hence can be viewed as a hybrid LiFi-WiFi network. In the system model, it is assumed that there are $M$ low-energy LED lamps in the indoor space considered. Moreover, regardless of their positions, the $N$ mobile devices are capable of accessing any of the $M$ indoor LED lamps and of downloading packets from the Internet via VLC. When a decision round is due, the access control strategy obeys the decision probability distribution of $P=\{p_{1}, p_{2}, ..., p_{M}\}$. And it has $\sum\limits_{m=1}^{M}{{{p}_{m}}=1}$, where $p_{m}$ denotes the probability of accessing the $m$th LED lamp. Furthermore, the service time of each LED lamp obeys the negative exponential distribution with a departure rate $\varsigma$, while the interval between system access requests, in the same way, obeys the negative exponential distribution with an arrival rate $\lambda$.
The VLC DL channel is characterized by a diffuse link, where the light beam is radiated within a certain angle. Thus, the indoor VLC channel can be modelled by combining the line of sight (LOS) path (Fig.~\ref{vlcmodel} (a)) as well as a single one-hop reflected path (Fig.~\ref{vlcmodel} (b)).

The expectation of the accumulated reward gap function is defined as the
metric for characterizing the performance of our AP selection
scheme, which represents the difference between the maximum
theoretical reward and the actually acquired reward after sequential
decision making experiments relying on the system's decision
probability distribution, which is formulated as.
\begin{equation}\label{15-1}
R_{P}(K)=\underset{{{i}_{1}},{{i}_{2}},...,{{i}_{K}}}{\mathop{\max }}\, E\left[\sum\limits_{k=1}^{K}{Q(i_{k},{{t}_{k}})}\right]-E\left[\sum\limits_{k=1}^{K}{Q({{a}_{k}},{{t}_{k}})}\right],
\end{equation}
where $Q({{i}_{k}},{{t}_{k}})$ denotes the user rate associated with the $k$th decision round in terms of the access decision ${i}_{k}$ at the instant ${t}_{k}$, with ${a}_{k}$ being the actual access decision.

Furthermore, in~\cite{wang2017learning} a pair of multi-armed bandit learning techniques, i.e. the `exponential weights for exploration and exploitation' (EXP3) as well as the `exponentially-weighted algorithm with linear programming' (ELP), were advocated for updating the AP-assignment decision probability distribution of each AP at each time instant for the sake of improving the link throughput based on the probability distribution $R_{P}(K)$ of~(\ref{15-1}). More explicitly, in contrast to the trial-and-error EXP3 algorithm, the ELP based AP selection algorithm was constructed for taking into account both the partially observed conditions of the APs as well as the network topology.

\begin{figure}[!t]
  \centering
  \includegraphics[width=0.48\textwidth]{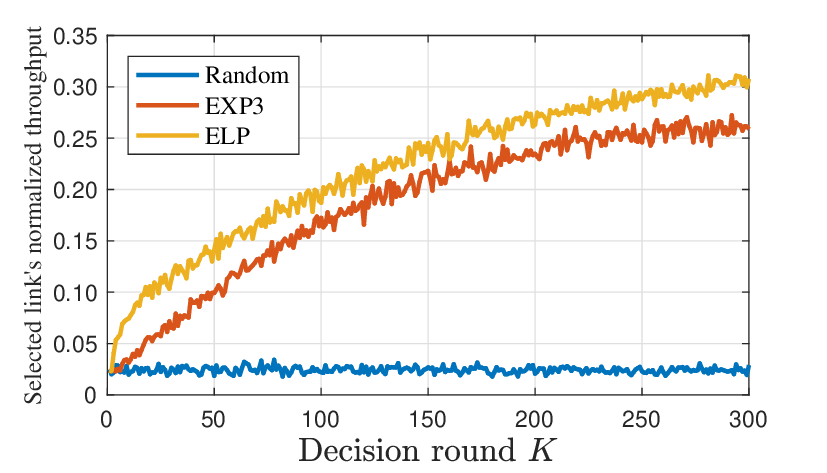}\\
  \caption{The normalized throughput of the selected VLC link versus $K$ for different LED AP selection schemes.~\copyright IEEE}
  \label{figure3a}
\end{figure}

\begin{figure}[!t]
  \centering
  \includegraphics[width=0.48\textwidth]{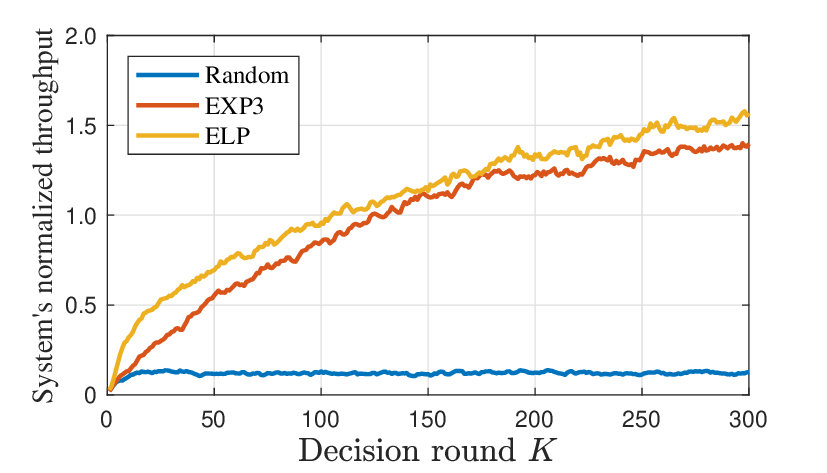}\\
  \caption{The system's normalized throughput versus $K$ for different LED AP selection schemes.~\copyright IEEE}
  \label{figure3b}
\end{figure}

The theoretical upper bound of the expected value of the accumulated
reward gap function of the EXP3- and ELP-based multi-armed bandit learning algorithms was also derived in~\cite{wang2017learning}. In Fig.~\ref{figure3a} and Fig.~\ref{figure3b}, the normalized throughput of the selected VLC links and of the whole system relying on the EXP3-based, ELP-based as well as on random LED AP selection schemes was compared. By contrast, the random selection scheme granted an identical decision probability of accessing any of the $M$ LEDs, namely $1/M$, for each lamp at each decision-making time instant. It was assumed that the negative exponential departure probability of each downloading service was $\varsigma=0.2$. Moreover, the initial state of the number of downloading services supported by each lamp was randomly chosen between $[1, 30]$. Upon increasing the number of decision rounds $K$, the EXP3- and ELP-based selection schemes had a higher accumulated normalized throughput than random selection. Furthermore, relying on more neighbor observation information as well as by exploiting the connection of the LED lamps, the ELP-based AP-selection scheme was shown to outperform that based on EXP3.

\subsection{MDP \& POMDP and Their Applications}
\subsubsection{Methods}
The classic Markov decision process (MDP)~\cite{Puterman1994Markov} constitutes a framework of making decisions in the context of a discrete-time stochastic environment of Markov state transitions, which provides the decision maker with the optimal actions to opt for at each state. It has been used in a wide range of disciplines, especially in automatic control~\cite{Cao2008Event}.
The goal of the decision maker, generally speaking, is to maximize the cumulative reward received over a long run and to find the corresponding optimal policy $\pi^{\ast}$ which represents a mapping from each state to the specific probabilities of choosing each legitimate action.

In an MDP model, the system's state transition follows the Markovian property, where the system's response at time epoch $(t+1)$ depends exclusively on the current state and on the agent's action at time epoch $t$.
Mathematically, at time epoch $t$, the system is in a certain state $S$, where the agent selects a legitimate action $A$ that is available in the state $S$. As a result, the system then acts at the next time epoch $(t+1)$ by moving into a new state $S'$ relying on the system's state transition probability of $p(S'|S,A)$. At the same time, the decision maker receives the corresponding reward $r(S,A)$.
The associated value function is then defined for quantifying how well the agent carries our its action over a long run commencing from the initial state $S_0$, which can be formulated as:
\begin{equation}\label{mdp1}
v_{\pi}=E_{\pi}\left[\sum\limits_{t=0}^{\infty}{\gamma^t r_{t+1} | S_t}\right],
\end{equation}
where $\gamma$ represents the discount factor and the mapping $\pi (A |S )$ represents the probability
of opting for action $A$ in the state $S$. Hence, the optimal policy $\pi^{\ast}$ can be formulated by maximizing the value function considered, i.e. we have $\pi^{\ast}=\argmin\limits_{\pi} v_{\pi}$.
The maximization of the value function can reformulated as an iterative equation with the aid of Bellman's optimality theorem~\cite{Crespo2003Stochastic}, which is given by:
\begin{equation}\label{mdp2}
\begin{aligned}
\pi^\ast(A \mid S)&=\argmax_{A} v^\ast(S)\\&=\argmax_{A}\sum_{S'} p(S' \mid S, A)\left[r(S, A)+\gamma v^\ast(S')\right].
\end{aligned}
\end{equation}

By contrast, as an extension of MDP, the partially observable Markov decision process (POMDP) only relies on partial knowledge about the hidden Markov system which is eminently suitable for scenarios, where the agent cannot directly observe the underlying system's state transitions. Hence, the agent has to constitute belief states and the associated belief transition function by relying on a set of observations instead of the real system states.
In a nutshell, the POMDP framework can be formulated as a quintuple of $\langle S, b, A, B, r \rangle$, i.e.
\begin{itemize}
  \item \emph{System's State $S$:} The system's state $S$ represents the system's legitimate state;
  \item \emph{Belief State $b$:} The belief state $b$ benchmarks the degree of the similarity
    between each of the system's legitimate state $S$ and the state estimated by the agent;
  \item \emph{Action $A$:} The action $A$ denotes the specific action that can be selected in the given state;
  \item \emph{Belief Transition Function $B$:} The belief transition function $B(b'\mid b, A)$ represents the probability of the belief state traversing from $b$ to $b'$ conditioned on selecting action $A$;
  \item \emph{Reward Function $r$:} The reward function $r(b,A)$ quantifies the immediate reward received by performing the selected action.
\end{itemize}
Similarly, the optimal policy $\pi^{\ast}$ can be obtained by solving the optimization problem of:
\begin{equation}\label{mdp3}
\begin{aligned}
\pi^\ast(A \mid b)&=\argmax_{A} v^\ast(b)\\&=\argmax_{A}\sum_{b'} B(b' \mid b, A)\left[r(b, A)+\gamma v^\ast(b')\right].
\end{aligned}
\end{equation}
\subsubsection{Applications}
As another important decision-making tools, which is different from the multi-armed bandit solutions, MDP/POMDP should firstly model the environment relying on either fully or partially observed knowledge.
To elaborate a little further, Massey \textit{et al}.~\cite{massey2004scheduling} proposed an MDP based downlink service scheduling policy for wireless service providers. Considering the time-sensitive nature of wireless tele-traffic patterns, their proposed scheduling policy was capable of maximizing the expected reward for the wireless service provider in the context of a multiplicity of services.
In~\cite{tang2013selfish}, Tang \textit{et al}. resorted to the MDP approach for enhancing a basic node-misconduct detection method, where a novel reward-penalty function was defined as a function of both correct and wrong decisions. The resultant adaptive node-misconduct detector maximized this reward-penalty function in diverse network states.
Moreover, Kong \textit{et al}. conceived a discrete-time MDP (DTMDP) aided mechanism~\cite{kong2014optimal} for dynamically activating and deactivating certain resources of the BS in the context of time-varying network traffic. More explicitly, at each decision round, the DTMDP had the option of activating a new resource module, deactivating the currently active resource module and no operation. The proposed switching mechanism reduced the power consumption, i.e. improved the energy efficiency at the BS.

As a further development, relying on the POMDP paradigm, Tseng \textit{et al}.~\cite{tseng2014pomdp} designed a cell selection scheme for improving the network's capacity, where the full cell loading status was not observable. Hence, it predicted the unavailable cell loading information from set of non-serving base stations and then took actions for improving the various performance metrics, including the system's capacity, the handover time as well as the mobility management as a whole.
Moreover, the belief state was defined for representing the
state uncertainty in terms of the statistical probability of a cell's specific loading state. The simulation results of Tseng \textit{et al}.~\cite{tseng2014pomdp} showed that their solution outperformed the conventional signal-strength aided and load-balancing based methods.
In order to save the energy of sensors, Fei \textit{et al}.~\cite{fei2010pomdp} proposed a POMDP aided K-sensor scheduling policy, which guaranteed the sensors' high-quality coverage and reduced the total energy consumption. Similarly, by striking a trade-off between the detection performance and energy consumption, Zois \textit{et al}.~\cite{zois2012pomdp} designed a POMDP aided sensor node selection scheme for WBANs by maximizing the system's lifetime as well as optimizing the physical state detection accuracy.
The main goal of the sensor node selection was to devise a schedule under which the sensors alternated between the active state and the dormant state relying on the specific network activity.
Upon relying on the decentralized POMDP (DEC-POMDP), Pajarinen \textit{et al}.~\cite{pajarinen2014optimizing} proposed a MAC solution, which promptly adapted both to the spatial and temporal opportunities facilitated by the wireless network dynamics, which yielded an increased throughput and reduced latency compared to the traditional carrier-sense multiple access relying on conventional collision avoidance (CSMA/CA) methods. Here, the POMDP tackled the uncertainty both in the environment's evolution and in the associated inaccurate observations.
Thanks to the cross layer optimization employed, more information can be gleaned
from the lower layers for enhanced network condition estimation. Then, Xie \textit{et al}.~\cite{xie2012novel} used the POMDP model for solving the frame size selection problem of the ubiquitous transmission control protocol (TCP) with the objective of improving the total estimated throughput by striking a tradeoff between the contention
probability and back-off time based on the current network condition.
Furthermore, Michelusi and Mitra~\cite{michelusi2015cross} conceived a cross-layer framework for jointly
optimizing the spectrum-sensing and access processes of cognitive wireless networks with the objective of
maximizing the throughput of the SU under a strict constraint on the maximal performance degradation imposed on the PU. Furthermore, the high complexity of the POMDP formulation
was mitigated by a low-dimensional belief representation, which was achieved by
minimizing the Kullback-Leibler divergence defined in~\cite{Kullback1951On}.

\subsubsection{An Example}
\begin{figure}[!t]
  \centering
  \includegraphics[width=0.48\textwidth]{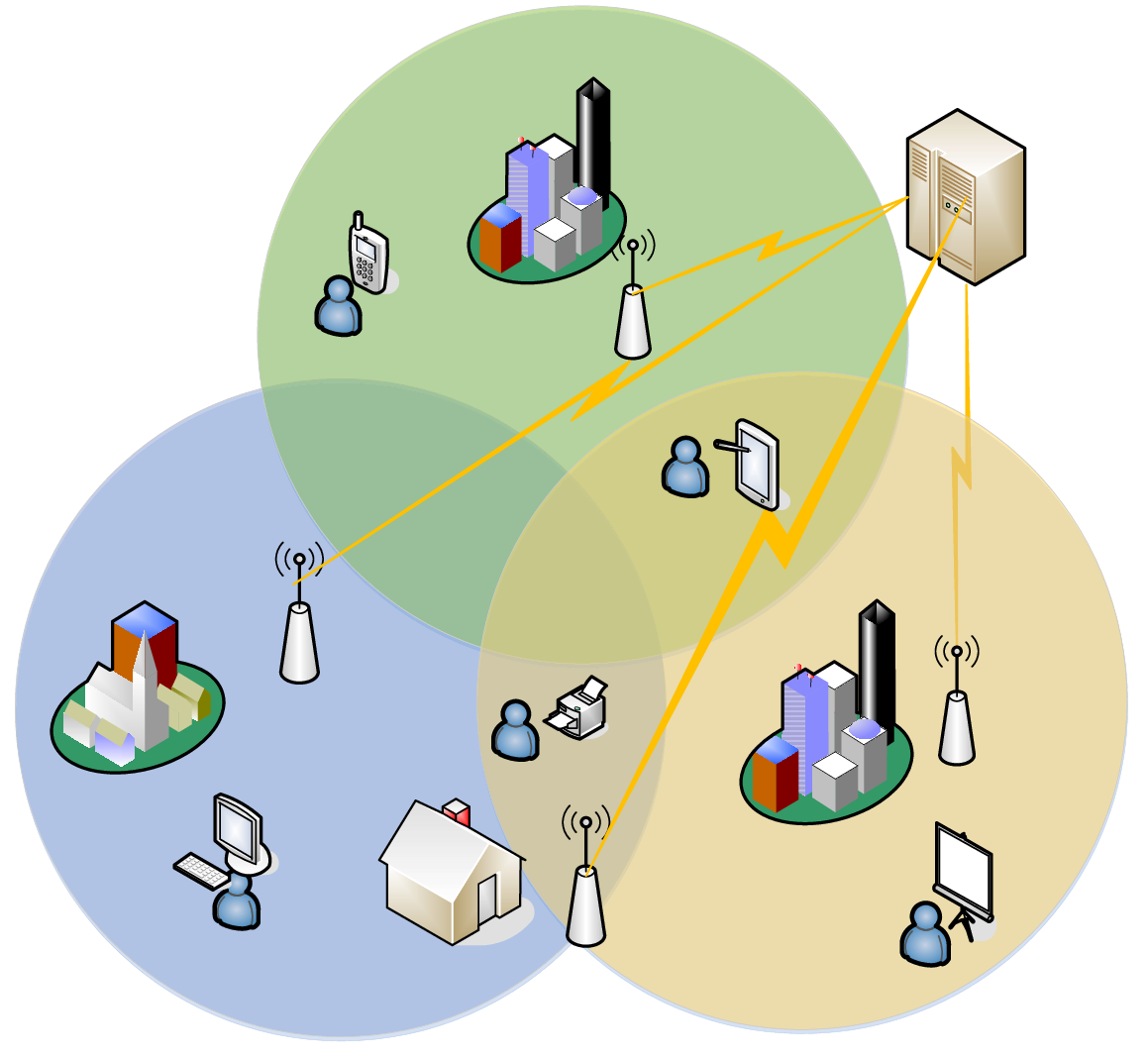}\\
  \caption{The scenario of the `super-WiFi' network~\cite{WangJSAC1}.~\copyright IEEE}
  \label{superwifi}
\end{figure}

As shown in Fig.~\ref{superwifi}, the `super-WiFi' network concept has been originally proposed for nationwide Internet access in the USA. However, the traditional mains power supply is not necessarily ubiquitous in this large-scale wireless network. Furthermore, the non-uniform geographic distribution of both the BSs and of the tele-traffic requires carefully
considered user-association. Relying on the rapidly developing energy harvesting techniques, in~\cite{WangJSAC1}, a POMDP-based access point selection strategies was conceived for an energy harvesting aided super-WiFi network.

It was assumed that both the battery states as well
as the user access states were completely observable. However, in practice the solar radiation intensity changes over time in a year, as influenced by the weather conditions. Furthermore, the radiation sensors have a limited sampling rate, which makes it hard to simultaneously record the solar radiation intensity and to accurately
estimate the system's battery state. Fortunately, relying on historical solar radiation observation data provided by the University of Queensland, Australia~\cite{solar}, in a short period of time, say, within an hour, the real-time harvested solar power can be modeled as $P_{H}=\theta+\kappa$, where $\theta$ is constant for an hour, while $\kappa$ is a small perturbation. Moreover, multiple factors, such as the effective irradiation area, the clouds' distribution, the sensors' operating status, etc. may independently affect the harvested power. Relying on the central-limit theorem, the perturbation $\kappa$ can be regarded as being Gaussian distributed. Hence, the distribution of $P_{H}$ can be written as $P_{H}\thicksim\mathcal{N}(\theta,\sigma^{2})$, where $\theta$ and $\sigma^{2}$ can be learned from the harvested data set.

Moreover, a queue-based user-association state model as well as a dynamic battery state model was established. Hence, the system's state having $K$ APs is constituted by both the user-association states as well as by the battery states. Let $U=({{U}_{1}},{{U}_{2}},...,{{U}_{K}})$ denote the user-association states, while $B=({{B}_{1}},{{B}_{2}},...,{{B}_{K}})$ represent the AP battery states, where ${U}_{k}\leq{U}_{M}$ and ${B}_{k}\!\leq\!{B}_{M}-1$. Furthermore, the super-WiFi system state can be written as a $2K$-element vector $S=(U_{1},B_{1},U_{2},B_{2},...,U_{K},B_{K})$, which includes both the $K$ APs' user-association states and the $K$ APs' battery states. Assuming the independence of each AP's two sub-states, the system's state transition probability can be expressed as:
\begin{equation}
P({{S}^{'}}|S)=\prod\limits_{k=1}^{K}{{{\Phi }_{k}}(U_{k}^{'}|{{U}_{k}},A)}{{\Psi }_{k}}(B_{k}^{'}|{{B}_{k}},A),
\end{equation}
where $A$ represents the users' actions in terms of which available APs they request association with.

Since the requesting users only have partial knowledge of the entire super-WiFi system's state, relying on the above definitions and hypotheses, we construct the POMDP decision-making model in terms of a quintuple of $\langle S, b, A, B, r \rangle$ as mentioned above.
The POMDP formulation can be reduced to a belief MDP with the aid of the belief state vector.
Therefore, the expected reward of the system relying on strategy $\Pi$ after an infinite number of time slots can be written as:
\begin{equation}\label{31}
{{V}^{\Pi }}(S\text{ }\!\!|\!\!\text{ }{{S}^{0}})=E[\sum\limits_{t=0}^{\infty }{{{\gamma }^{t}}{{r}^{t}}({{S}^{t}},\Pi ({{S}^{t}}))b(S^{t})}|{{S}^{0}}],
\end{equation}
where $S^{0}$ is the initial system state, while $b(S)$ is the belief state vector reflecting the grade of similarity between the current estimated state and the legitimate system state $S$. Moreover, $r$ is the immediate reward of the system and $\gamma$ represents the discount rate.
Then, the optimal strategy can be constructed by invoking dynamic programming aided iterative algorithms for maximizing the expected reward function.

Bearing in mind the large values of $K$, $U_{M}$ and $B_{M}$, as well as the users' rapidly fluctuating arrival rate $\lambda$ and departure rate $\mu$, obtaining the optimal POMDP solution may face the curse of dimension disaster. In order to reduce the computational complexity, a suboptimal algorithm was proposed in~\cite{WangJSAC1}. Explicitly, Algorithm 2 of~\cite{WangJSAC1} aimed for maximizing the expectation of the system's energy function, which was defined as:
\begin{equation}\label{a22}
H[b(S)] = \sum\limits_{S \in \mathbb{S}} b(S)\sum\limits_{i=1}^{K}\underset{\Pi }{\mathop{\min}}\,\{E_{iR}+E_{iH}-E_{iC}, E_{max}\},
\end{equation}
where $E_{iR}$ represents the residual energy of AP $i$, while $E_{iH}$ is its energy harvested under the assumption that the harvested power level remains quasi-static during the information transmission interval and $E_{iC}$ denotes the energy consumption. Finally, $E_{max}$ is the capacity of the AP's battery.

\begin{figure}[!t]
  \centering
  \includegraphics[width=0.48\textwidth]{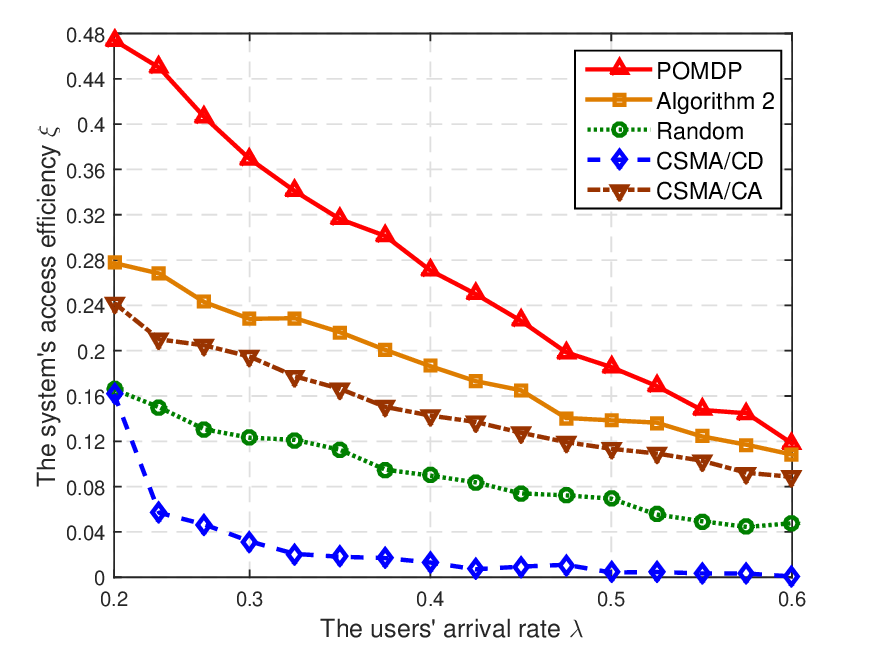}\\
  \caption{The system access efficiency versus users' arrival rate for different access schemes ($K=2$, $U_{M}=1$, $B_{M}=3$).~\copyright IEEE}
  \label{333a}
\end{figure}

\begin{figure}[!t]
  \centering
  \includegraphics[width=0.48\textwidth]{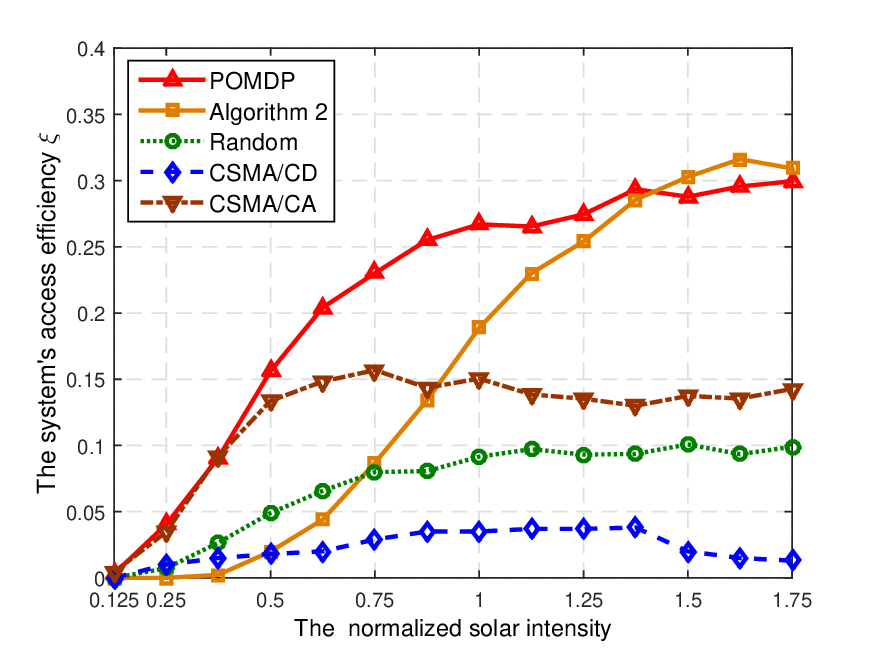}\\
  \caption{The system access efficiency versus solar radiation intensity for different access schemes ($K=2$, $U_{M}=1$, $B_{M}=3$).~\copyright IEEE}
  \label{333b}
\end{figure}

The efficiency of the AP selection algorithms proposed in~\cite{WangJSAC1} was compared in terms of the system's access efficiency defined as $\xi=N_{S}/T_{S}$, where $N_{S}$ is the total number of successful access attempts during the entire simulation time $T_{S}$.
In Fig.~\ref{333a} and Fig.~\ref{333b}, multiple APs ($K=2$) are considered with the maximum number of admitted users being $U_{M}=1$, while having a maximum number of battery states given by $B_{M}=3$. Moreover, the departure rate is $\mu = 0.05$. We may conclude from Fig.~\ref{333a} that a highly loaded system makes the carrier-sense multiple access with collision detection (CSMA/CD) method almost useless, when the users' arrival rate reaches a certain value. As shown in Fig.~\ref{333b}, where $\lambda=0.4$, the system's access efficiency recorded for all the AP selection algorithms only increases with the solar radiation intensity in a relatively small range. However, the performance of the CSMA/CD, CSMA/CA\footnote{Strictly speaking, the CSMA/CD and CSMA/CA in this paper are different from the Ethernet's data link layer protocols. Here, both of them represent the access control mechanisms. We use the same acronym CSMA/CD and CSMA/CA for convenience.}, as well as of the random selection algorithm remains unchanged, regardless of the increase in solar radiation intensity. Moreover, the suboptimal Algorithm 2 of~\cite{WangJSAC1} is capable of outperforming the POMDP method at a strong solar radiation intensity, which may be deemed to be the result of the approximations and hypotheses inherent in the POMDP model.

\subsection{Temporal Difference Learning and Its Applications}
\subsubsection{Methods}
\begin{figure*}
  \centering
 \includegraphics[width=0.90\textwidth]{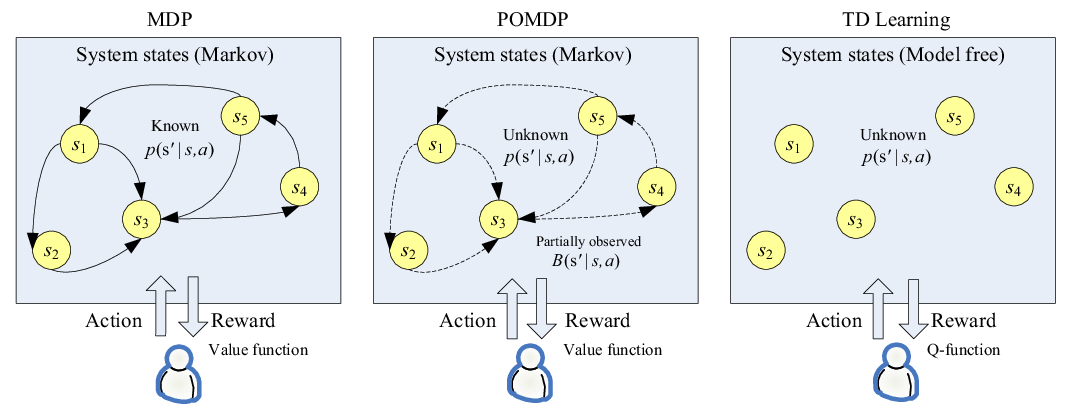}\\
  \caption{The comparison between the MDP, POMDP and TD learning.}\label{reinforcementmodel}
\end{figure*}

Temporal-difference (TD) learning is a model-free reinforcement learning method, which is capable of directly gleaning knowledge from raw experience without a model of the environment or receiving delayed reward, which can be typically viewed as a combination of Monte Carlo methods and of dynamic programming. More specifically, it samples the environment like the Monte Carlo methods, and then updates the corresponding parameters relying on current estimates like dynamic programming does. By contrast, TD learning operates in an on-line fashion by relying on the result of a single time step, rather than waiting for the final outcome until the end of an episode of the Monte Carlo method.
Moreover, it has an advantage over the dynamic programming methods since it does not require a model of the state transition probabilities as shown in Fig.~\ref{reinforcementmodel}. TD learning can be readily invoked for finding an optimal action policy for any finite MDP associated with an unknown system model.
Fig.~\ref{reinforcementmodel} shows the difference between the MDP, POMDP and TD learning.

A pair of popular representatives of the TD learning family are constituted by the Q-learning and by the ``state-action-reward-state-action'' (SARSA) technique, which interacts with the environment and updates the state-action value function, namely the Q-function, based on the action it takes. In contrast to SARSA, Q-learning updates the Q-function relying on the \textit{maximum reward} provided by one of its available actions.
Specifically, the update of the Q-function in SARSA can be formulated as~\cite{kaelbling1996reinforcement}:
\begin{equation}\label{td1}
Q(S, A)\leftarrow (1-\alpha)Q(S, A)+\alpha \left[r+ \gamma Q(S', A')\right],
\end{equation}
while in Q-learning, the update of the Q-function can be cast as~\cite{watkins1992q}:
\begin{equation}\label{td2}
Q(S, A)\leftarrow (1-\alpha)Q(S, A)+\alpha \left[r+ \gamma \max_{A \in \mathbb{A}} Q(S', A)\right],
\end{equation}
where $S$ represents the system's state and $A$ is the action selected by the agent, whilst $\mathbb{A}$ represents the available set of actions. Moreover, $\alpha$ is the update weighting coefficient and $\gamma$ denotes the discount factor.
As for the convergence analysis, SARSA is capable of converging with probability $1$ to an optimal policy as well as to an optimal state-action value function, provided that all the state-action pairs are visited a sufficiently high number of times. However, because of the independence of making an action and that of updating the Q-function, Q-learning has no delayed reward as TD-learning, which tends to facilitate an earlier convergence than SARSA~\cite{sutton1998reinforcement}.

\subsubsection{Applications}
As a benefit of being free from modeling the environment, TD learning is capable of providing competent decisions even in unknown environments. Table~\ref{tbqlearning} summarizes a variety of compelling applications found in wireless networks for both SARSA and Q-learning along with their brief description.

\begin{table*}[t!]
\begin{center}
\scriptsize
\caption{Compelling Applications of the SARSA Algorithm and Q-learning in Wireless Networks}
\label{tbqlearning}
\begin{tabular}{ | c | l | c | r |}
\hline
Paper  & Method & Scenario & Application \& Description \\ \hline
\cite{lilith2004dynamic} & reduced-state SARSA  &  cellular network &  dynamic channel allocation considering both mobile traffic and call handoffs.\\
\hline
\cite{lunden2011reinforcement} & on-policy SARSA &  CR network & distributed multiagent sensing policy relying on local interactions among SU\\ \hline
\cite{chettibi2012adaptive} & on-policy SARSA &  MANET & energy-aware reactive routing protocol for maximizing network lifetime\\ \hline
\cite{suga2013joint} & on-policy SARSA &  HetNet & resource management for maximizing resource utilization and guaranteeing QoS\\ \hline
\cite{ortiz2016reinforcement} & approximate SARSA &  P2P network & energy harvesting aided power allocation policy for maximizing the throughput\\ \hline
\cite{kazemi2011dynamic} & Q-learning &  WBAN & power control scheme to mitigate interference and to improve throughput\\ \hline
\cite{leite2012flexible} & Q-learning &  OFDM system & adaptive modulation and coding not relying on off-line training from PHY\\ \hline
\cite{jung2012relay} & Q-learning &   cooperative network  & efficient relay selection scheme meeting the symbol error rate requirement\\ \hline
\cite{galindo2010distributed} & decentralized Q-learning &   CR network& aggregated interference control without introducing signaling overhead\\\hline
\cite{prashanth2014two} & convergent Q-learning &   WSN  & sensors' sleep scheduling scheme for minimizing the tracking error\\\hline
\end{tabular}
\end{center}
\end{table*}

\section{Deep Learning in Wireless Networks}
\label{Deep Learning in NGWN}
\subsection{Deep Artificial Neural Networks and Their Applications}
\subsubsection{Methods}
Artificial neural networks~\cite{zurada1992introduction} constitute a set of algorithms conceived by imitating the interaction amongst neurons in the human brain, which are designed to extract features for clustering and classification tasks.

In a common artificial neural network (ANN) model~\cite{haykin1994neural}, the input of each artificial neuron is a real-valued signal, where the output of each artificial neuron is subjected to by some non-linear functions, namely the so-called activation functions. This is formulated as:
\begin{equation}\label{dl1}
y_k=\psi(u_k+b_k),
\end{equation}
where
\begin{equation}\label{dl2}
u_k=\Sigma_{i=1}^{m}w_{ki}x_i.
\end{equation}
Here, $x_i$ and $y_k$ represent the input signal and output signal, respectively, while $w_{ki}$ is the associated weight and $b_k$ is the bias. Moreover, $\psi(\bullet)$ is the activation function.
Artificial neurons and their connections typically use a weighting factor for adjusting the ``speed'' of the learning process. Moreover, artificial neurons are organized in the form of layers. Different layers perform different transformations of their inputs. Basically, the input signals travel from the first layer to the last layer, possibly via multiple hidden layers.

\begin{figure*}
  \centering
 \includegraphics[width=0.90\textwidth]{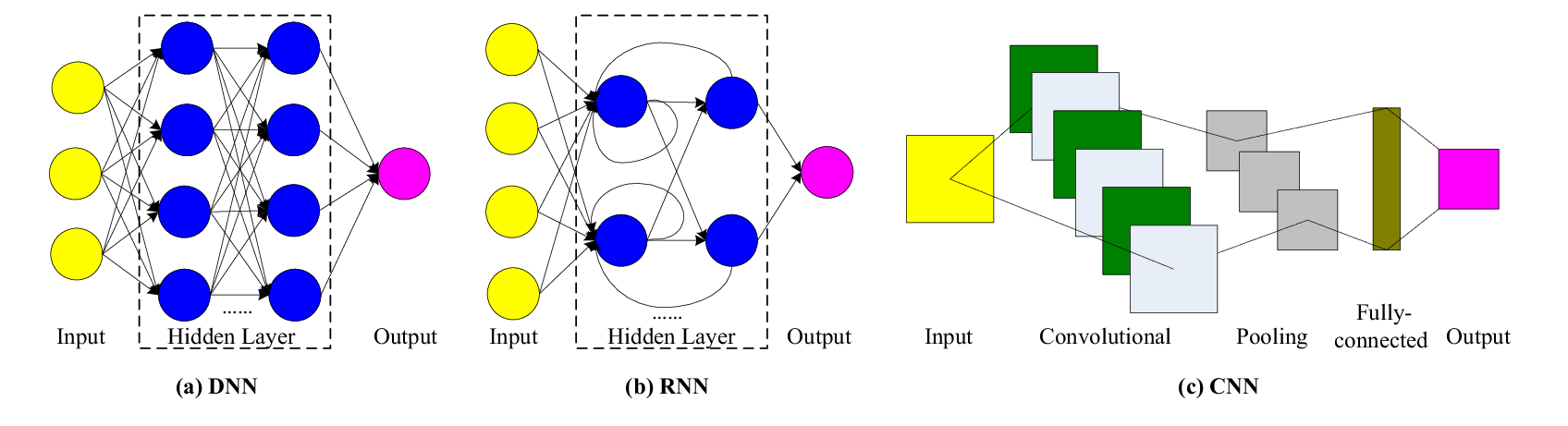}\\
  \caption{The basic architecture of DNN, RNN and CNN.}\label{deep learning}
\end{figure*}

The general deep neural network (DNN) is characterized by multiple hidden layers between the input and output layers as shown in Fig.~\ref{deep learning}~(a), which is capable of modeling complex relationships of the processed data with the aid of multiple non-linear transformations. In a DNN, the provision of extra layers facilitates the composition of features from lower layers, which is beneficial in terms of more accurately modeling complex data than a `shallow' network having a single hidden layer. In comparison to `shallow' networks, DNNs are capable of attaining a performance, despite requiring less training. Furthermore, DNNs may be viewed as a type of feed-forward network, where the processed data flows in the direction from the input layer to the output layer without looping back. The forward propagation algorithms and backward propagation algorithms represent a step-by-step problem solving process. Specifically, the forward propagation algorithms apply the carefully trained weight matrixes and bias vectors for carrying out the associated linear and activation operations. By contrast, the backward propagation algorithms, which are widely used in the industrial field define a so-called loss function for quantifying the difference between the output produced by the training samples' and the real output. Then, by minimizing the loss or error we can optimize the connection weights of the neural network. Although there are
impressive applications of DNN, improving their convergence behavior requires further research.

By contrast, in a recurrent neural network (RNN) a neuron in one layer is capable of connecting to the neurons of previous layers. Therefore, a RNN is capable of exploiting the dynamic temporal information hidden in a time sequence and it beneficially exploits its ``memory'' inherited from the previous layers for processing the future inputs, as shown in Fig.~\ref{deep learning}~(b). RNNs have the advantage of requiring reduced  training and impose a reduced computational complexity, because fewer tensor operations have to be calculated.
Popular algorithms used for training the RNN include the real time recurrent learning technique of~\cite{williams1989experimental}, the causal recursive backpropagation algorithm of~\cite{campolucci1999line} and the backpropagation through time algorithm of~\cite{werbos1990backpropagation}, just to name a few.

The convolutional neural network (CNN) constitutes a popular class of feed-forward deep artificial neural networks, which only require modest preprocessing.They are capable of reducing the number of parameters step-by-step as we move from the input layer to the hidden layers. As seen in Fig.~\ref{deep learning}~(c), a basic CNN architecture is composed of an input layer, an output layer as well as multiple hidden layers, which are often referred to as convolutional layers, pooling layers and fully connected layers. More particularly, the convolutional layers carry out the convolution operation, also termed as the cross-correlation operation, which generates a multi-dimensional feature map relying on a number of so-called filters. The CNN has been successfully used both in image and video recognition~\cite{simonyan2014very}, in natural language processing~\cite{young2018recent}, in recommender systems~\cite{wang2015collaborative}, etc. Fig.~\ref{deep learning} contrasts the basic architecture of DNN, RNN and CNN, respectively.

\subsubsection{Applications}

In this subsection, we will consider the benefits of deep artificial neural network algorithms in a variety of wireless networking scenarios. As mentioned before, deep artificial neural networks are capable of capturing the non-linear and often dynamically varying relationship between the inputs and outputs. Hence they have a powerful prediction, inference and data analysis capability by exploiting the vast amount of data generated both by the environment and by the users. As for \emph{learning from the environment}, we are able to harness DNNs trained by the data gleaned over the air for the sake of channel estimation~\cite{ye2018power}, interference identification~\cite{schmidt2017wireless}, localization~\cite{wang2015phasefi,wang2016csi,wang2017csi,wang2017device,zhang2016deep}, etc. By contrast, with regard to \emph{learning from the users or devices}, DNN algorithms can also be used for predicting the users' behaviors, such as their content interests~\cite{chang2018learn}, mobility patterns~\cite{ouyang2016deepspace}, etc. in order to beneficially design the dynamic content caching of BSs and to efficiently allocate wireless resources, for example.

Traditional signal processing approaches supported by statistics and information theory in communication systems substantially rely on accurate and tractable mathematical models. Unfortunately, however, practical communication systems may have a range of imperfections and non-linear factors, which are difficult to model mathematically. Given that DNN algorithms do not require a tractable model, they are capable of \emph{remedying the imperfections in the physical layer} by learning both from the environment and from previous inputs relying on a specific hardware configuration. To elaborate, Ye \textit{et al}.~\cite{ye2018power} proposed a DNN aided channel estimation method for learning the wireless channel characteristics, such as the nonlinear distortion, interference and frequency selectivity. The DNN aided channel estimation method was shown to be more robust than traditional methods, especially in the context of having fewer training pilots, in the absence of cyclic prefix, as well as in the face of nonlinear clipping noise.
Apart from estimating the channel characteristics, DNNs can also be used for classifying modulated signals in the physical layer. Rajendran \textit{et al}.~\cite{rajendran2017distributed} conceived a date-driven automatic modulation classification (AMC) scheme hinging on the long short term memory (LSTM) aided RNN, which captured the time domain (TD) amplitude
and phase information of modulation schemes carried in the training data without expert knowledge. Their simulations showed that the novel AMC
had an average classification accuracy of about $90\%$ in the context of time-varying SNR ranging from 0dB to 20dB.
As for signal detection, Farsad and Goldsmith~\cite{farsad2017detection} developed a deep learning aided signal detector, where the transmitted signal can be efficiently estimated from its corrupted version observed at the receiver. The detector was trained relying on known transmitted signals, but without any knowledge of the underlying wireless channel model and estimated the likelihood of each symbol, which was beneficial for carrying out soft decision error correction afterwards.
In the application of interference identification, Schmidt \textit{et al}.~\cite{schmidt2017wireless} proposed a $64$-feature-map assisted CNN based wireless interference identification scheme. The CNN model learned the relevant features through self-optimization during the GPU based training process, which was first designed in~\cite{o2016convolutional}. By carefully considering the realistic capability of wireless sensors, the model relied on the time- and frequency-limited sensing snapshots having the duration of 12.8 $\mu\mathrm{s}$ as well as the bandwidth
of 10MHz. The proposed CNN based wireless interference identifier was shown to have a higher identification accuracy than the state-of-the-art schemes in the context of low SNRs, such as $-5$dB, for example.

Furthermore, we can use DNNs for \emph{modelling the entire physical layer of a communication system} without any classic components such as source coding, channel coding, modulation, equalization, etc.
In~\cite{o2017introduction}, O'Shea \textit{et al}. used a DNN to represent a simple communication system with one transmitter and one receiver that can be trained as a so-called auto-encoder without knowing the accurate channel model. Moreover, a CNN algorithm was conceived for modulation classification based on both sampled radio frequency time-series data and expert knowledge integrated by radio transformer networks (RTN). Additionally, O'Shea \textit{et al}.~\cite{o2017deep} extended the DNN aided auto-encoder to a single user MIMO communication scenario, where the physical layer encoding and decoding processes were jointly optimized as a single end-to-end self-learning task. Their simulation results showed that the auto-encoder based system outperformed the classic space time block code (STBC) at 15dB SNR. Furthermore, D{\"o}rner \textit{et al}.~\cite{dorner2018deep} also developed a DNN based prototype system solely composed of two unsynchronized off-the-shelf software-defined radios (SDR). This prototype system was capable of mitigating the current restriction on short block lengths.

DNNs also play a critical role in supporting a variety of compelling upper layer applications, such as traffic prediction~\cite{wang2017spatiotemporal}, packet routing~\cite{kato2017deep} and control~\cite{tang2018removing}, traffic offloading~\cite{li2018learning}, resource allocation~\cite{sun2017learning}, attack detection~\cite{he2017real}, just to name a few. For instance, Wang \textit{et al.}~\cite{wang2017spatiotemporal} presented a hybrid deep learning aided structure for spatio-temporal traffic modeling and prediction in cellular networks by mining information from the China Mobile dataset. It used a novel deep learning aided auto-encoder for modeling the spatial features of wireless traffic, while using LSTM units for temporal modeling. Additionally, Kato \textit{et al.}~\cite{kato2017deep} proposed a supervised DNN aided traffic routing scheme, which outperformed the classic open shortest path first (OSPF) scheme in terms of requiring a lower overhead, whilst maintaining a higher throughput and lower delay.
By contrast, a real-time deep CNN based traffic control mechanism learning from previous network anomalies was conceived by Tang \textit{et al.}~\cite{tang2018removing}, which substantially reduced the average delay and packet loss rate. Hence, deep learning aided traffic control may indeed constitute a potential candidate for gradually replacing traditionally routing protocols in future wireless networks. Furthermore, Li \textit{et al.}~\cite{li2018learning}
integrated both the DNN structure and the edge computing technique into the multimedia IoT, which was able to improve the efficiency of multimedia processing.
Sun \textit{et al.}~\cite{sun2017learning} treated the power control problem in interference-limited wireless networks as a `black box'. They proposed an `almost-real-time' power control algorithm relying on a DNN structure trained by simulated data. In comparison to traditional mathematical tools, the approximation error of the DNN aided algorithm is closely related to the depth of the DNN considered. As for network security issues, for example, He \textit{et al.}~\cite{he2017real} constructed a conditional deep belief network (CDBN) for the real-time detection of malicious false data injection (FDI) attacks in the smart grid, which was trained by historical measurement data.
The simulations conducted using the IEEE 118-bus test system and the IEEE 300-bus test system showed that the CDBN aided FDI detection scheme was resilient to the environmental noise and had a higher detection accuracy than its SVM aided counterparts.

As a successful example of learning from the environment, DNNs are beneficial in terms of extracting electromagnetic fingerprint information from the wireless channel for \emph{indoor localization}. In~\cite{wang2015phasefi,wang2016csi,wang2017csi}, Wang \textit{et al.} proposed a DNN having three hidden-layer for training the calibrated CSI phase data, where the fingerprint information was represented by the DNN's weights. Their experimental results showed that the DNN aided localization scheme performed well in different propagation environments, including an empty living room, and a laboratory in the presence of mobile users.
In~\cite{wang2017device}, Wang \textit{et al.} proposed a deep learning
method for supporting device-free wireless localization and activity recognition relying on learning from the wireless signals around the target, where a sparse auto-encoder network was used for automatically learning the discriminative features of wireless signals. Furthermore, a softmax-regression-based framework~\cite{heckerman1997models} was formulated for the location and activity recognition based on merged features. Moreover, in~\cite{zhang2016deep}, Zhang \textit{et al.} constructed a four-layer DNN for extracting reliable high level features from massive Wi-Fi data, which was pre-trained by the stacked denoising auto-encoder. Additionally, an HMM aided high-accuracy localization algorithm was proposed for smoothening the estimate variation. Their experimental results showed a substantial localization accuracy improvement in the context of a widely fluctuating wireless signal strength.

With regard to learning from users or devices, Ouyang \textit{et al.}~\cite{ouyang2016deepspace} conceived a CNN-aided online learning architecture for understanding human mobility patterns relying on analyzing continuous mobile data streams. Al-Molegi \textit{et al.}~\cite{al2016stf} integrated both the spatial features gleaned from GPS data and the temporal features extracted from the associated time stamps for predicting human mobility based on a RNN. Moreover, Song \textit{et al.}~\cite{song2016deeptransport} proposed an intelligent deep LSTM RNN based system for predicting both human mobility and the specific transportation mode in a large-scale transportation network, which was beneficial in terms of providing accurate traffic control for intelligent transportation systems (ITS). Additionally, a mobility prediction technique relying on a complex extreme learning machine (CELM) was developed by Ghouti \textit{et al.}~\cite{ghouti2014mobility} in order to jointly optimize both the bandwidth and the power MANETs. In~\cite{agarwal2016learning}, both the multi-layer perception and RNN models were employed by Agarwal \textit{et al.} for characterizing the activity of primary users in CR networks, where three different traffic distributions, namely Poisson traffic, interrupted Poisson traffic and self-similar traffic were used for training the related models.

Table~\ref{tbdeep} lists a range of typical applications of DNNs along with a brief description.

\begin{table*}[t!]
\begin{center}
\scriptsize
\caption{Compelling Applications of Deep Artificial Neural Networks in Wireless Networks}
\label{tbdeep}
\begin{tabular}{ | c | l | c | r |}
\hline
Paper & Application  &  Method & Description  \\ \hline
\cite{ye2018power} & channel estimation & DNN & learn nonlinear distortion, interference and frequency selectivity of wireless channels\\ \hline
\cite{rajendran2017distributed} & modulation classification & RNN & capture amplitude and phase information without expert knowledge\\ \hline
\cite{farsad2017detection} &  signal detection & DNN & transmit signal detection from noisy and corrupted signals without underlying CSI\\ \hline
\cite{schmidt2017wireless} &  interference identification & CNN & learn features through self-optimization during the GPU based training process\\ \hline
\cite{o2017introduction} &  PHY representation & DNN & represent simple system having one transmitter and receiver without accurate CSI\\ \hline
\cite{o2017deep} &  PHY representation & DNN & represent single user MIMO system relying on DNN aided auto-encoder\\ \hline
\cite{dorner2018deep} &  software-defined radio & DNN & be capable of easing the current restriction on short block lengths\\ \hline
\cite{wang2017spatiotemporal} &  traffic prediction & DNN & deep auto-encoder and LSTM for modeling spatial and temporal features\\ \hline
\cite{kato2017deep} &  packet routing & DNN & traffic routing scheme with little signal overhead, large throughput and small delay\\ \hline
\cite{tang2018removing} &  traffic control & CNN &  consider previous network abnormalities, lower average delay and packet loss rate\\ \hline
\cite{li2018learning} &  traffic offloading & DNN & integrate both DNN structure and edge computing technique into multimedia IoT\\ \hline
\cite{sun2017learning} &  power control & DNN & an almost-real-time power control algorithm in interference-limited wireless networks\\ \hline
\cite{he2017real} &  network security & DBN &  a real-time detection of malicious false data injection attack in smart grid\\ \hline\
\cite{wang2015phasefi,wang2016csi,wang2017csi,wang2017device,zhang2016deep} &  indoor localization & DNN & device-free wireless localization and recognition by learning from ambient wireless signals\\ \hline
\cite{ouyang2016deepspace} &   mobility prediction & CNN &  learn human mobility pattern relying on analyzing continuous mobile data stream\\ \hline
\cite{al2016stf} &  mobility prediction & RNN & integrate spatial feature from GPS and temporal feature from associated time stamps \\ \hline
\cite{song2016deeptransport} &  transportation mode & RNN & predict both human mobility and transportation mode for large-scale transport networks \\ \hline
\cite{agarwal2016learning} &  activity prediction & RNN & characterize primary users' activity in CR with different traffic distribution \\ \hline
\cite{zhang2019deep} & traffic prediction & CNN &  traffic modeling and prediction based on a convolutional long short-term memory network \\ \hline
\cite{eisen2019learning} & resource allocation & DNN & universal DNN framework for a number of common wireless resource allocation problems \\ \hline
\cite{zhu2019joint} & PHY representation & CNN & modulation, equalization, and demodulation with a convolutional autoencoder \\ \hline
\cite{cui2019spatial} & link scheduling & CNN & optimal scheduling of interfering links in dense wireless networks \\ \hline
\end{tabular}
\end{center}
\end{table*}
\subsection{Deep Reinforcement Learning and Its Applications}
\subsubsection{Methods}
The deep reinforcement learning technique is constituted by the integration of the aforementioned DNNs and reinforcement learning. Explicitly, in deep reinforcement learning methods, DNNs are used for approximating certain components of reinforcement learning, including the state transition function, reward function, value function and the policy.
These components can be viewed as a function of the weights in these DNNs, which can be updated with the aid of the classic stochastic gradient descent.

In particular, the deep Q-Network (DQN) constitutes the first deep reinforcement learning solution, which was proposed by Mnih \textit{et al.} in 2015~\cite{mnih2015human}, which avoids the instability of the reinforcement learning algorithm, which may even become divergent when its action-value function is approximated relying on a non-linear function.
To elaborate a little further, DQN stabilizes the training process of the action-value function approximation by relying on experience replay. Furthermore, DQN only requires modest domain knowledge. The deep Q-learning algorithm in DQN is a variant of the classical Q-learning algorithms, which is integrated with the deep CNN model, where the convolutional filters seen in Fig.~\ref{deep learning}~(c) are used for representing the effects of receptive fields. One of the outputs of the deep CNN involved yields the specific value of the Q-function for a possible action.
Beyond the DQN, substantial efforts have also been invested in improving the performance and stability, as exemplified by the double DQN~\cite{van2016deep} and the dueling DQN~\cite{wang2015dueling}.
Thanks to the powerful feature representation capability of DNNs and of the reinforcement learning algorithms, DQN performs well in a range of compelling applications as exemplified by the AlphaGo, which is the first super-human program to defeat a professional human chess-player.

\subsubsection{Applications}
Deep reinforcement learning is eminently suitable for supporting the interaction in autonomous systems in terms of a higher level understanding of the visual world, which can be readily applied to a diverse analytically intractable problems in future wireless networks.

Given the intrinsic advantages of the reinforcement learning in environment in interactive decision making, it may play a significant role in the field of control decision~\cite{zhang2017learning,zhu2017communication}. Specifically, Zhang \textit{et al.}~\cite{zhang2017learning} proposed a model-free UAV trajectory control scheme relying on deep reinforcement learning for data collection in smart cities, where a powerful deep CNN was used for extracting the necessary features, while a DQN model was used for decision making.
Given the sensing region and the related tasks, this algorithm supported efficient route planning for both the UAVs and mobile charging stations involved.
In~\cite{zhu2017communication}, a deep reinforcement learning aided communication-based train control system was conceived by Zhu \textit{et al.} which jointly optimized the communication handoff strategy and the control functions, while reducing the energy consumption. Real channel measurements and real-time train position information were used for training the DQN model, which resulted in optimal communication and control decisions.

Furthermore, the resource allocation problems of wireless networks, such as energy scheduling, traffic scheduling, caching decisions, user association, etc. can be efficiently solved by deep reinforcement learning at a low computational complexity~\cite{zhang2017energy,xu2017deep,zhu2017new,he2018green,he2017deep,he2017optimization,he2018integrated}.
For example, Zhang \textit{et al.}~\cite{zhang2017energy} proposed a deep Q-learning model for system's dynamic energy scheduling, which relied on the amalgamated stacked auto-encoder and Q-learning model. More specifically, the stacked auto-encoder was used for learning the state-action value function of each strategy in any of the available system states. Moreover, Xu \textit{et al.}~\cite{xu2017deep} proposed a deep reinforcement learning framework for power-efficient resource allocation in CRANs, which optimized the expected and cumulative long term power consumption, including the transmit power consumption, the sleep/active transition power consumption as well as the RRU's power consumption. A two-step deep reinforcement learning aided decision making scheme was conceived, where the learning agent first decides on activating/deactivating the sleeping mode of each RRU, and then determines the optimal beamformer's power allocation.
As for traffic scheduling, Zhu \textit{et al.}~\cite{zhu2017new} designed a stacked auto-encoder assisted deep learning model for packet transmission planning in the face of multiple contending channels in cognitive IoT networks, which aimed for maximizing the system's throughput. In this architecture, MDP was used for modelling the system states. Given the large state-action space of the system, the stacked auto-encoder was used for constructing the mapping between the state and the action for accelerating the process of optimization.
Furthermore, a deep Q-learning algorithm was conceived for designing both the cache allocation and the transmission rate in content-centric IoT networks for the sake of maximizing the long-term QoE~\cite{he2018green}, where He \textit{et al.} considered both the networking cost as well as the users' mean opinion score. In~\cite{he2017deep} and~\cite{he2017optimization}, He \textit{et al.} proposed a DQN based user scheduling scheme for a cache-enabled opportunistic interference alignment (IA) assisted wireless network in the context of realistic time-varying channels formulated as a finite-state Markov model. More specifically, the DQN was constructed by relying on a sophisticated action-value function for the sake of reducing the computational complexity. Their simulation results demonstrated that the DQN aided IA assisted user scheduling was beneficial in terms of substantially improving the network's throughput vs energy efficiency trade-off.
To elaborate a little further, He \textit{et al.}~\cite{he2018integrated} utilized deep reinforcement learning for constructing their resource allocation policy relying on a joint optimization problem, which considered the programmable networking, information-centric caching as well as mobile edge computing in the context of connected vehicular scenarios. Moreover, the $\varepsilon$-greedy policy was utilized for striking an attractive trade-off between exploration and exploitation.

In order to curb the potentially excessive computational complexity resulting from having a large state space and to deal with its partial observability in cognitive radio networks, Naparstek and Cohen developed a distributed dynamic spectrum access scheme relying on deep multi-user reinforcement learning, where each user maps his/her current state to spectrum access actions with the aid of a DQN for the sake of maximizing the network's utility which was achieved without any message exchanges~\cite{naparstek2017deep}. Additionally, Wang \textit{et al.}~\cite{wang2018deep} proposed an adaptive DQN algorithm for dynamic multichannel access, which was capable of achieving a near-optimal performance outperforming the Myopic policy~\cite{ccetinkaya1998optimal}\footnote{The Myopic policy is one that simply optimizes the average immediate reward. It is called `myopic' in the sense that it merely considers the single criterion, while it has the advantage of being easy to implement.} and the Whittle's Index-based heuristic algorithms~\cite{ansell2003whittle}\footnote{The Whittle's Index-based algorithm is one of index heuristic algorithms which is designed to solve a problem in a more efficient way than traditional methods often used for solving NP-hard problems. More explicitly, the Whittle's Index policy is a low-complexity heuristic policy.} in complex scenarios.

Table~\ref{tbdeeprl} lists some typical applications of deep reinforcement learning in wireless networks.

\begin{table*}[t!]
\begin{center}
\scriptsize
\caption{Compelling Applications of Deep Reinforcement Learning in Wireless Networks}
\label{tbdeeprl}
\begin{tabular}{ | c | l | c | r |}
\hline
Paper &  Scenario &  Application & Description  \\ \hline
\cite{zhang2017learning} & UAV network &  trajectory control  &  a model-free UAV trajectory control scheme in smart cities relying on DQN\\ \hline
\cite{zhu2017communication} & ITS & train control &  jointly optimize the communication handoff strategy and control performances\\ \hline
\cite{zhang2017energy} &  energy-aware network  &energy scheduling & associate the stacked auto-encoder and the deep Q-learning model\\ \hline
\cite{xu2017deep} &  CRAN & power allocation & decide RRU's sleeping mode and the optimal beamformer's power allocation\\ \hline
\cite{zhu2017new} &  cognitive IoT &  traffic scheduling& construct the mapping between states and actions relying on stacked auto-encoder\\ \hline
\cite{he2018green} &  content-centric IoT & cache allocation & jointly design cache allocation and transmission rate for maximizing long-term QoE\\ \hline
 \cite{he2017optimization} &  IA network & user scheduling & obtain the action-value function relying on DQN for lowering complexity\\ \hline
\cite{he2018integrated} &   vehicular network & resource allocation & consider programmable SDN, information-centric caching and mobile edge computing\\ \hline
\cite{naparstek2017deep} &  CR network & spectrum access &  distributed spectrum access for maximizing network utility without message exchanges\\ \hline
\cite{wang2018deep} &  CR network & multichannel access &  adaptive DQN aided multichannel access yielding a near-optimal performance \\ \hline
\cite{yang2019deep} &  MEC network & resource allocation & DQN assisted cache, computation and power resources joint association\\ \hline
\cite{he2019deep} &  CR network & energy scheduling & energy efficiency optimization for distributed cooperative spectrum sensing \\ \hline
\cite{dai2019artificial} & vehicular network & resource allocation &  edge computing and caching resources dynamic association for IoV\\ \hline
\end{tabular}
\end{center}
\end{table*}

\section{Future Research and Conclusions}
\label{Future Research and Conclusions}
\subsection{Future Machine Learning Aided Network Applications}
It is anticipated that the next-generation network infrastructure will be migrating to a service-oriented architecture by taking advantage of both NFV and SDN for supporting eMBB, mMTC, and uRLLC along with attractive social media and augmented reality (AR) aided entertainment, smart manufacturing, smart energy, e-health and autonomous driving.

Given that ML can be readily executed both by agents, as well as by edge- and cloud-computing, it is eminently suitable for supporting sophisticated networking functionalities. We can categorize ML-based networking into:
\begin{itemize}
  \item \emph{ML-Aware Networking}: The networking entities are aware of the availability of ML functionalities and optionally might exploit advantages of ML, for example, when the user load escalates.
  \item \emph{On-/Off-line ML-Aided Networking}: The network functionalities must resort to invoking ML, for example, because the search-space of joint scheduling and resource-allocation is excessive for full search. Depending on the specific application, this may rely either on off-line ML or on on-line ML.
  \item \emph{On-line ML-Aided Networking}: The network functionalities must resort to on-line ML, for example, anticipatory mobility management requires on-line ML.
\end{itemize}

To facilitate prompt ML-aided network control, new tele-traffic samples may potentially be included for near-instantaneous learning according to the following use cases:
\begin{itemize}
  \item In contrast to traditional networking, near-instantaneous mobility trajectory inference relies on on-line datasets for training the ML module.
  \item Context information relying on location information inference or on user experience as a function of latency, jitter, packet dropping probability can be taken into account.
  \item Planning and policy relying on ML functionalities for improving a certain work flow, an agent's specific policy and/or rewards, along with the extracted features or learning models may be exploited, where the agent may be constituted either by a network entity or a smart UE.
\end{itemize}

To expound a little further, ML can be used for diverse application scenarios, such as:
\begin{itemize}
  \item \emph{Channel State Estimation}: Having accurate CSI is critical for all air-interface techniques, which has been inferred/estimated with the aid of deep learning.
  \item \emph{User Behavior Inference}: User behavior may be characterized for example in terms of their mobility patterns, while supporting high-quality network management and mobility management with the aid of big data analysis relying on ML.
  \item \emph{Traffic Prediction}: Network intelligence and user mobility patterns can be used for predicting the wired/wireless network traffic in support of efficient network/radio resource allocation.
  \item \emph{Network Security}: ML is also capable of enhancing the network security, whilst detecting attacks and intrusions.
  \item \emph{Anticipatory Networking Mechanism}: Online learning can be used to develop new predictive networking mechanism. For example, anticipatory mobility management (AMM) using Naive Bayesian and recursive belief update was conceived for ultra-low latency wireless networking~\cite{lin2018anticipatory,chen2018ultra}.
\end{itemize}

Based on the aforementioned future network architectural features, we list a range of research ideas on promising applications of ML in future wireless networks.
\begin{itemize}
  \item \emph{UAV-aided networking}: Given the agility of UAV nodes as well as the bursty and often unpredictable nature of terrestrial wireless traffic, ML models can be used both for predicting the traffic demand and for adaptively adjusting the UAVs' location.
  \item \emph{mMTC and uRLLC network}: While wireless networks have primarily served communications among individuals, in the era of the IoT, wireless networking also supports myriads of machines and intelligent devices. In this era a pair of 5G operational modes - namely mMTC and uRLLC - are expected to play key roles~\cite{soldani20185g}. ML is capable of enhancing conventional networks designed for mMTC, for example by invoking reinforcement learning to appropriately select the access points of MTC~\cite{liu2017enb}. The uRLLC mode of operation constitutes a rather young technological territory, which can be jointly designed with mMTC~\cite{lien2017efficient}. To reduce the network's latency from hundreds of milliseconds as experienced in state-of-the-art mobile communications to the desirable range of just a few milliseconds, ML is capable of supporting so-called anticipatory mobility management, which integrates the naive Bayesian classification of the previously used APs and geographical regression for the predictive analysis of data. Another example of disruptive technical trend is to investigate how wireless networking impacts on the smart agents operated by ML.
  \item \emph{Narrow band IoT (NB-IoT)}: NB-IoT allows a large number of low-power devices to connect to the cellular network, where the devices require long-term Internet access and dense wireless coverage. ML algorithms are capable of supporting intelligent resource allocation, optimal AP deployment and efficient access.
  \item \emph{Socially-aware wireless networking}: The operation of socially-aware wireless networks relies on a variety of social attributes, where ML schemes are beneficial in terms of facilitating feature extraction, social group-of-interest formation, classification and prediction of these social attributes, such as human mobility, social relations, behavior preference, etc.
  \item \emph{Wireless Virtual reality (VR) networks}: VR networks facilitate for the users to experience and interact with immersive environments, which requires flawless audio and video data processing capability. ML algorithms have the potential of circumventing the conception of complex joint source- and channel-coding schemes by further developing the auto-encoder principles.
  \item \emph{Network integration, representation and design}: ML may provide an alternative for network representation, where we can integrate each classic communication-theoretic blocks including source- and channel-encoding, modulation, demodulation, decoding, etc. into a ``black box''. By simply learning from and processing previous input and output signals, the receivers become capable of adaptively understanding the operational mechanism of the ``black box'' considered.
  \item \emph{Wireless network tomography}: State-of-the-art wireless networks support a vast number of nodes, such as those of the IoT, where the provision of global information is practically impossible for each node. Hence a new class of problems arises in the context of distributed wireless networks, which is related to the acquisition of network-related information. Classic network tomography~\cite{castro2004network} defines the problem as: ${\bf Y} = {\bf AX}$, where $\bf X$ is a $J$-dimensional vector of the network's dynamically fluctuating parameters, such as the link delay or traffic activity, $\bf Y$ is the $I$-dimensional vector of measurements and $\bf A$ is known as the routing matrix that is likely to define a rank-deficient scenario in network tomography. Such inverse matrix problems of inference may be solved by EM or Markov Chain Monte Carlo methods, and they are mathematically similar to the regression problems of ML. Network tomography can be applied to the inference of a wireless network's status and the subsequent optimization of the network's operation~\cite{sagduyu2017wireless}. Further generalization using statistical inference and singular value decomposition for statistically characterizing and enhancing the radio resource utilization relies on accurate spectrum sensing in cognitive radio networks~\cite{yu2010cognitive}. Another interesting application is to infer activities beyond sight (i.e. see-through walls) by applying variance-based radio tomography, Tikhonov least-squares regularization, and Kalman tracking~\cite{wilson2011see}.
  \item \emph{Large-scale Pareto-optimization}: In the above-mentioned optimization problems the designer is routinely faced by having to strike a trade-off amongst conflicting factors, as exemplified by the ubiquitous bandwidth- vs. power-efficiency trade-off in wireless communications. However, there are numerous other trade-offs to be taken into account, such as for example the data-security vs. integrity trade-off or the BER vs. delay trade-off, as well as the BER vs. complexity trade-off, just to mention a few. Hence it is desirable to find all the Pareto-optimal operating points of future wireless systems, which jointly constitute the optimal Pareto-front of multi-component optimization problems~\cite{fei2017survey}. Again, none of the solutions on the optimal Pareto-front may be improved without degrading at least one of the parameters. Naturally, as we mentioned, the optimization search-space is extended every time, when a new parameter is incorporated. Hence ML constitutes a promising tool for the fertile ground of multi-component Pareto-optimization of a plethora of systems.
\end{itemize}

\subsection{Future Wireless Networking Techniques for Multi-Agent Systems Using Machine Learning}
In addition to applying ML in wireless networking, the question of how to design a wireless network for example for multi-robot systems or for multi-agent systems arises.

In this context, Legg and Hutter~\cite{legg2007universal} gave an informal definition of machine intelligence: ``Intelligence measures an agent's ability to achieve goals in a wide range of environments.'' Distributed artificial intelligence has been brought into the lime-light of  artificial intelligence research over three decades ago~\cite{bond2014readings}, which has a pair of popular sub-disciplines: distributed problem solving and multi-agent systems. Distributed problem solving typically decomposes tasks into several not completely independent sub-problems that can be executed on different processors and then synthesizes a solution. On the other hand, multi-agent systems are typically constituted by robots, having certain goals and actions in challenging operating environments. Although the communication supporting the actions of agents has been studied for a while, under idealized simplifying assumptions, the features of realistic wireless communications and networking have not been taken into consideration~\cite{stein2012decentralised, ge2017distributed, zhou2016multiagent}.

An interesting study disseminated in~\cite{ko2018wireless} has considered the collective behavior of autonomous vehicles moving across Manhattan in New York city. Each autonomous vehicle acted as an intelligent agent using reinforcement learning. In contrast to human-to-human personal communication, the reward map and policy of another autonomous vehicle within the interaction range would be useful information to exchange. In this wireless networking scenario, the freshness of such information is critical hence ultra-low-latency wireless networking is preferred. In fact, for collaborative robots relying on their own machine learning algorithms while aiming for achieving a common goal, ultra-low-latency wireless networking is of paramount importance. Given the plethora of open research problems in networked multi-agent systems, such as their network topology, this is a compelling research area.

\subsection{Conclusions}
In summary, we have reviewed the thirty-year history of ML. We surveyed a range of typical algorithms and models conceived for supervised learning, unsupervised learning, reinforcement learning as well as deep learning, respectively. Furthermore, we have highlighted the development tendency of wireless network techniques and a variety of representative scenarios for future wireless networks as seen in Fig.~\ref{wireless develop} and Fig.~\ref{wireless network}. We also have provided a case-by-case description of numerous compelling applications relying on ML algorithms in wireless networks as shown in Table~\ref{allapplications}, followed by a pair of detailed application examples relying on our recent research results. In comparison with state-of-the-art survey papers seen in Fig.~\ref{timeline}, our paper overviews all the four popular kinds of learning schemes and their applications in future wireless networks, which has a full scope of how ML algorithms bear fruits in the past decades in wireless networks.
Indeed, powerful ML methods are poised to occupy an important position in tackling intractable and hitherto uncharted scenarios and applications in wireless networks.

\appendix
The applications of ML in wireless networks are summarized in Table~\ref{allapplications}.
\begin{table*}[htbp]
\renewcommand\arraystretch{0.89}
\scriptsize
  \centering
  \caption{Applications of ML in Wireless Networks}
\begin{tabular}{|l|l|l|l|c|}
    \hline
    \multicolumn{1}{|c|}{\textbf{Scenario}} & \multicolumn{1}{c|}{\textbf{Application}} & \multicolumn{1}{c|}{\textbf{\hspace{2mm} Learning Technique \hspace{2mm}}} & \multicolumn{1}{c|}{\textbf{Specific Method}} & \textbf{Rreference} \\
    \hline
    Ad hoc network & reactive routing & reinforcement learning & on-policy SARSA & [296] \\
    \hline
    body area network & power control & reinforcement learning & Q-learning & [299] \\
    \hline
    cellular network & dynamic channel allocation & reinforcement learning & reduced-state SARSA & [294] \\
    \hline
    cellular network & traffic prediction & deep learning & DNN   & [327] \\
    \hline
    cellular network & traffic prediction & deep learning & CNN   & [338] \\
    \hline
    cellular network & relay selection & unsupervised learning & K-means & [244] \\
    \hline
    cellular network & interference alignment  & deep learning & DQN   & [350] \\
    \hline
    centralized CR network & channel selection & supervised learning & SVM   & [227] \\
    \hline
    cognitive femtocell network & interference control & reinforcement learning & decentralized Q-learning & [302] \\
    \hline
    CR ad hoc network & signal detection & unsupervised learning & ICA   & [265] \\
    \hline
    CR ad hoc network & multi-agent sensing  & reinforcement learning & on-policy SARSA & [295] \\
    \hline
    CR network & data recovery & unsupervised learning & ICA \& PCA & [261] \\
    \hline
    CR network & relay selection & reinforcement learning & Q-learning & [301] \\
    \hline
    CR network & activity prediction & deep learning & RNN, CNN & [337] \\
    \hline
    CR network & spectrum access & deep learning & deep reinforcement learning & [352] \\
    \hline
    CR network & channel access & deep learning & DQN   & [353] \\
    \hline
    CR network & energy scheduling & deep learning & deep reinforcement learning & [357] \\
    \hline
    CR network & PHY representation & deep learning & DNN   & [324] \\
    \hline
    CRAN  & power allocation & deep learning & deep reinforcement learning & [346] \\
    \hline
    dense cellular network & link scheduling & deep learning & CNN   & [341] \\
    \hline
    energy-aware network & energy scheduling & deep learning & DQN   & [153] \\
    \hline
    heterogeneous network & resource management  & reinforcement learning & on-policy SARSA & [297] \\
    \hline
    heterogeneous network & packet routing & deep learning & DNN   & [328] \\
    \hline
    heterogeneous network & data estimation & supervised learning & H-SVM, SVM & [221] \\
    \hline
    heterogeneous network & gateway deployment & unsupervised learning & K-means & [243] \\
    \hline
    heterogeneous network & data prediction & supervised learning & SVM   & [224] \\
    \hline
    IoT network & traffic scheduling & deep learning & deep Q-learning & [347] \\
    \hline
    IoT network & cache allocation & deep learning & deep reinforcement learning & [348] \\
    \hline
    IoT network & traffic estimation & supervised learning & regression & [199] \\
    \hline
    IoT network & traffic offloading & deep learning & DNN   & [330] \\
    \hline
    M2M network & user detection & unsupervised learning & EM algorithm, Bayes method & [255] \\
    \hline
    MANET & behavior learning & supervised learning & SVM   & [225] \\
    \hline
    MANET & anomaly detection & unsupervised learning & K-means & [246] \\
    \hline
    MEC network & resource allocation & deep learning & DQN   & [356] \\
    \hline
    MIMO  & blind transceiver & unsupervised learning & K-means & [248] \\
    \hline
    MIMO  & channel estimation & unsupervised learning & EM algorithm, Bayes method & [250] \\
    \hline
    MIMO  & symbol detection & unsupervised learning & EM algorithm, Bayes method & [253] \\
    \hline
    MIMO  & channel estimation & unsupervised learning & ICA   & [262] \\
    \hline
    MIMO  & blind receiver & unsupervised learning & ICA   & [263] \\
    \hline
    MIMO  & PHY representation & deep learning & DNN   & [325] \\
    \hline
    MIMO  & modulation and coding  & reinforcement learning & Q-learning & [300] \\
    \hline
    mobile computing network & user location & supervised learning & SVM   & [223] \\
    \hline
    multiantenna CR network & primary user detection & unsupervised learning & EM algorithm & [251] \\
    \hline
    multiantenna CR network & channel state detection & unsupervised learning & EM algorithm & [252] \\
    \hline
    OFDM network  & channel estimation & deep learning & DNN   & [312] \\
    \hline
    P2P network & power allocation  & reinforcement learning & SARSA & [298] \\
    \hline
    smart grid network & network security & deep learning & DBN   & [332] \\
    \hline
    software-defined network & reosurce allocation & deep learning & DNN   & [326] \\
    \hline
    UAV network & trajectory control & deep learning & DQN   & [344] \\
    \hline
    vehicular network & traffic control & deep learning & deep reinforcement learning & [345] \\
    \hline
    vehicular network & traffic prediction & supervised learning & KNN   & [203] \\
    \hline
    vehicular network & resource allocation & deep learning & deep reinforcement learning & [351] \\
    \hline
    vehicular network & resource allocation & deep learning & deep reinforcement learning & [358] \\
    \hline
    Wi-Fi network & interference elimination & supervised learning & KNN   & [208] \\
    \hline
    WI-Fi network & indoor localization & deep learning & DNN   & [314] \\
    \hline
    Wi-Fi network & attacker counting & supervised learning & SVM   & [228] \\
    \hline
    Wi-Fi network & indoor location & unsupervised learning & PCA   & [258] \\
    \hline
    Wi-Fi network & wireless coexistence & supervised learning & logistic regression & [197] \\
    \hline
    wireless communication & signal detection & unsupervised learning & K-means & [249] \\
    \hline
    wireless communication & modulation classification & supervised learning & KNN   & [207] \\
    \hline
    wireless communication & interference cancellation & unsupervised learning & ICA   & [264] \\
    \hline
    wireless communication & modulation classification & deep learning & RNN, CNN & [321] \\
    \hline
    wireless communication & signal detection & deep learning & DNN   & [322] \\
    \hline
    wireless communication & interference identification & deep learning & CNN   & [313] \\
    \hline
    wireless communication & mobility prediction & deep learning & CNN   & [320] \\
    \hline
    wireless communication & mobility prediction & deep learning & RNN   & [334] \\
    \hline
    wireless communication & transportation mode & deep learning & RNN, DNN & [335] \\
    \hline
    wireless communication & PHY representation & deep learning & CNN   & [340] \\
    \hline
    wireless mesh networks & traffic control & deep learning & CNN   & [329] \\
    \hline
    wireless network & antenna selection & supervised learning & Bayes method, SVM & [232] \\
    \hline
    wireless network & QoE prediction & supervised learning & Bayes, SVM & [236] \\
    \hline
    wireless network & intrusion detection & unsupervised learning & K-means & [247] \\
    \hline
    wireless network & power control & deep learning & DNN   & [331] \\
    \hline
    wireless network & resource allocation & deep learning & DNN   & [339] \\
    \hline
    wireless sensor network & sensor partitioning  & unsupervised learning & \textit{K-mean} & [245] \\
    \hline
    wireless sensor network & interference estimate & supervised learning & regression & [195] \\
    \hline
    wireless sensor network & spectrum sensing & supervised learning & regression & [196] \\
    \hline
    wireless sensor network & PHY authentication & supervised learning & logistic regression & [198] \\
    \hline
    wireless sensor network & map reconstruction & supervised learning & regression & [200] \\
    \hline
    wireless sensor network & wireless localization & supervised learning & regression, SVM & [201] \\
    \hline
    wireless sensor network & anomaly detection & supervised learning & KNN   & [204] \\
    \hline
    wireless sensor network & missing data estimation & supervised learning & KNN   & [206] \\
    \hline
    wireless sensor network & localization estimation & supervised learning & SVM   & [222] \\
    \hline
    wireless sensor network & signal classification & supervised learning & SVM,DNN & [226] \\
    \hline
    wireless sensor network & network association & supervised learning & Bayes & [233] \\
    \hline
    wireless sensor network & network state detection & unsupervised learning & EM algorithm & [254] \\
    \hline
    wireless sensor network & source localization & unsupervised learning & EM algorithm & [256] \\
    \hline
    wireless sensor network & data aggregation & unsupervised learning & PCA   & [259] \\
    \hline
    wireless sensor network & data recovery & unsupervised learning & PCA   & [260] \\
    \hline
    wireless sensor network & sleeping scheduling & reinforcement learning & Q-learning & [303] \\
    \hline
    WLAN  & indoor location & supervised learning & Bayes method, SVM & [235] \\
    \hline
    WLAN  & anomaly detection & supervised learning & naive Bayes method & [234] \\
    \hline
  \end{tabular}
  \label{allapplications}%
\end{table*}%


%



%
%

\ifCLASSOPTIONcaptionsoff
  \newpage
\fi



\bibliographystyle{IEEEtran}
\bibliography{ref}

\vspace{-5mm}

\begin{IEEEbiography}[{\includegraphics[width=1in,height=1.25in]{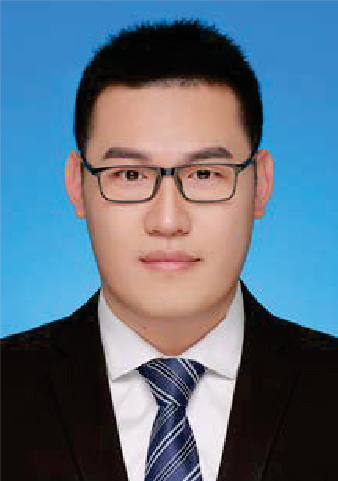}}]{\textbf{Jingjing Wang}} (S'14-M'19) received his B.S. degree in Electronic Information Engineering from Dalian University of Technology, Liaoning, China in 2014 and the Ph.D. degree in Information and Communication Engineering from Tsinghua University, Beijing, China in 2019, both with the highest honors. From 2017 to 2018, he visited the Next Generation Wireless Group chaired by Prof. Lajos Hanzo, University of Southampton, UK. Dr. Wang is currently a postdoc researcher at Department of Electronic Engineering, Tsinghua University. His research interests include resource allocation and network association, learning theory aided modeling, analysis and signal processing, as well as information diffusion theory for mobile wireless networks. Dr. Wang was a recipient of the Best Journal Paper Award of IEEE ComSoc Technical Committee on Green Communications \& Computing in 2018, the Best Paper Award from IEEE ICC and IWCMC in 2019.
\vspace{-5mm}
\end{IEEEbiography}

\begin{IEEEbiography}[{\includegraphics[width=1in,height=1.25in]{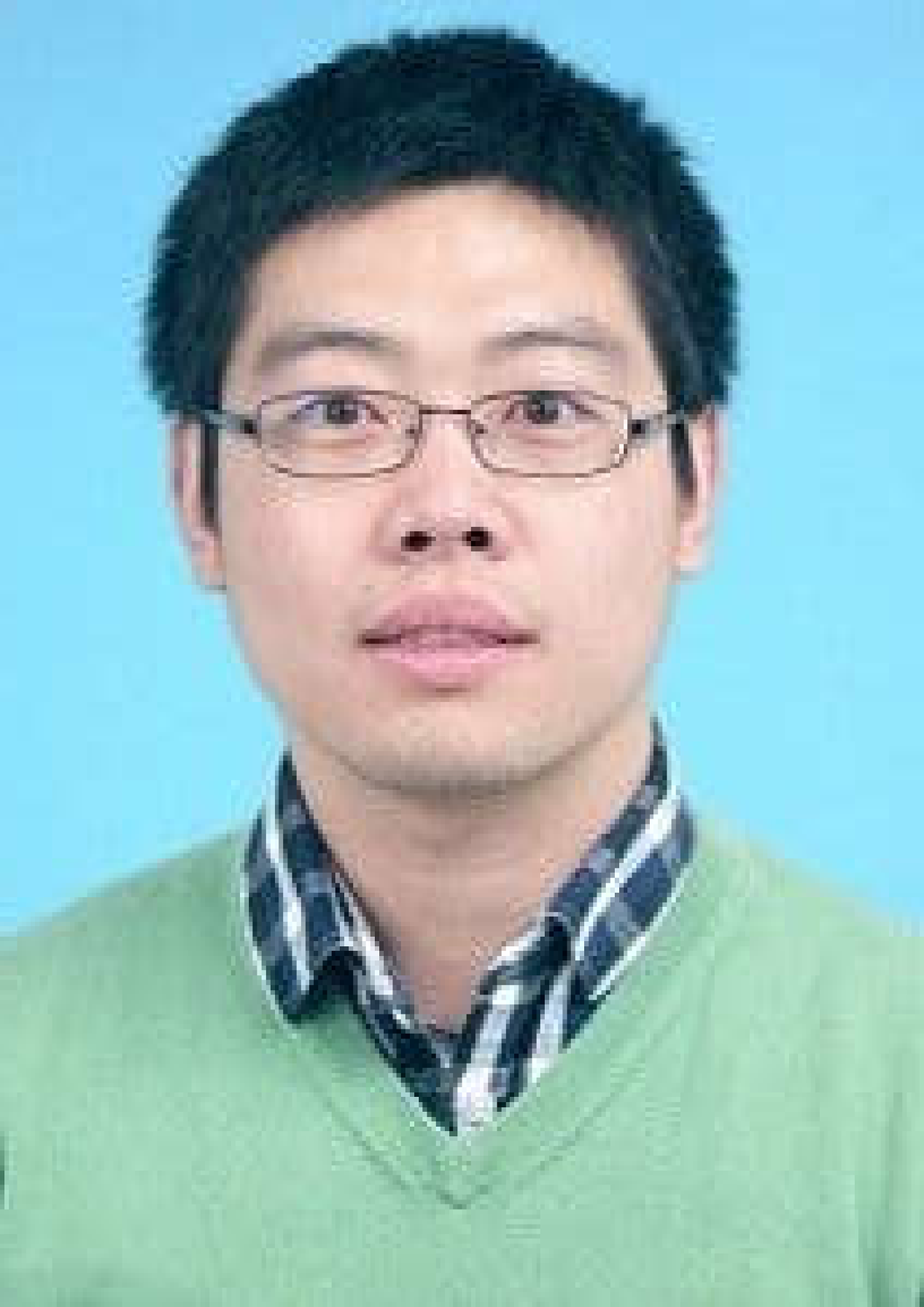}}]{\textbf{Chunxiao Jiang}} (S'09-M'13-SM'15) received the B.S. degree from Beihang University, Beijing in 2008 and the Ph.D. degree in electronic engineering from Tsinghua University, Beijing in 2013, both with the highest honors. From 2011 to 2012, he visited University of Maryland, College Park as a joint PhD supported by China Scholarship Council. From 2013 to 2016, he was a postdoc with Tsinghua University, during which he visited University of Maryland, College Park and University of Southampton. Since July 2016, he became an assistant professor in Tsinghua Space Center, Tsinghua University. His research interests include space networks and heterogeneous networks. Dr. Jiang is the recipient of the Best Paper Award from IEEE GLOBECOM in 2013, the Best Student Paper Award from IEEE GlobalSIP in 2015, IEEE Communications Society Young Author Best Paper Award in 2017, the Best Paper Award IWCMC in 2017, the Best Journal Paper Award of IEEE ComSoc Technical Committee on Communications Systems Integration and Modeling in 2018.
\vspace{-5mm}
\end{IEEEbiography}

\begin{IEEEbiography}[{\includegraphics[width=1in,height=1.25in]{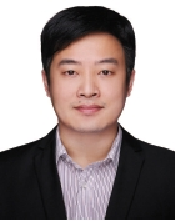}}]{\textbf{Haijun Zhang}} (M'13, SM'17) is currently a Full Professor in University of Science and Technology Beijing, China. He serves as an Editor of IEEE Transactions on Communications, IEEE Transactions on Green Communications Networking, and IEEE Communications Letters. He received the IEEE CSIM Technical Committee Best Journal Paper Award in 2018, IEEE ComSoc Young Author Best Paper Award in 2017, and IEEE ComSoc Asia-Pacific Best Young Researcher Award in 2019.
\vspace{-5mm}
\end{IEEEbiography}

\begin{IEEEbiography}[{\includegraphics[width=1in,height=1.25in]{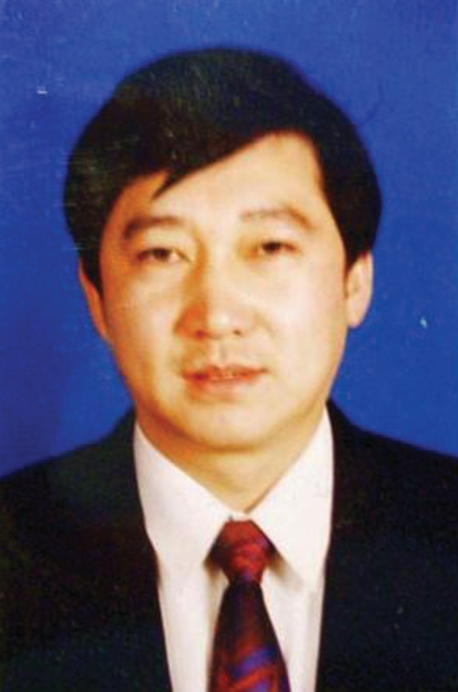}}]{\textbf{Yong Ren}} (M'11-SM'16) received his B.S, M.S and Ph.D. degrees in electronic engineering from Harbin Institute of Technology, China, in 1984, 1987, and 1994, respectively. He worked as a post doctor at Department of Electronics Engineering, Tsinghua University, China from 1995 to 1997. Now he is a professor of Department of Electronics Engineering and the director of the Complexity Engineered Systems Lab in Tsinghua University. He holds 60 patents, and has authored or co-authored more than 300 technical papers in the behavior of computer network, P2P network and cognitive networks. He has serves as a reviewer of IEICE Transactions on Communications, Digital Signal Processing, Chinese Physics Letters, Chinese Journal of Electronics, Chinese Journal of Computer Science and Technology, Chinese Journal of Aeronautics and so on. His current research interests include complex systems theory and its applications to the optimization and information sharing of the Internet, Internet of Things and ubiquitous network, cognitive networks and Cyber-Physical Systems.
\vspace{-5mm}
\end{IEEEbiography}

\begin{IEEEbiography}[{\includegraphics[width=1in,height=1.25in]{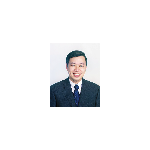}}]{\textbf{Kwang-Cheng Chen}} (F'07) received the B.S. degree from the National Taiwan University in 1983, and the M.S. and Ph.D. degrees from the University of Maryland, College Park, MD, USA, in 1987 and 1989, all in electrical engineering. From 1987 to 1998, he was with SSE, COMSAT, IBM Thomas J. Watson Research Center, and National Tsing Hua University, working on mobile communications and networks. From 1998 to 2016, he was a Distinguished Professor with the National Taiwan University, Taipei, Taiwan. He also served as the Director of the Graduate Institute of Communication Engineering, the Director of the Communication Research Center, and the Associate Dean for Academic Affairs with the College of Electrical Engineering and Computer Science, from 2009 to 2015. Since 2016, he has been a Professor of electrical engineering with the University of South Florida, Tampa, FL, USA. His recent research interests include wireless networks, artificial intelligence and machine learning, IoT and CPS, social networks, and cybersecurity. He has been actively involved in the organization of various IEEE conferences as the General/TPC Chair/Co-Chair, and has served in editorships with a few IEEE journals. He also actively participates in and has contributed essential technology to various IEEE 802, Bluetooth, LTE and LTE-A, 5G-NR, and ITU-T FG ML5G wireless standards. He has received a number of awards, including the 2011 IEEE COMSOC WTC Recognition Award, the 2014 IEEE Jack Neubauer Memorial Award, and the 2014 IEEE COMSOC AP Outstanding Paper Award.
\vspace{-5mm}
\end{IEEEbiography}

\begin{IEEEbiography}[{\includegraphics[width=1in,height=1.25in]{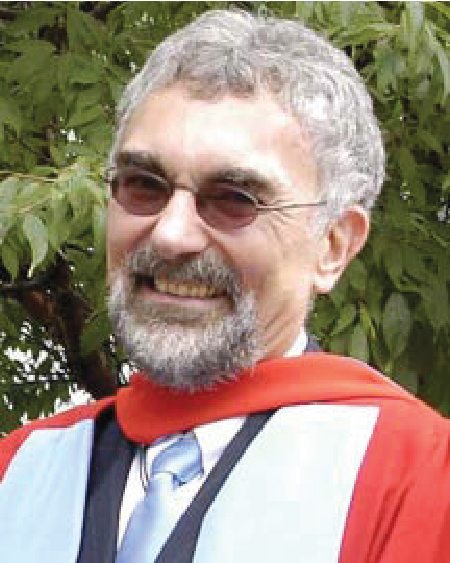}}]{\textbf{Lajos Hanzo}}
(\url{http://www-mobile.ecs.soton.ac.uk}, \url{https://en.wikipedia.org/wiki/Lajos_Hanzo})
FREng, FIEEE, FIET, Fellow of EURASIP, DSc received his Master degree
in electronics in 1976 and his doctorate in 1983. He holds an
honorary doctorate by the Technical University of Budapest (2009) and
by the University of Edinburgh (2015). He is a Foreign Member of the
Hungarian Academy of Sciences and a former Editor-in-Chief of the IEEE
Press.  He has served as Governor of both IEEE ComSoc and of VTS.  He
has published 1900+ contributions at IEEE Xplore, 18 Wiley-IEEE Press
books and has helped the fast-track career of 119 PhD students.
\vspace{-5mm}
\end{IEEEbiography}
\end{document}